\documentclass[11pt, english]{report}

\usepackage[margin=1in]{geometry}

\usepackage[utf8]{inputenc}
\usepackage[T1]{fontenc}
\usepackage{lmodern}
\usepackage[english]{babel}

\usepackage{amsmath}
\usepackage{amssymb}
\usepackage{amsthm}
\usepackage{mathtools}

\usepackage{graphicx}
\usepackage{float}
\usepackage{subfig}
\usepackage{booktabs}
\usepackage{tabularx}

\newtheorem{theorem}{Theorem}[chapter]
\numberwithin{theorem}{chapter}
\newtheorem{definition}{Definition}[chapter]
\newtheorem{lemma}{Lemma}[chapter]

\newcommand{\quotes}[1]{``#1''}

\title{Statistical Arbitrage in Polish Equities Market Using Deep Learning Techniques%
\thanks{This research was supported by the Polish National Science Centre (NCN) Grant 2019/35/D/ST6/03060.}
}

\author{
  Marek Adamczyk\\
  University of Wrocław\\
  \texttt{marek.adamczyk@cs.uni.wroc.pl}
  \\[0.8em]
  Michał Dąbrowski\\
  University of Wrocław\\
  \texttt{michaldabrowski1998@gmail.com}   
}

\date{} 

\begin{document}
\maketitle
\begin{abstract}
{We study a systematic approach to a popular Statistical Arbitrage technique of Pairs Trading. Instead of relying on $2$ highly correlated assets, the latter one is substitute with the most accurate replication of the first with the use of so called \textit{risk-factors}. Such factors can be determined by: Principal Components Analysis (PCA), actual market exchange traded funds (ETFs) or, as a authorial technique and thus our contribution to the literature, Long short-term memory networks (LSTMs). Residuals between the main asset and its replication' returns are analysed on a basis of their potential mean-reversion properties. Trading signals are later generated for sufficiently fast mean-reverting portfolios to profit from any technical mispricings.\\
 Besides the introduction of a new deep-learning based method, paper re-defines methods already presented by authors of 2008's paper \textit{Statistical Arbitrage in the U.S. Equities Market} to match conditions of the polish stock exchange market. For that reason, instead of \textit{SP500} stocks', components of \textit{WIG20} and \textit{mWIG40} combined are in scope of trading activities with an addition of polish sector indices. Overall market factors such as the risk free rate or transaction costs are also adjusted from mentioned paper for better reality matching.\\
 After setting up the scope, all details of the strategy are explained: from the theory behind risk-factors representation, through the modelling of residuals with Ornstein-Uhlenbeck process till trading signals generation procedure. They are followed by a separate section concerning specifics of each replicating technique with a general overview of the method and its application for our purposes. Throughout the entire thesis various examples are graphically made for better understanding of discussed topics. The final part of the paper concerns testing of the overall Pairs Trading strategy and of its presented variations.\\
 To keep the results relevant and tested in different economic conditions, two backtesting periods are distinguished: 2017-2019 and a highly recessive 2020. All strategies manage to profit during the first interval with the PCA approach achieving around $20\%$ of combined return and even up to $2.63$ annualized Sharpe ratio (in 2017). Even though a lot of assumptions is changed in comparison to Avellaneda and Lee' 2008 paper, received results and main conclusions are highly comparable. During the COVID-19 recession, ETFs technique are the only profitable one achieving annual return of $5\%$- both the PCA and LSTM methods fail to produce any profits. All LSTM results can be seen as promising and should be optimized in future works, especially since it is possibly the first take on such application of recurrent neural networks.}
\end{abstract}
\tableofcontents
\clearpage

\chapter{Introduction}
No matter if you are an amateur investor or own a hedge fund: the main goal of trading can be reduced to increasing the initial capital invested (or at least not losing it in comparison to the overall market movement). Then the actual process may be seen as a sequence of transactions following the motto of \quotes{buy low, sell high}. Obviously, the complicated part resolves around the actual decision-making: what equity should I purchase? How much of it should I have? Is now a good moment to get rid of it? Well, it is good if the price is higher than when I bought it but perhaps it is better to wait even longer. To somehow take those decisions off the shoulders of pure randomness and help making up ones mind $2$ main types of analysis were established. The first one, which will not be the area of our interest, is the \textit{Fundamental Analysis}- it resolves around using the knowledge on the current economic status, industry conditions and financial strength of individual companies. The aim here is to measure the intrinsic (i.e. \quotes{fair}) values of stocks and compare those with the actual prices on the market. Based on whether they are seen as undervalued or overvalued transactions can be determined. But, as mathematicians and not economists we are actually going to focus on the second type of financial analysis- \textit{Technical Analysis}. It was introduced in the late 1800s by Charles Dow- founder and first editor of the \textit{Wall Street Journal}. Dow believed that all necessary info is discounted into the actual price (and the volume of transactions) so there is no need to seek for qualitative factors of companies' conditions. He suggested that stocks' movements can be decomposed based on what period they reflect: besides the main trend expressing the events of last years, shorter ones, which are sensitive even up to days, can be distinguished. Dow stated that such trends exist despite the presence of so-called \quotes{market noise}- unpredictable behaviour independent of the overall drifts. His theory also introduced what we now understand as market indices- a way of looking at stocks' prices through companies' sectors. If particular stock's movement is diverging from the overall sector drift one may expect an incoming change. In accordance with Dow's theory, we can describe \textit{Technical Analysis} as the study of trends. The aim is to use historical data and statistical methods to separate trends from the noise and predict recessions before they actually happen. It is clear then that Charles Dow did not actually introduce any methods of technical analysis- he prepared the necessary fundamentals to justify appropriateness of such approaches.\\
We usually associate term \textit{arbitrage} with the gap between single asset prices on different stock exchange markets. The classical economical theory states that if markets are efficient there should be no possibility of arbitrage but in reality it is not always the case. To profit from arbitrage is then to buy such asset for a lower price and instantly sell it for the higher one. Such behaviour of many traders (\textit{arbitrageurs}) leads to practically automatic correction of any mispricings. Therefore, one does not actually needs to consider two separate markets- it is sufficient enough to seek arbitrage opportunities on a single market and profit from the corrections. At the same time, if there is no \quotes{anchoring} price to compare with one needs to construct it in a more theoretical way. Techniques of doing it are referred to as \textbf{Statistical Arbitrage}. To somehow illustrate it, the simplest approach would be to compare the current price with a possibly weighted average of historical prices. If the overall trend is not strong, any significant deviations from such average may be seen as arbitrage opportunity. In the presence of a trend, the mean can be replaced with a sloping line representing the drift. These are not so sophisticated methods because it is usually not the case that trend of the market is a linear one. But since we are considering trends and historical prices to reflect the future, Statistical Arbitrage is seen as a great example of Technical Analysis.\\
The following paper aims to present $3$ techniques of Statistical Arbitrage. They will all focus on constructing portfolios that separate market and sector trends from movements that are unique to a given stock. With such portfolios arbitrage opportunities can be found with the assumption of their mean-reversion properties. The main difference between discussed techniques is going to be the way the market components are constructed- we will consider \textbf{Exchange Traded Funds (ETFs)} representing market indices together with factors made artificially (since the first ones may not be sufficient enough). Such factors are going to be constructed with the use of \textbf{Principal Components Analysis} of the returns and, in the final and most sophisticated strategy, based on \textbf{Recurrent Neural Networks} simulations. Then, the ultimate goal will be to rate the performance of Statistical Arbitrage in general and compare different approaches by trading on historical data of the polish equities market- especially in the context of the general market recession following the events of the COVID-19 pandemic.\\
Even though the actual results of considered strategies will pose as the final part of the paper it is important to set up the economic scope of our considerations. We are also going to use real examples to illustrate theoretical concepts throughout the entire paper. Therefore, let us start with an overall insight into the polish equities market.
\section{Polish equities market in the scope of our consideration}
Polish equities market is centred around Warsaw Stock Exchange Market (GPW). Total capitalization, which is the summed value of all outstanding stocks on the market, resolves around $300$ bln EUR- this makes GPW the biggest stock exchange in central and eastern part of Europe. As of July 2023, Warsaw Stock Exchange Market consisted of $415$ domestic and foreign companies. It is not only dependent on polish economy but also on global markets- especially American and German (NYSE and FWB consecutively) with the latter one being the biggest in European Union. In contradiction to those markets there are not so many indices on GPW. They are all variations of what will become the main subject of validating our strategies- \textit{WIG}.
\subsection{\textit{WIG} and its variations}
\begin{figure}[H]
\centering
\includegraphics[scale=0.5]{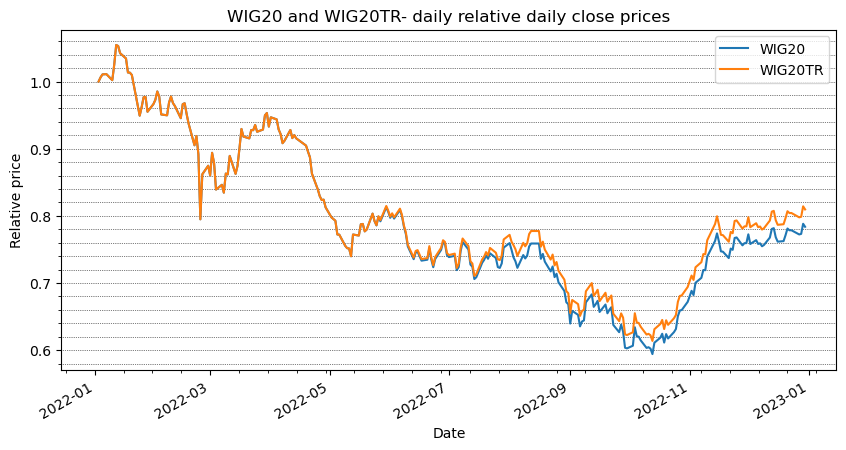}
\caption{\textit{WIG20} and Total Return {WIG20TR} daily close prices throughout 2022}
\label{fig:wig20tr-years}
\end{figure}

\textit{WIG} (whose acronym stands for \textit{Warszawski Indeks Giełdowy}) is the oldest index on the Warsaw Stock Exchange Market dating back in April 1991. It consists of all companies on the main market with at least $10\%$ of the shares trading for not less than $1$ mln EUR. Because of that the index serves as an indicator of Poland's overall condition. \textit{WIG} is calculated as the total value of its companies' stocks relative to day the index was issued (when it was priced as $1000$ points).  One can consider its value as worth of \quotes{the entire main market} portfolio and for that reason all cash payments such as dividends and interest are included and assumed to be reinvested. That makes \textit{WIG} the Total Return Index- a more appropriate way of expressing possible profits coming from investing in index's participants than a Price Index which only includes stocks' raw prices.
\begin{figure}[H]
\centering
\includegraphics[scale=0.5]{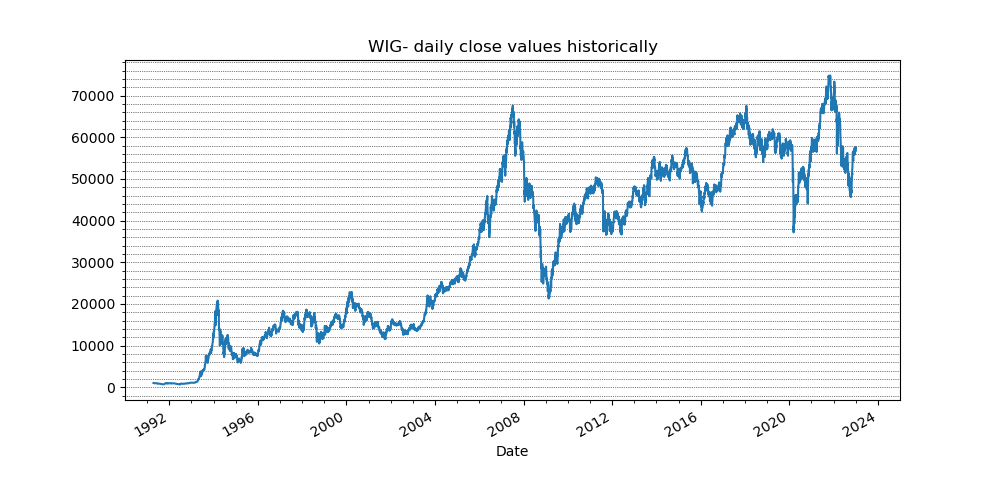}
\caption{\textit{WIG} daily close prices throughout years}
\label{fig:wig-years}
\end{figure}
\noindent Figure \ref{fig:wig-years} shows entire history of the index. We can clearly identify the positive impacts such as the start of Poland's European integration in 1994 or an exponential growth after joining EU, but also the negative ones: 2008 global crisis and COVID-19 pandemic.\\
As mentioned above \textit{WIG} is constructed out of practically all members present on the polish equities market. Additionally, the influence of a single company cannot exceed $10\%$ and each sector can participate only up to $30\%$- to keep such requirements (and to include new participants- their number is unlimited) index is updated quarterly. Even with such restrictions \textit{WIG} is mainly influenced by the \quotes{big sharks}- top $5$-$10$ biggest capitalization companies on the market. For that reason index finds it hard to follow movements of sectors that are not so crucial to the domestic economy status- the perspective is just too broad. There are actually no exchange traded funds (they will be described in greater detail later- for now one may assume that these are portfolios of stocks that you can purchase) tracking \textit{WIG}- this makes it even more of theoretical indicator rather than an actual portfolio to follow. As a way of \quotes{fixing} mentioned weaknesses \textit{WIG}'s variations were introduced. They can be separated into 2 categories: rating indices and sector indices. The first one focuses on indices gathering \textit{WIG}'s companies based on a predefined, quarterly adjusted ranking calculated with current capitalization and yearly turnover total value. Here are the $3$ most popular ones:
\begin{itemize}
\item \textit{WIG20};
\item \textit{mWIG40};
\item \textit{sWIG80}.
\end{itemize}
\textit{WIG20} gathers first $20$ places in the ranking, \textit{mWIG40}- places from $21$ to $60$ and \textit{sWIG80} consists of biggest $80$ companies that did not make it to \textit{WIG20} or \textit{mWIG40}. In contrast to \textit{WIG} these are all Price Indices and for that reason they are accompanied with Total Return equivalents (\textit{TR} name extension- see Figure \ref{fig:wig20tr-years}). Ranking indices are way easier to replicate than \textit{WIG} because of smaller components' number. Although \textit{WIG20}'s behaviour is very similar to \textit{WIG}'s- considering for example \textit{sWIG80} can introduce a new perspective of smaller companies' current status. In each index, every company has a certain share that reflects its capitalization and therefore directly corresponds to position in the ranking. Then, the final index's value can be seen as a weighted average of participants' stock prices. Since structurally speaking ranking indices have a lot in common, let us just focus on the most common one- \textit{WIG20}.\
Like \textit{WIG}, \textit{WIG20} is also relative to its initial $1000$ points value dating back in 1994. It reflects weighed average price of top $20$ biggest, most liquid polish companies. 
\noindent Figure \ref{fig:wig20tr-years} shows \textit{WIG20}'s historical behaviour together with its Total Return version which is being calculated since 2012. Dividends are usually paid mid-year, that is why \textit{WIG20TR} diverges from \textit{WIG20} in the second half of 2022. Coming back to the previous plot, it is also noticeable that \textit{WIG20} behaves very similarly to \textit{WIG}- only the magnitudes are significantly different. \textit{WIG20}'s components are revised quarterly together with a new ranking version.
\begin{figure}[H]
\centering
\includegraphics[scale=0.7]{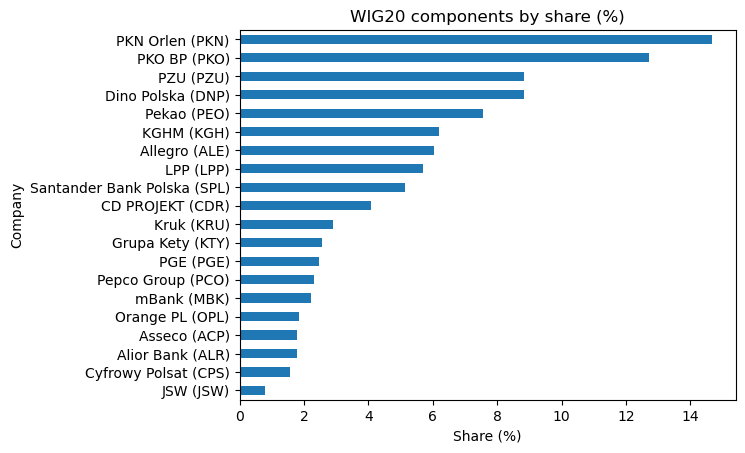}
\caption{\textit{WIG20} companies by share in index (as of July 2023)}
\label{fig:wig20-share}
\end{figure}
\noindent On Figure \ref{fig:wig20-share} we can see index's participants distribution. Share of a single company cannot exceed $15\%$. Top $5$ companies in total have more than $50\%$ share in the index- they can be seen as the main trend drivers. Quarterly corrections of the participants usually concern lower places' companies with very small shares thus some of them can be skipped while replicating the index. Since the beginning of 2022 the only changes were: the inclusions of  Alior Bank (ALR), Tauron Polska (TPE), Mercator Medicalthe (MRC)  for mBank (MBK), Pepco (PCO) and CCC (CCC) consecutively. Below we are going to consider plots analysing \textit{WIG20} and its components based on a full year circle of 2022- for simplicity and based on the reasoning above current squad of the index will be considered throughout the entire period.
\begin{figure}[H]
\centering
\includegraphics[scale=0.6]{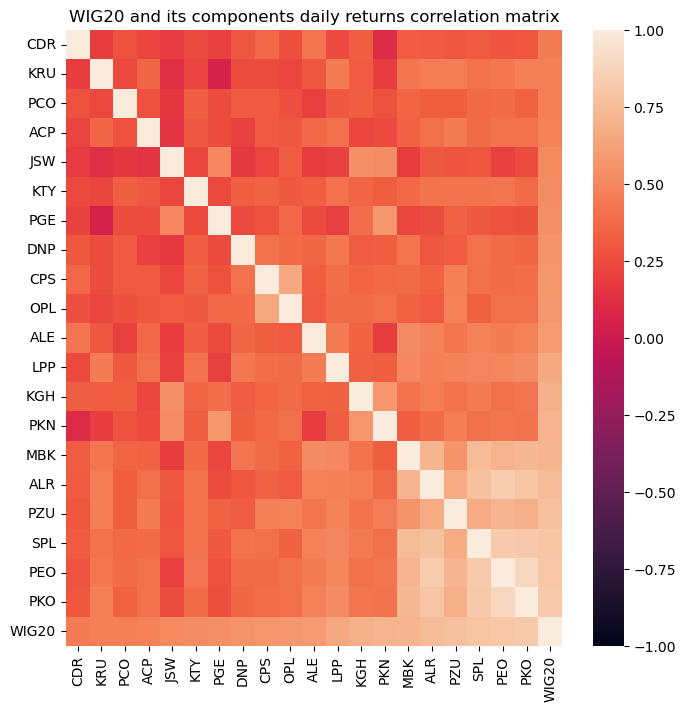}
\caption{\textit{WIG20} and its components' daily returns correlation matrix}
\label{fig:wig20-corr}
\end{figure}
\noindent Returns understood as close raw price difference quotients with $1$ trading day as $dt$ for both the index and its companies were gathered to calculate a correlation matrix between them (returns are usually assumed to be Gaussian which means that their similarities can be explained in such way). It can be seen on Figure \ref{fig:wig20-corr} with color gradings representing magnitudes- components are ordered based on their correlation to the index itself. It is clear that all correlations are either strongly positive- assets are making profits in parallel; or almost zero- there is no similarity between companies' trends. It is worth noticing that the strongest \textit{WIG20}-correlated participants are banks. All of them are also highly correlated with each other and non-bank members. The main reason is that all companies are heavily dependent on financial institutions providing them loans and deposits. Because of these and similar connections, transition between share in index and the impact of a single company is not quantitatively ideal (yet still visible). The most uncorrelated company here is CD Project Red (CDR)- games developer more sensitive to game release dates or media trends rather than on the domestic economy.
\begin{table}[H]
\centering
\begin{tabular}{lrrr}
\toprule
Close price \\
Stat &    Mean &     Std & Relative Std \\
Ticker &         &         &              \\
\midrule
PKN    &   62.67 &    5.02 &         0.08 \\
PKO    &   28.87 &    4.27 &         0.15 \\
PZU    &   32.77 &    5.43 &         0.17 \\
DNP    &  374.86 &   44.45 &         0.12 \\
PEO    &   80.80 &   12.89 &         0.16 \\
KGH    &  114.58 &   16.19 &         0.14 \\
ALE    &   27.77 &    4.27 &         0.15 \\
LPP    & 9827.32 & 1644.26 &         0.17 \\
SPL    &  277.85 &   52.22 &         0.19 \\
CDR    &  118.34 &   20.52 &         0.17 \\
KRU    &  310.97 &   49.45 &         0.16 \\
KTY    &  533.77 &   42.65 &         0.08 \\
PGE    &    7.18 &    1.29 &         0.18 \\
PCO    &   38.19 &    3.62 &         0.09 \\
MBK    &  295.38 &   61.10 &         0.21 \\
OPL    &    6.56 &    0.59 &         0.09 \\
ACP    &   72.97 &    5.55 &         0.08 \\
ALR    &   35.16 &    7.64 &         0.22 \\
CPS    &   17.69 &    1.07 &         0.06 \\
JSW    &   47.68 &    7.68 &         0.16 \\
\bottomrule
\end{tabular}
\caption{\textit{WIG20}'s components daily close prices statistics (based on 2022's data)}
\label{tab:1}
\end{table}
\noindent Stocks of index's members are very hard to compare price-wise because of different magnitudes. Table \ref{tab:1} shows their basic sample statistics from the last year. Relative standard deviation is defined as
$$\sigma_{\text{rel}}=\frac{\sigma}{\mu},$$
where $\mu,\sigma$ are mean and standard deviation. Its value gives us a better understanding of data's instability. LPP's stock stands out as far the most expensive one- number of shares needed for index's participation at around $6\%$ is thus much lower than for other companies. Relative standard deviations are mostly low for all members- because of their sizes and liquidity they are usually stable. To compare components' prices movement throughout the last year we are going to consider daily prices in a relative way i.e. considering changes of $1$ PLN worth of every asset purchased at the beginning of January 2022. Dividends nor any other stock ownership's benefits are not going to be taken into account here.
\begin{figure}[H]
\centering
\includegraphics[scale=0.4]{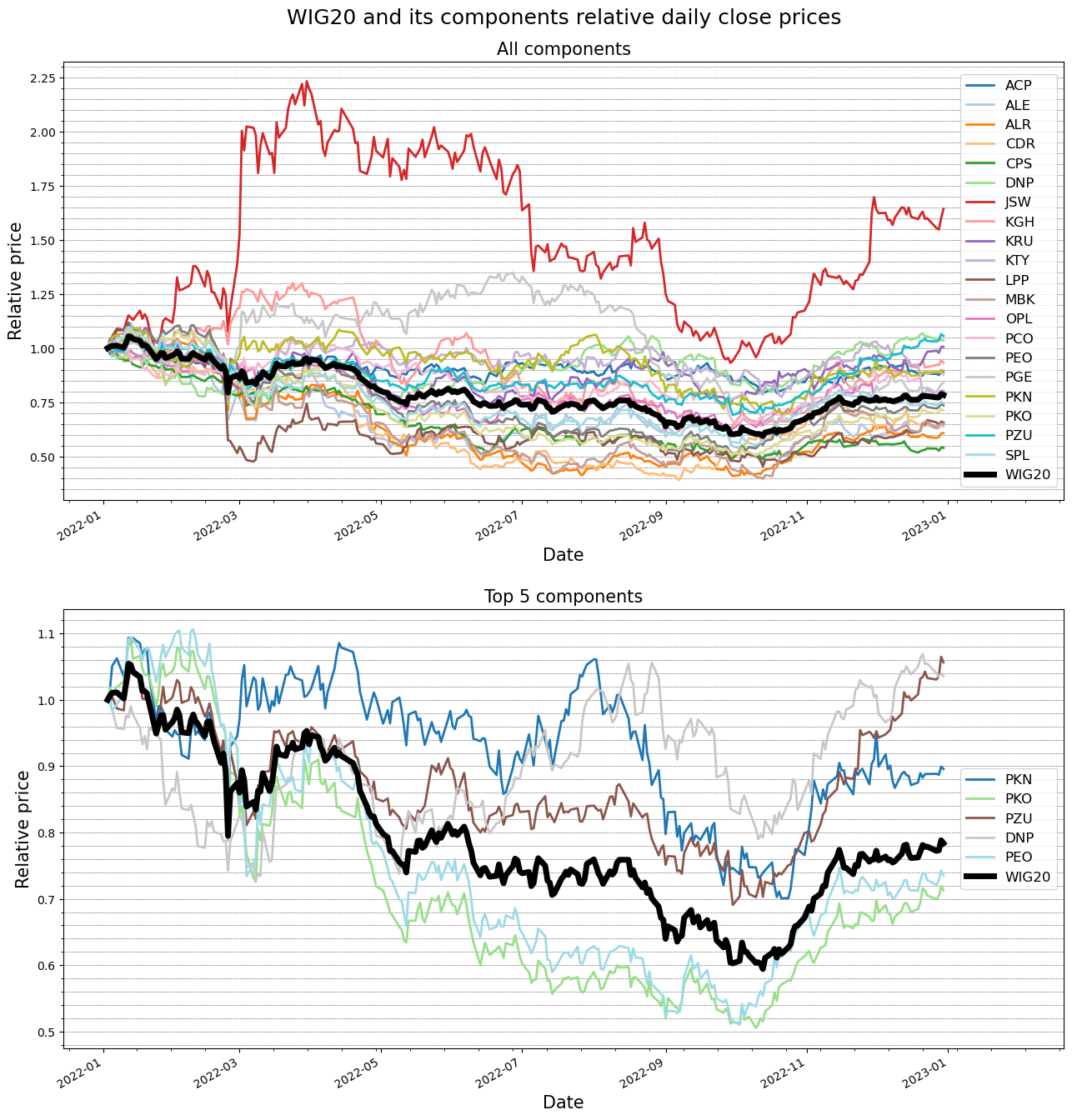}
\caption{\textit{WIG20} and its components' relative daily close prices}
\label{fig:wig20-years}
\end{figure}
\noindent Figure \ref{fig:wig20-years} consists of $2$ plots: the first one aims to present all components together with the index itself. We can see that \textit{WIG20}'s relative value lays in the middle of components' ones. It can be viewed as quite intuitive since the index was suppose to average all prices. This is also the reason why index' movement is not as volatile as for the components. Latter plot focuses only on the top $5$ companies of the index. Although as we already noticed their returns are not necessary as strong \textit{WIG20}-correlated as the share would suggest, relative prices are following very similar movements to the index. At the same time, each stock has its own unique \quotes{noise} not explainable by \textit{WIG20}.\\
\textit{WIG20} as a Price Index does not have an Exchange Traded Fund but there exists one covering \textit{WIG20TR}- \textit{BETA ETF WIG20TR}. It tracks the Total Return Index value by directly replicating its portfolio, reinvesting all received benefits and adding a small share of derivative instruments for additional stability.
\begin{figure}[H]
\centering
\includegraphics[scale=0.6]{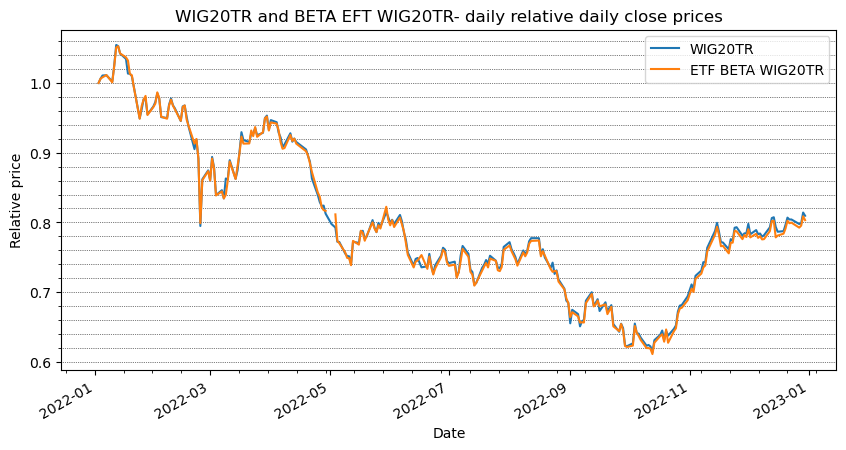}
\caption{\textit{WIG20TR} and its tracking index relative daily close prices}
\label{fig:wig20etf-years}
\end{figure}
\noindent Following the same approach as with presenting \textit{WIG20} and its components' prices, Figure \ref{fig:wig20etf-years} aims to show how close \textit{BETA ETF WIG20TR} is from the actual index. Other $2$ main ratings indices mentioned above follow the same pattern- they do have ETFs tracking their Total Return versions.\\
Let us recall that on the polish stock exchange market we can distinguish $2$ main types of indices. We just described rating indices together with their most common example- \textit{WIG20}. The second group gathers so called sector indices. As the name suggests they focus on all companies from given sector. For each industry subindex shares' packages of its participants are identical to ones of \textit{WIG}. Another similarity with the main index is that sector indices are all Total Return ones. Number of participating companies is changing with inclusions and exclusions dependent only on presence in \textit{WIG}.
\begin{table}[H]
\scriptsize
\centering
\begin{tabular}{llr}
\toprule
{} &        sector &  number of members \\
name       &               &                    \\
\midrule
\textit{WIG-BANKI}  &         banks &                 13 \\
\textit{WIG-BUDOWN} &  architecture &                 34 \\
\textit{WIG-CHEMIA} &     chemistry &                  5 \\
\textit{WIG-ENERG}  &        energy &                 12 \\
\textit{WIG-GORNIC} &        mining &                  5 \\
\textit{WIG-GRY}   &         games &                 20 \\
\textit{WIG-INFO}   &   informatics &                 26 \\
\textit{WIG-LEKI}   &        pharma &                  8 \\
\textit{WIG-MEDIA}  &         media &                 12 \\
\textit{WIG-MOTO}   &          moto &                  4 \\
\textit{WIG-NRCHOM} &   real estate &                 22 \\
\textit{WIG-ODZIEZ} &       clothes &                 15 \\
\textit{WIG-PALIWA} &         fuels &                  3 \\
\textit{WIG-SPOZYW} &          food &                 19 \\
\bottomrule
\end{tabular}
\caption{Main polish sector indices with their components' numbers (as of July 2023)}
\label{tab:sectors}
\end{table}
\noindent Table \ref{tab:sectors} presents a representative mixture of sector indices for different aspects of polish economy. As can be seen their members' number varies from just $3$ (\textit{WIG-PALIWA}) to $34$ (\textit{WIG-BUDOWN}). We will not analyse sector indices separately- instead let us consider them jointly.\\
Since exact squads of each index are not going to be considered we can look at the last $3$ calendar years of historical data instead of just $1$ year like we did with \textit{WIG20}. This should give as more insight in subindices' similarities and unique behaviours.
\begin{figure}[H]
\centering
\includegraphics[scale=0.5]{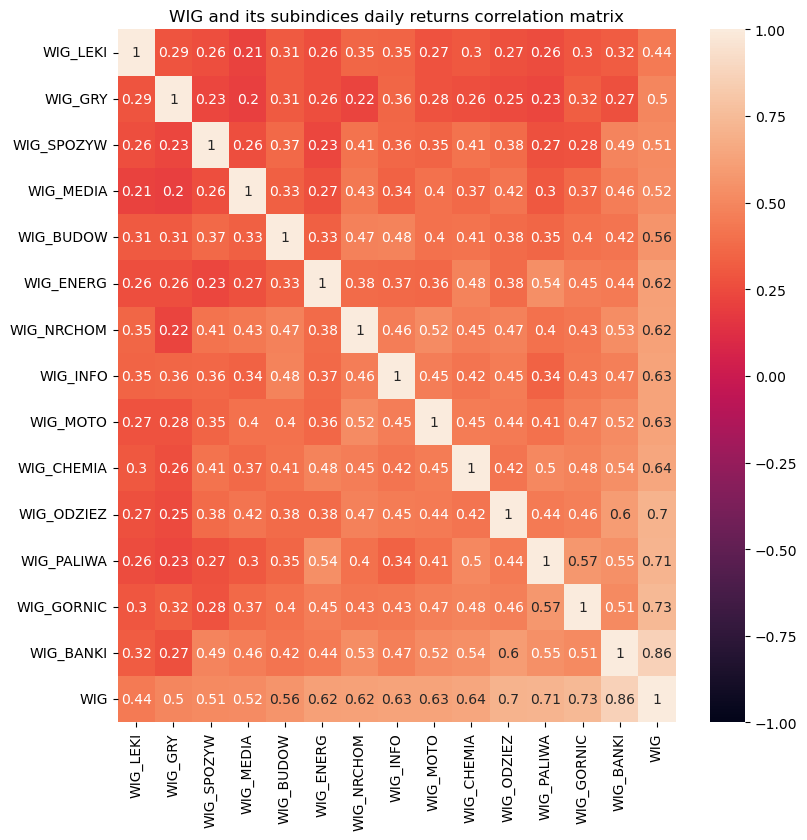}
\caption{\textit{WIG} and its subindices' daily returns correlation matrix}
\label{fig:wigsector-corr}
\end{figure}
\noindent Figure \ref{fig:wigsector-corr} presents correlations between subindices' daily returns. \textit{WIG} was also included and sector indices were sorted based on their correlations with it. Correlations around $0.4-0.5$ suggest a basic, common trend between indices interrupted by unique sector movements. \textit{WIG-BANKI} is by far the most correlated one with \textit{WIG}. It is not a surprise since financial institutions gathered in this subindex have one of the biggest capitalizations on polish exchange market (most of them were included in \textit{WIG20}). Some intuitive correlations can be spotted such as slightly higher one between \textit{WIG-PALIWA} (fuels) and \textit{WIG-ENERG} (energy) indices. \textit{WIG-GRY} (games) and \textit{WIG-LEKI} (pharma) are by far the most independent ones in comparison to the others. The first one, as already mentioned in terms of its leading company CD Project Red, is not so sensitive to polish domestic economy, while the latter one was heavily influenced by COVID-19 vaccines production during 2020-2022.
\begin{figure}[H]
\centering
\includegraphics[scale=0.4]{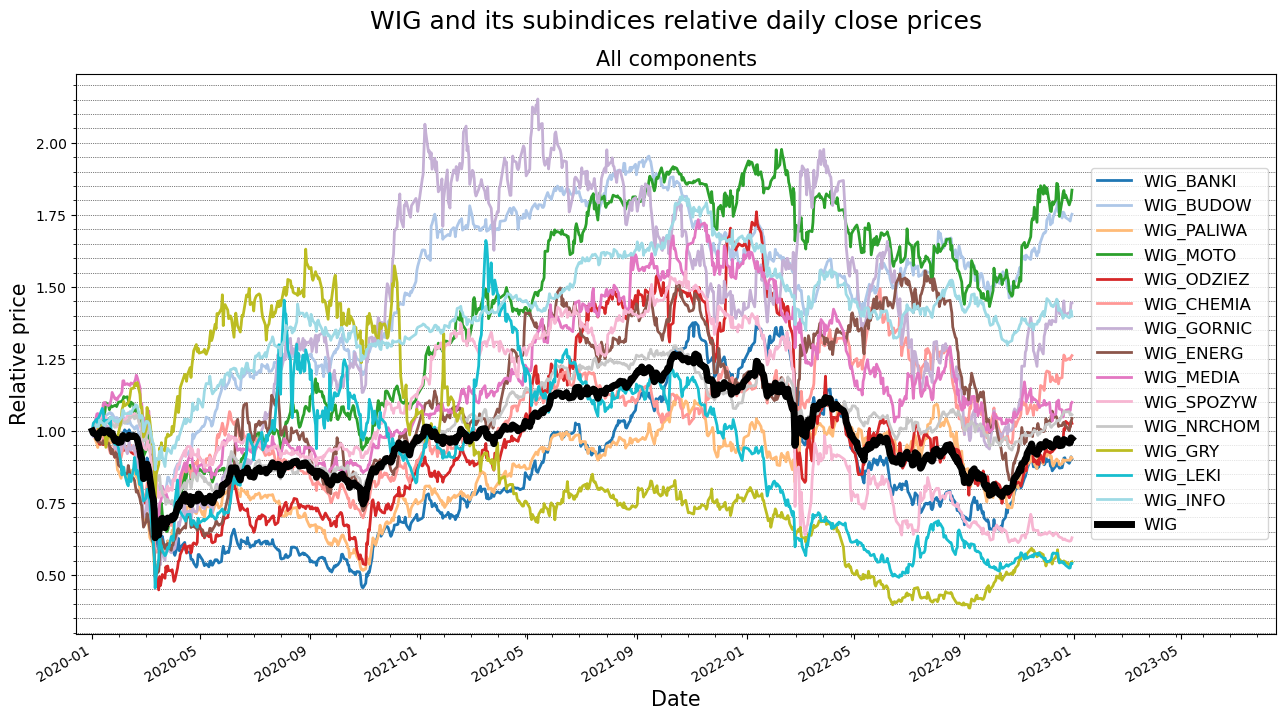}
\caption{\textit{WIG} and its subindices' relative daily close prices}
\label{fig:wigsector-years}
\end{figure}
\noindent Relative close prices were again calculated, this time for sector indices- results can be seen on Figure \ref{fig:wigsector-years}. All sector indices are less stable than \textit{WIG}- each of them has unique deviations which can be interpreted analysing historical events- nevertheless some major, common trends are visible for the entire spectrum. Considered sector indices do not have ETFs tracking them. For that reason artificial funds are going to be used assuming that portfolio replication relies on us.\\
After setting up all the necessary characteristics of polish equities market, let us now consider how are we planning to profit from it.
\chapter{Theory of Statistical Arbitrage}
The following section aims to give an insight into statistical arbitrage strategy of \textbf{Pairs Trading}. Firstly, we are going to consider the basic concept of the technique together with our way of tackling it. Then, details of each strategy's step will be discussed on a theoretical and practical level.
\section{Pairs trading- what is it?}
Recall that we considered \textit{arbitrage} as an inconsistency of asset's price between $2$ markets. Assuming there exists what we would call a \quotes{fair} price of an asset, at least one of the markets is over- or undervaluing it. Even in absence of such \quotes{fair} value, one can still profit from buying with the lower price and selling with the higher one on the second market (which overvalues considered asset relatively to the first one). So, are we going to observe different stock exchanges as our sophisticated strategy? Well, not really- asset price on different markets is just the simplest example of $2$ highly correlated time series that in principle should converge. If instead of same asset's second market price, one used price of a different asset that behaves very similarly to the first one, any major gaps between them would be seen as temporary. Profiting from corrections of such discrepancies, no matter how the whole market performs, is where arbitrage meets the idea of \textbf{Pairs Trading}.\\
Concept of looking for equities that \quotes{move together} and trading them whenever their prices diverge was first introduced by Wall Street quant Nunzio Tartaglia. He developed multiple statistical trading strategies and pairs trading turned out to be one of the most successful ones- Morgan Stanley, the company that he worked in at that time, made a $50$ million \$ profit by using it in 1987\cite{pt_history}. Even though the following years were not so profitable for the fund, pairs trading gained popularity as a counter to mainstream, \quotes{intuition} trading. Additionally, due to its market-neutrality (independence or low correlation with current market trends) together with other statistical methods pairs trading was one of the best performing strategies during the \quotes{Black Monday} market crash in 1987 (one of the biggest crises in American's history). Next $30$ years developed Tartaglia's concept both on theoretical level- inserting a high mathematical theory behind it, and by the introduction of Deep Learning- on a practical one. Till this day pairs trading remains one of the most popular technique among quantitative traders.\\
Pairs trading approach questions efficient-market hypothesis (EMH) that in its strongest form states that asset's price always reflects all public and private information and therefore profiting from fundamental or technical analysis is impossible. Formally speaking, for a given moment in time $t>0$ the current price of an asset- $S_t$ is a conditional expectation of stochastically discounted (with factor $Q_{t+1}$) future price $S_{t+1}$ increased by future dividend $D_{t+1}$ with respect to a filtration $\mathcal{F}_t$ representing all available information till $t$:
$$S_t = \mathbb{E}[Q_{t+1}(S_{t+1} + D_{t+1}) | \mathcal{F}_t ].$$
EMH was put together by Eugene Fama\cite{emh} and have gained mixed reviews since then. Current discourse consists of more criticism to the theory: articles presenting repeatable profits from applying methods of fundamental and technical analyses in different market scenarios can be classified as empirical evidences of hypothesis's incorectness. Another perspective questioning EMH is one of behavioural economists claiming that even if the market is efficient investors aren't. From behaviourists point of view traders make mistakes and are exposed to biases such as overconfidence or overreaction leading to arbitrage opportunities. In his book\cite{pedersen} L. H. Pedersen gives a thoughtful summary of the discourse: he claims that market is \quotes{efficiently inefficient} meaning that even though mispricings can occur they are quickly closed and therefore can only be exploited by best traders. This paper does not aim to disprove Fama's theory, although will present empirical results that are closer to Pedersen's perspective.\\
As described above, in a classical approach of pairs trading $2$ similarly behaving asset $A$ and $B$ are picked. Then a portfolio $P$ of $1$ PLN worth of $A$ and $-1$ PLN (shorted) worth of $B$ is constructed. Note that it's initial cost is $0$. Let $A_t$ and $B_t$ represent values of $1 PLN$ invested in both assets after $t>0$. Let $M_t$- be the common part of $A_t$ and $B_t$- usually called the market component. In other words:
$$\begin{cases}
A_t = M_t + \epsilon^A_t\\
B_t = M_t + \epsilon^B_t\\
\end{cases},$$
where $\epsilon^A_t, \epsilon^B_t$ are market-independent, unique price components of $A_t$ and $B_t$. Then, we can see our portfolio value $P_t$ as:
$$P_t = A_t - B_t = (M_t + \epsilon^A_t) - (M_t + \epsilon^B_t) = \epsilon^A_t - \epsilon^B_t.$$
Such portfolio does not include any market components and is therefore assumed to be market-neutral. Since the market tends to quickly correct any arbitrage opportunities, one should also expect that both unique price components are relatively small and circulate around $0$- from a non-economical perspective they can be viewed as \quotes{noise}. No matter if both assets are increasing or decreasing in value- every discrepancy between $A_t$ and $B_t$ where the first one is higher can be used by selling the portfolio with profit, and every reverse situation ($B_t > A_t$) should only be temporary and thus not affect $P$ for the long-term. Such pair can also be monitored without actually owning it until a \quotes{good} moment of $B_t$ being significantly larger than $A_t$ occurs- then a reversed scenario is not even needed to profit, even a correction to $A_t \approx B_t$ would be enough to sell. For better understanding let us consider an example concerning highly correlated polish energy providers (top companies of \textit{WIG-ENERG} index): Tauron Polska and ENEA SA. Based on the last year (July 2022 to July 2023) correlation between their stocks' daily returns was equal to $\rho = 0.7269$.
\begin{figure}[H]
\centering
\includegraphics[scale=0.55]{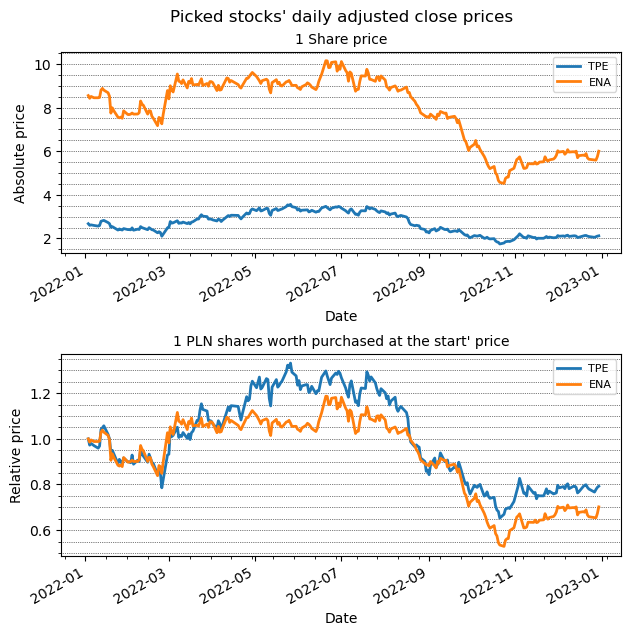}
\caption{Tauron Polska (TPE) and ENEA (ENA) absolute and relative daily close prices}
\label{fig:pair_years}
\end{figure}
\noindent Figure \ref{fig:pair_years} shows $2$ plots: the upper one presents adjusted (with additional payments included) daily close prices of considered stocks while the bottom one aims to show the development of investing $1$ PLN in both stocks at the beginning of the period. Latter plot visualizes a common trend between $2$ companies together with unique components' influence making the relative prices circulate around each other. A qualitative reasoning behind such oscillation may be that these companies are in a sphere of interest for similar investors- prices drops of one makes them switch to the other. We will now consider a pair portfolio constructed at the start of the period with long $1 PLN$ worth of TPE and short $1 PLN$ worth of ENA. Portfolio is not going to be adjusted throughout the period.
\begin{figure}[H]
\centering
\includegraphics[scale=0.55]{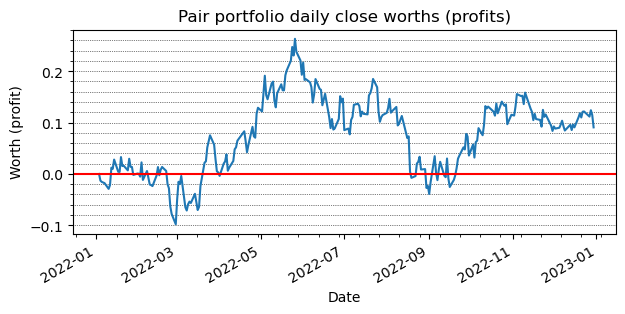}
\caption{Portfolio of long TPE and short ENA daily close worth}
\label{fig:portfolio-years}
\end{figure}
\noindent As can be seen on Figure \ref{fig:portfolio-years}, portfolio is circulating around $0$ depending on which stock's relative price is currently higher. Bearish (downward) market trend of stocks' prices during the last $4-5$ months is \quotes{cancelled} in the portfolio. Sell of the portfolio at any moment when its value is higher than $0$ is a profit for the investor since the entering cost of purchasing it was $0$. As mentioned before, portfolio can be also tracked theoretically and purchased later, when its value is negative. Then trader would hope for a correction of the discrepancy and profit by selling afterwards.\\
A major drawback of using $2$ stocks as a pair is that we are exposed to each one's noise. It is then harder to model pairs' portfolio since those unique parts may f.e. have different oscillation frequencies. Perfect solution would then be to consider pure market component $M_t$ instead of the second stock. Then, a modified portfolio value $P'_t$ would be:
$$P'_{t} = A_t - M_t = M_t + \epsilon^A_t - M_t = \epsilon^A_t.$$
This way there is no need for pairs selection, every company can be described in terms of its price unique component's parameters. The only selection needed would then be picking stocks with best $\epsilon_t$ \quotes{circulating} properties (we will explain it in more details in further sections). But how can you trade on the market? This is the exact moment where indices can be introduced. From the definition, by gathering many similar companies, they are supposed to represent current market's trends. Unique trends of included stocks are smoothened and the final result gives a reasonable indicator of $M_t$ for each participant. Although trading with indices directly is obviously impossible, there are exchange traded funds- dynamically updating stock portfolios imitating a given index. Such portfolios can be purchased directly on stock exchange market- it takes work off the investor who does not want to buy each participant's stocks separately and keep their number relevant to index's shares. Since polish indices' spectrum is not so wide (especially when excluding those not backed by corresponding ETF) own indices can also be developed without limiting oneself to what market offers. Using statistical and Deep Learning methods for any group of stocks multiple market components can be considered without actually matching them to any concrete economic aspect. Influences of each stock in such artificial factors would then be transferred into specific amounts to purchase in owned portfolios. Following M. Avellaneda and J.H. Lee's 2008 paper\cite{main_paper}, we are going to use real and artificial indices as market components to subtract unique price movements with oscillation properties. There are $3$ questions still left unanswered:
\begin{itemize}
\item Does every stock have a market component in its price? What is the theory behind it?
\item Does every stock's price unique \quotes{leftover} circulates? How fast will it come back to $0$?
\item When is the best moment to trade with pair portfolio? On what basis and how to identify the signals?
\end{itemize}
Fortunately, each of these questions has a separate section dedicated to answer it comprehensively. Thus, in the following chapters we will consider every aspect of pairs trading theory gathering and combining various papers' discourses. Let us start with the theory of stocks returns' $\beta$ models.
\section{Multi-factor $\beta$ model of returns}
For the basic definition of pairs trading we proposed that $A_t$: value of $K^A_{1\text{ PLN}}$ company $A$'s stocks (worth $1$ PLN at the start) after $t>0$ can be decomposed to $2$ components: market factor $M_t$ representing part of the market \quotes{contained} in $A_t$ and independent part $\epsilon_{A,t}$- unique to chosen company. In other words:
$$A_t = M_t + \epsilon^A_t.$$
Although this way of thinking is generally appropriate and intuitive, used notation may be seen as a bit oversimplified- it does not incorporate how strong the correlation between $A_t$  and the market actually is. We will therefore formalize and generalize our assumptions. Let us consider stochastic process $S_t, t>0$ modelling close prices of a given company in time (years). For simplicity we are going to assume that such process is continuous. Then, for a time interval $dt$ (f.e. $\frac{1}{252}$ representing one trading day) let $R_t = \frac{S_{t + dt} - S_t}{S_t}$ represent stock's returns. 
Similarly, let us consider $S^F_t, t>0$- process of capitalization-weighted market index and its returns $F_t, t>0$. Under an assumption that for any $t_0$ both $R_t$ and $F_t$ come from a normal distribution let 
\begin{equation}
\beta = \frac{\text{Cov}(R_t,F_t)}{\text{Var}(F_t)},
\label{eq:B}
\end{equation}
be a time-constant, scaled covariance between processes. Then we can construct:
\begin{equation}
R_t = \alpha dt + \beta F_t + \epsilon_t,
\label{eq:R}
\end{equation}
which is a simple regression model (commonly called $2$-parameters model) decomposing stock's return into deterministic drift $\alpha dt$, market (systematic) $\beta F_t$ factor and a stock-unique, mean $0$, uncorrelated (idiosyncratic) process $\epsilon_t$ which is also normal for given $t$. Taking the expected values in Equation \ref{eq:R} we get:
\begin{equation}
\mathbb{E}[R_t] = \alpha dt + \beta \mathbb{E}[F_t].
\end{equation}
Parameter $\beta$ defined as in Equation \ref{eq:B} actually minimizes the squared residuals sum of fitted model (like in standard OLS). Equation \ref{eq:R} was initially fully defined by E. F. Fama (1973)\cite{fama} and heavily based on W. F. Sharpe \textit{Capital Asset Pricing} model (1964)\cite{capm}. Fama considered it in relation to his efficient market theory. Through empirical tests he  proved that on average there are no other measures of risks affecting stocks' returns than ones of market and thus confirmed the correctness of proposed model. Sharpe thought of $F_t$ as of a diversified portfolio returns and therefore studied the theoretical expected return a given stock should have to be included in the portfolio. Although proposed model is very simple, concepts of $\beta$ and $\alpha$ for given stock are still used as basic indicators of its market-dependence and of how it is performing compared to the market consecutively. 
\begin{figure}[H]
\centering
\includegraphics[scale=0.6]{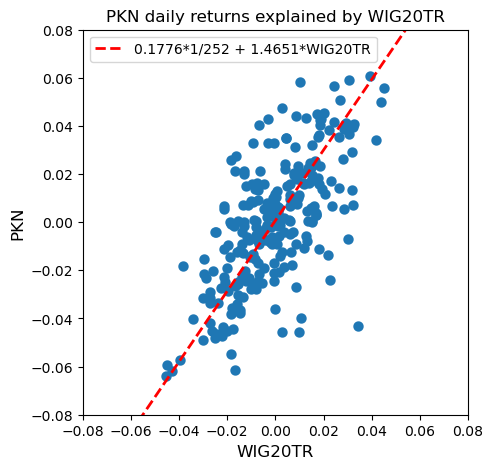}
\caption{PKN Orlen daily returns explained by $2$-parameters model with \textit{WIG20} serving as the market}
\label{fig:regression}
\end{figure}
\noindent Figure \ref{fig:regression} shows how PKN Orlen daily returns from July 2022 to July 2023 are explained by \textit{WIG20}'s returns from analogous period. Parameter $\beta = 1.4651$ was calculated based on the same time-window using empirical estimator of covariance, $\alpha = 0.1776$ was then derived as a mean of $(R^{PKN}_t - \beta R^{\textit{WIG20}}_t)$ scaled by $dt = \frac{1}{252}$. In this example PKN is giving larger returns relative to the \quotes{market} with an additional, constant overperformence. Interestingly, we are actually going to focus on the part that was omitted by both Sharpe and Fama because of their (approximate) market's efficiency assumption- the idiosyncratic, market-neutral factor. Since our model is based on returns $\epsilon_t$ is not actually modelling the accumulated worth of such portfolio. Therefore we will switch to more appropriate naming of $\epsilon_t = dI_t$ where $I_t$ represents relative value of the oscillating, idiosyncratic, pair portfolio. It is also stationary as a process (although its mean does not need to be $0$) and normal for any given $t_0$. Such portfolio consists of a given stock and regression model's replication. It can be also seen as an indicator what is the relation between a \quotes{fair} price and an actual one.
\begin{figure}[H]
\centering
\includegraphics[scale=0.8]{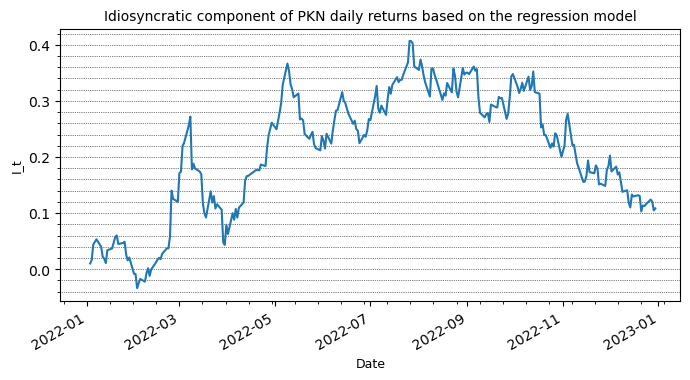}
\caption{Idiosyncratic component $I_t$ of PKN Orlen daily returns throughout training period}
\label{fig:idio-years}
\end{figure}
\noindent Continuing with the same case, Figure \ref{fig:idio-years} presents how the idiosyncratic part of PKN returns behaves on the training period. Assuming that the residuals serve as increments of $I_t$, we summed them up cumulatively to get $I_t$. Cumulative sums of $dI_t$ are clearly circulating around a constant without any significant drift- this agrees with $I_t$'s stationarity assumptions. We can also see it as a value of a simple pairs portfolio consisting of PKN shares together with shorted index's Exchange Traded Fund- $\alpha$ does not need to be cancelled in practice since we aim to profit from corrections and not the trend itself.\\
Today's economy is so complex that even though it is commonly used, decomposition into a single market factor and the remaining part may not be sufficient enough. For that reason Arbitrage Pricing theory was introduced\cite{apt} as a generalized and thus improved version of Capital Asset Pricing model we introduced earlier. New model states that for given asset $i$ its returns $R^i_t$ can be decomposed as follows:
\begin{equation}
R^i_t = \alpha_i dt + \Sigma_{j=1}^r \beta_{ij} F_t^j + dI^i_t,
\label{eq:apm}
\end{equation}
where $F_t^1,\ldots F_t^r$ are returns of $r$ different systematic factors and $\beta_{i1},\ldots \beta_{ir}$ represent asset $i$ sensitivities to a given factor (they are not necessary as in Equation \ref{eq:R} because the factors may also be correlated with each other). Such factors might represent different sector indices correlated with considered asset. It is assumed that both $R^i_t$, $F_i, i = 1\ldots r$ and $dI^i_t$ for given $t_0$ are normal and $dI_t$ is again an increments process of fluctuating process $I_t$. Model can be viewed as multi-factors regression equation, therefore $\beta$s can be derived by minimizing its mean squared error.
\begin{figure}[H]
\centering
\includegraphics[scale=0.8]{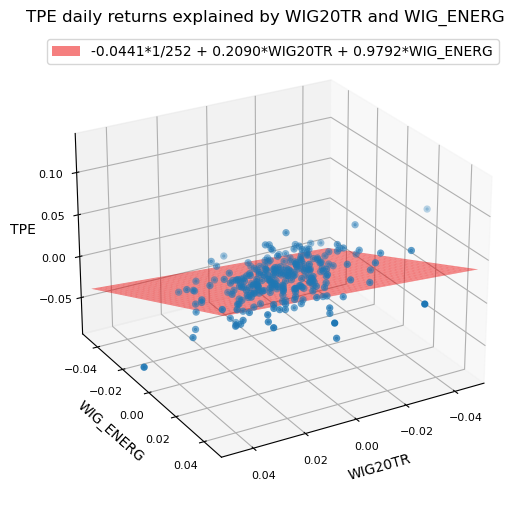}
\caption{Tauron Polska daily returns explained by multi-factor $\beta$ model with \textit{WIG20} and \textit{WIG-ENERG} serving as systematic factors}
\label{fig:regression3d}
\end{figure}
\noindent Following the same methodology as with PKN $2$-parameters model example, Figure \ref{fig:regression3d} shows multi-factor $\beta$ model fit for Tauron Polska returns taking \textit{WIG20} and sector index- \textit{WIG-ENERG}. Coefficients were selected as in a standard OLS model. It is important to mention that such explanatory variables are not fully practical since there is no existing ETF of \textit{WIG-ENERG}.\\
As mentioned, Arbitrage Pricing theory is a generalization of Fama's $2$-parameters model. It is then more flexible with its variables and therefore has more explanatory power than CAPM. Assumption about one universal factor explaining the entire market is abandoned and replaced with more reasonable one of multiple economic risk drivers. Still, similarly to CAPM, Arbitrage Pricing theory assumes that the market is approximately efficient with the fair asset price being represented on average and thus explained by the model but it also enables traders to seek for arbitrage opportunities throughout the analysis of remaining residuals. Theory behind such analysis is going to be the topic of the following section.
\section{Mean-reverting stochastic processes of stocks' residuals}
We are considering a framework where stocks' prices can be considered \quotes{fair} and therefore not misvalued on average. Still, on daily basis arbitrage opportunities of over- or underpricing are possible. They happen because of investors' mistakes and biases such as overreaction or \quotes{herd following} but are quickly corrected- traders, f.e. arbitrage seeking ones \quotes{move} prices back to their \quotes{fair} value by changing trading demand. If underpriced stock is identified, buying volume grows which then leads to price increase forming almost a self-fulfilling prophecy. If \quotes{fair} value is indicated by a model, cumulative residuals represent potential mispricings and therefore should be approximated by processes of similar \quotes{autocorrection} properties. Additionally, Arbitrage Pricing theory expects all components to be normal for any given time moment, especially $I_t$ and its increments $dI_t$ (for Gaussian $I_t$ latter one comes automatically from normal distribution properties). Summing up those assumptions and switching to a more theoretical \textit{jargon} we seek for Gaussian, unconditionally stationary, mean-reverting processes that circulate around their means. For clarity, \quotes{unconditional stationarity} means that without conditioning on its initial value (not assuming it is known) process is stationary. \textbf{Ornstein–Uhlenbeck process} is going to be our natural candidate fulfilling all the requirements- this approach is consistent with Avellaneda's paper\cite{main_paper} and commonly used in field's literature.\\
Ornstein-Uhlebeck process was introduced by physicists Leonard Ornstein and George Eugene Uhlenbeck in 1930s. Let us consider its formal definition.
\begin{definition}
The \textit{Ornstein-Uhlenbeck} (OU) process is a stochastic, unconditionally stationary, Gaussian process satisfying the following stochastic differential equation:
\begin{equation}
dX_t = \kappa (\mu - X_t)dt + \sigma dB_t,
\label{eq:ou}
\end{equation}
where $B_t, t>0$ is a standard Brownian motion. Constant parameters represent:
\begin{itemize}
\item $\mu \in \mathbb{R}$- long term mean of the process,
\item $\kappa \in \mathbb{R}_+$- speed of process's mean-reversion,
\item $\sigma \in \mathbb{R}_+$- volatility of the process.
\end{itemize}
\end{definition}
\noindent Considering deterministic process $X'_t$ where Brownian motion increment (the only source of randomness) in the equation is ignored we have:
\begin{align*}
&dX'_t = \kappa (\mu - X'_t) dt\\
&\frac{dX'_t}{dt} = \kappa \cdot \mu - \kappa X'_t
\end{align*}
Introducing $X'_0 = x'_0$ it can be easily solved with the solution being
$$X'_t = \mu + (x'_0 - \mu)\exp{(-\kappa t)}.$$
As time goes to infinity, $X'_t \longrightarrow \mu$. Process converges to its mean exponentially with rate $\kappa$ and magnitude proportional to the difference between its current value and $\mu$. Similar properties can be transferred to $X_t$ whose solution goes as follows.\\
Let $X_0 = x_0$ and $Y_t = X_t - \mu$- a centred version of $X_t$. Then Equation \ref{eq:ou} can be re-written as:
$$\begin{cases}
dX_t = dY_t = -\kappa Y_t + \sigma dB_t,\\
Y_0 = y_0 = x_0 - \mu.
\end{cases}
$$
To get rid of the drift $\kappa Y_t$ we will consider process $Z_t = \exp{(\kappa t)}Y_t$. Then from a Leibnitz product rule:
\begin{align*}
dZ_t =& \exp{(\kappa t)}dY_t + \kappa \exp{(\kappa t)}Y_t =\\
=& \exp{(\kappa t)}(-\kappa Y_t + \sigma dB_t + \kappa Y_t) = \exp{(\kappa t)}\sigma dB_t.
\end{align*}
Using Ito's integral notation and substituting $Z_0 = \exp{(\kappa t)}Y_0$, $Z_t$ becomes:
$$Z_t = Z_0 + \sigma \int_0^t \exp{(\kappa s)}dB_s.$$
Coming back to initial $X_t$ process our formula is:
\begin{equation}
X_t = \mu + (x_0 - \mu)\exp{(-\kappa t)} + \sigma \int_0^t \exp{(-\kappa (t-s))}dBs.
\label{eq:ou_sol}
\end{equation}
The only part which can be further transformed is the integral. Let us consider the following lemma.
\begin{lemma}
For a deterministic function $f \in \mathcal{L}_2$ and $0\leq s < t$
$$\int_s^t f(u) dB_u \sim \mathcal{N}(0;\int_s^t f^2(u) du),$$
where $\int_s^t (\cdot) dBu$ is Ito's integral.
\end{lemma}
\noindent Having:
\begin{align*}
\int_0^t (\exp{(-\kappa (t-s))})^2 ds =& \int_0^t \exp{(-2\kappa (t-s))} ds = \exp{(-2\kappa t)}\int_0^t \exp{(2\kappa s)}ds =\\
=&  \frac{\exp{(-2\kappa t)}}{2\kappa}(\exp{(2\kappa t)} - 1) = \frac{1}{2\kappa}(1 - \exp{(-2\kappa t)}),
\end{align*}
we can re-write integral as $B_{\frac{1}{2\kappa}(1 - \exp{(-2\kappa t)})}$ and then Equation \ref{eq:ou_sol} becomes:
\begin{equation}
X_t = \mu + (x_0 - \mu)\exp{(-\kappa t)} + \sigma B_{\frac{1}{2\kappa}(1 - \exp{(-2\kappa t)})}.
\end{equation}
Now one can easily see that conditioning on $X_0=x_0$ for any $t_0>0$ $X_{t_0}$ follows normal distribution (which makes it a Gaussian process) with:
$$
\begin{cases}
\mathbb{E}[X_t | X_0 = x_0] = \mu + (x_0 - \mu)\exp{(-\kappa t)},\\
\text{Var}(X_t | X_0 = x_0) = \frac{\sigma^2}{2\kappa}(1 - \exp{(-2\kappa t)}).
\end{cases}
$$
Asymptotically, as time goes to infinity, $X_t \longrightarrow \mathcal{N}(\mu, \frac{\sigma^2}{2\kappa})$- parameters of such normal distribution are also the unconditional mean and variance of $X_t$- that is why we assume that OU process is stationary.
It is worth noticing that the mean exactly equals the value derived from the deterministic differential equation provided earlier.  Intuitively, $X_t$ is then like $X'_t$ with additional \quotes{noise} serving as a contrary force to the convergence. This way process gains its mean-reversion property that makes $X_t$ circulate around $\mu + (x_0 - \mu)\exp{(-\kappa t)}$ with speed determined by $\kappa$.
\begin{figure}[H]
\centering
\includegraphics[scale=0.4]{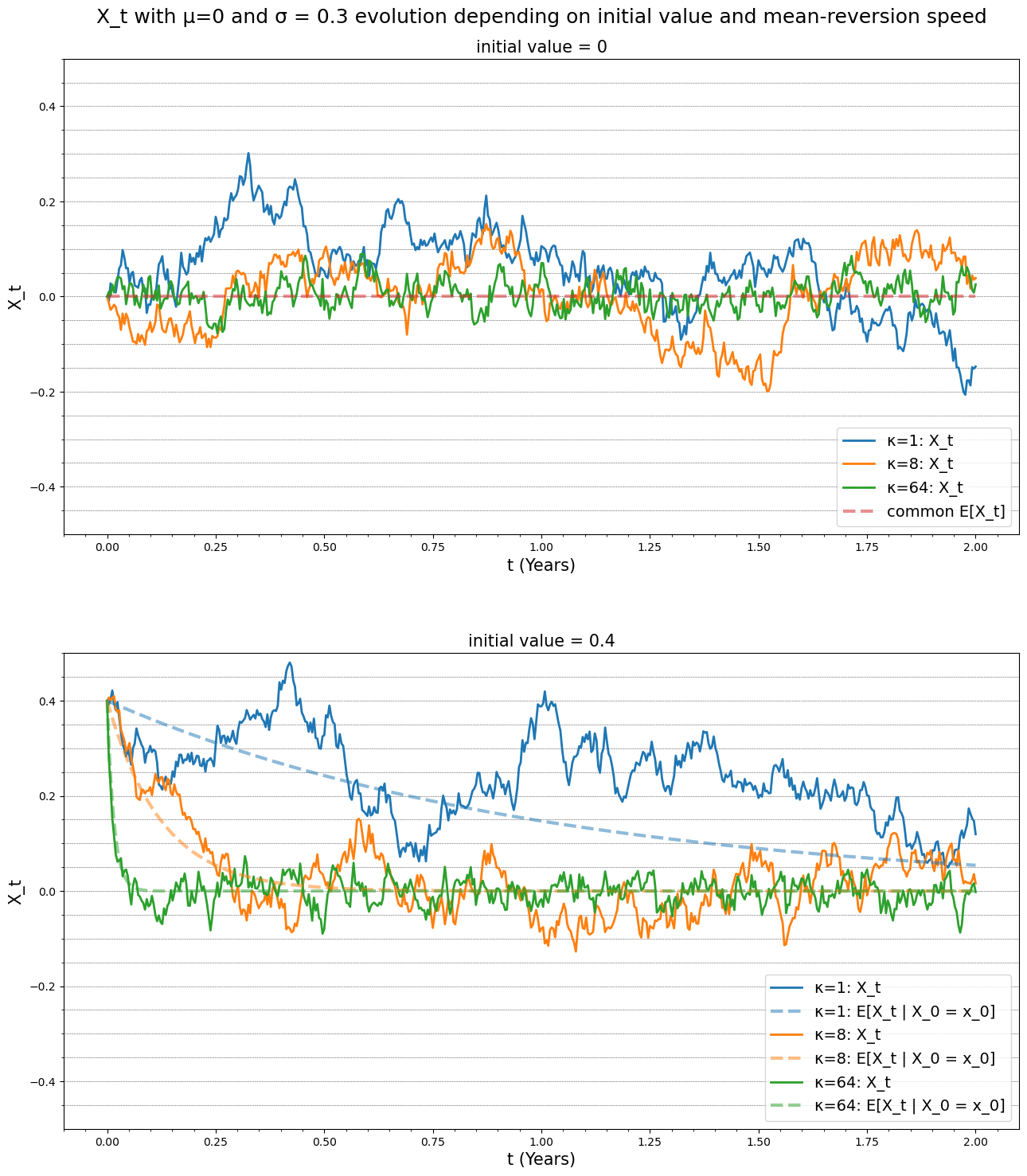}
\caption{Ornstein-Uhlenbeck process simulations depending on $x_0$ and $\kappa$ parameters}
\label{fig:ou-years}
\end{figure}
\noindent Figure \ref{fig:ou-years} shows how $X_t$ circulates around $\mathbb{E}[X_t | X_0 = x_0]$. On the upper picture we can see a constant conditional mean version of the process (where $\mu = x_0$), whereas the lower one shows $\mathbb{E}[X_t | X_0 = x_0]$ exponentially decreasing towards $\mu$ from higher $x_0$. In both cases for larger $\kappa$s there are more oscillation nodes (their frequency is higher)- it serves as a graphical explanation why $\kappa$ is associated with the speed of mean-reversion. We will seek for stocks for which residuals' $\kappa \gg 1$ so that transactions profiting from mispricings can be opened and closed multiple times before the end of trading period. Parameters $\mu$ and $\sigma$ are also important since asymptotic mean is an anchoring and volatility shows how big deviations should be expected. Therefore, they both indicate whether our idiosyncratic portfolio' value will rebound soon. Let us then consider how to extract such parameters from historical data that is assumed to represent Ornstein-Uchlenbeck process. Using Brownian motion properties, for given $dt$ we can re-write Equation \ref{eq:ou} in the following way:
\begin{align*}
&X_{t+dt} - X_t = \kappa \cdot \mu \cdot dt - \kappa \cdot dt \cdot X_t + \sigma (B_{t+dt} - B_t)\\
&X_{t+dt} =\kappa \cdot \mu \cdot dt + (1 - \kappa \cdot dt)X_t + \mathcal{N}(0;\sigma^2 dt)
\end{align*}
or equivalently:
\begin{equation}
X_{t+dt} = \phi_0 + \phi_1\cdot X_t + \zeta_{t+dt} \text{ where }\phi_0 = \kappa \cdot \mu \cdot dt, \phi_1 = (1 - \kappa \cdot dt) \in [0;1]\text{ and }\zeta_{t+dt} \sim \mathcal{N}(0;\sigma^2 dt).
\label{eq:ab}
\end{equation}
This way, OU process can be seen as a continuous extension of lag $1$ autoregressive model (AR1) defined below.

\begin{definition}
Autoregressive model is a time-series model where given state $X_k$ is dependent only on finite number of previous states and a possible constant drift. In other words, for $p$ previous states-dependence and :
$$X_k = \phi_0 + \Sigma_{j=1}^p \phi_j X_{k-j} + \zeta_k,$$
where $\phi_0, \ldots \phi_p$ are the coefficients and $\zeta_k$ are independent random variables with mean $0$ and constant variance $\sigma^2_{\zeta}$. The simplest auto regression model (AR(p)) is one with $p=1$:
$$X_k = \phi_0 + \phi_1 X_{k-1} + \zeta_k.$$
Such process is stationary only if $|\phi_1|<1$, then $\text{Corr}(X_k, X_{k-j})=\phi_1^j$. Additionally, partial correlation including only direct influences between states (conditioning on the ones in-between) is given by:
$$\text{PartCorr}(X_k, X_{k-j})=
\begin{cases}
\phi_1, j = 1,\\
0,\text{ otherwise}.
\end{cases}
$$
\end{definition}
\noindent Because of that, real data (discrete in its nature) can be fitted with OU process using similar techniques as in time-series analysis. Let us assume that based on $W$ trading days historical window we have extracted idiosyncratic parts of daily returns for a given stock: $dI_1, \ldots dI_W$ (simplifying the indexing for the sake of brevity). Keeping in mind that these are just the increments we calculate:
$$I_k = \Sigma_{j=1}^k dI_j, k = 1, \ldots W,$$
which can be seen as a discrete version of OU process that we are estimating. If the $\beta$ coefficients were based on the same data, due to regression principles the sum of all coefficients- $I_W$ is $0$. Then, we need to solve AR(1) equation for $I_k, , k = 1, \ldots W$. There are many ways of getting $\phi_0$ and $\phi_1$, one of them being the use of Yule-Walker equations. This is a method of moments- it translates empiric correlations within the sequence to model's parameters. After estimating AR(1) coefficients, having Equation \ref{eq:ab} formulas, $dt=\frac{1}{252}$ and $\widehat{\text{Var}(\zeta)}$ being the sample variance of $\zeta$- errors of the fitted model, we can derive $\kappa, \mu, \sigma$ as follows:
\begin{equation}
\begin{aligned}
&\kappa = -\log{(\phi_1)}\cdot 252,\\
&\mu = \frac{\phi_0}{1-\phi_1},\\
&\sigma = \sqrt{\frac{\widehat{\text{Var}(\zeta)}\cdot 2 \kappa}{1-\phi_1^2}}.
\end{aligned}
\label{eq:kms}
\end{equation}
If $I_k$ and $I_{k+1}$ are strongly correlated changes in time-series are relatively slow- it corresponds directly with mean-reversion speed parameter $\kappa$ which would be $\approx 0$ for $\phi_1$ close to $1$.
\begin{figure}[H]
\centering
\includegraphics[scale=0.6]{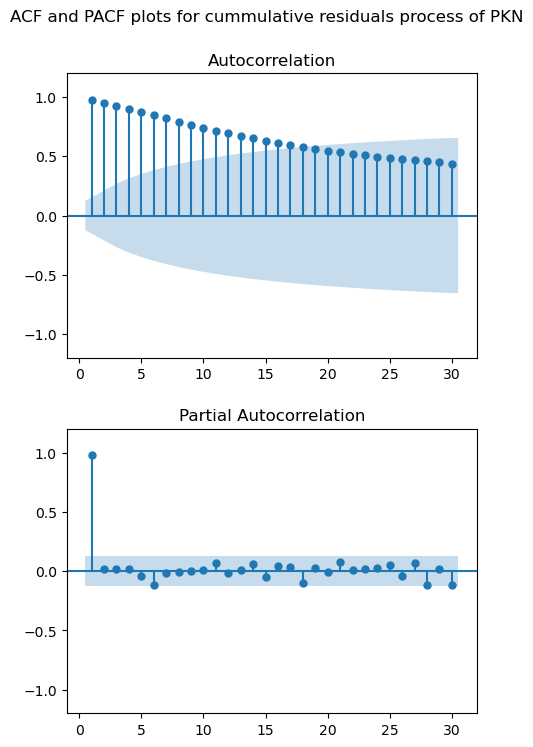}
\caption{Autocorrelation and partial autocorrelation plots for idiosyncratic component $I_t$ of PKN Orlen daily returns}
\label{fig:pacf}
\end{figure}

\begin{table}[H]
\centering
\begin{tabular}{rrr}
\toprule
$\kappa$ &$\mu$&$\sigma$ \\
\midrule
6.0443 & 0.2278 & 0.3459 \\
\bottomrule
\end{tabular}
\caption{OU process parameters' estimates for PKN Orlen}
\label{tab:2}
\end{table}
\noindent Figure \ref{fig:pacf} continues with the PLN Orlen example that we introduced earlier. Taking time-series presented on Figure \ref{fig:portfolio-years} into account, it shows empirical autocorrelations (ACF) and partial autocorrelations (PACF) for first $10$ lags. It can be seen that their structure follows AR(1) assumptions since ACF shrinks with increasing lags but does not disappear and PACF practically vanishes after lag $1$. Keeping in mind that $Corr(X_k, X_{k-1})=\phi_1$, this parameter is over $0.9$. It is not a surprise because $I_t$ was calculated as a cumulative sum and therefore heavily depends on closest predecessor. Table \ref{tab:2} shows $\kappa,\mu,\sigma$ derived from Equation \ref{eq:kms} formulas for considered example. Although we do not have a concrete comparison yet, $\kappa$ seems relatively high which is consistent with $I_t$ mean-reverting quite fast (Figure \ref{fig:portfolio-years}).
As mentioned before, all $3$ parameters are indicators of portfolio's relative performance. They, together with actual $I_t$ value, should be then used to derive signals for actual transactions. For example, if value of the idiosyncratic component is high relative to $\mu$ and $\sigma$ one should open a short trade expecting portfolio's value to drop generating potential profit. Next section will be dedicated to presenting signal function and thresholds activating specific actions.
\section{Signals generation}
We are going to trade on multiple companies' idiosyncratic components with all of them being modelled by the Ornstein-Uhlenbeck processes. For a given trading day, to decide whether a position on a certain company should be opened/closed we must look at how far $I_t$ currently is from its long-term mean. Typically, long/short position will be opened when process lies \quotes{far} (relative to its volatility) below/above average and then closed after it mean-reverses. Depending on stock's specificity parameters $\mu$ and $\sigma$ will be different- therefore to have unified thresholds we are going to normalize $I_t$ for each stock.\\
Recall that the asymptotic distribution (when $t \longrightarrow \infty$) of OU process is $\mathcal{N}(\mu, \frac{\sigma^2}{2\kappa})$. For a given stock $i$ and its idiosyncratic factor $I^i_t$ let us then consider normalized process $G^i_t$:
$$G^i_t = \frac{I^i_t - \mu_i}{\sqrt{\frac{\sigma_i^2}{2\kappa_i}}}.$$
It is clear that $G^i_t$ is also a Gaussian process and that as $t$ goes to infinity, $G^i_t \longrightarrow \mathcal{N}(0,1)$. Keep in mind that we are not actually interfering with the portfolio as it is still represented by $I^i_t$- $G^i_t$ is just a theoretical signal generator. Consider that today's $G^i_t$ value is $g^i_t$, then these are the exact rules to follow:
\begin{align*}
\text{open long position if }g^i_t &< -\overline{g}_{ol}\\
\text{open short position if }g^i_t &> +\overline{g}_{os}\\
\\
\text{close long position if }g^i_t &> -\overline{g}_{cl}\\
\text{close short position if }g^i_t &< +\overline{g}_{cs}\\
\end{align*}
where $\overline{g}_{ol},\overline{g}_{os},\overline{g}_{cl},\overline{g}_{cs}$ are the cut-offs. Now, even though the thresholds for triggering trades can be common for all stocks and throughout entire trading period, we still do not have a quantitative way of picking the most optimal ones. To narrow down the possibilities, from the $68$-$95$-$99.7$ rule we can propose that absolute values of position-opening thresholds should lay within $1$ and $2$- this way hitting them will not be so common but still achievable (landing in such interval on a specific side has around $14\%$ probability). When it comes to magnitudes of position-closing thresholds they should lay below $1$ and not be too close to the first ones (this way actual profit can be made). Avellaneda and Lee estimated the cut-offs empirically, based on simulating strategies on an additional training period from 2002 to 2004 in the ETFs market modelling approach\cite{main_paper}. Based on performance metrics they picked the following thresholds as optimal:
\begin{equation}
\overline{g}_{ol} = \overline{g}_{os} = 1.25; \overline{g}_{cl} = 0.5; \overline{g}_{cs} = 0.75
\label{eq:signals}
\end{equation}
We will follow a very similar approach, optimizing above barriers on an additional training period but separately for each market generation algorithm. 
\begin{figure}[H]
\centering
\includegraphics[scale=0.7]{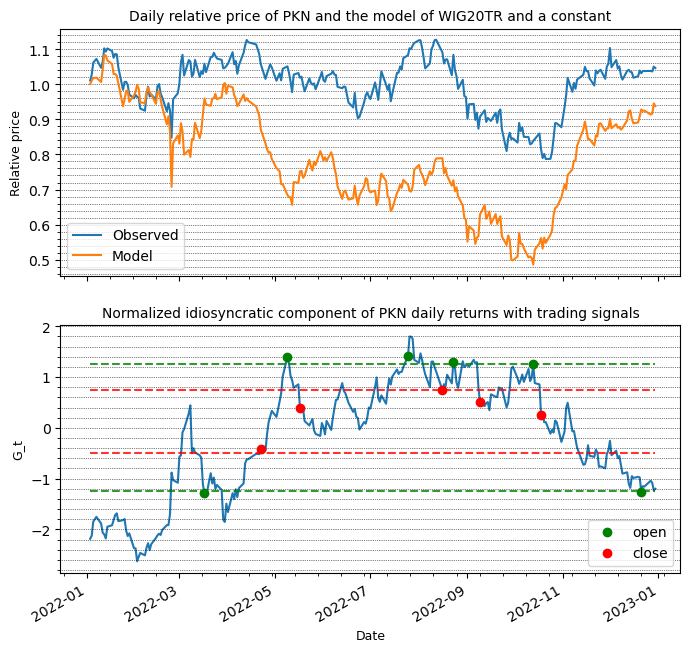}
\caption{Relative prices of PKN Orlen returns and their imitating portfolio together with trading signals based on Avellaneda's cut-offs}
\label{fig:signals}
\end{figure}
\noindent Let us revisit our PKN Orlen stocks' example one last time. Figure \ref{fig:signals} shows relative prices of PKN and their predictions based on \textit{WIG20} together with normalized $I_t$ process named $G_t$. Dashed lines on the lower plot represent Equation \ref{eq:signals}'s thresholds- green ones for opening position and red ones for closing. We assume that at given time trader can hold only one opened position of idiosyncratic portfolio. Thus, position has to be closed before opening another one (no matter if \quotes{long} or \quotes{short}). Dots show exact days where signals for opening (green) and closing (red) would appear. The longer the $y$-axis distance between two consecutive points, the bigger the profit achieved from the $2$-steps trade. Our every day goal is thus to update OU process parameters based on actualized historical time window, calculate current $G_t$ value and decide on a potential trade. It is clear that $2$ sub-plots are corresponding to each other: overpricings of PKN are reflected in high $G_t$ values indicating optimal moments for shorting the pair portfolio. Same can be said in a reverse situation where PKN is undervalued in comparison to the model. Keep in mind that when a signal gets generated, we do not actually know how $G_t$ will form in the following days. Decision is based only on how the idiosyncratic part was behaving in the past. In other words, same value of $I_t$ on $2$ separate days may lead to different $G_t$s because of updated $\kappa,\mu,\sigma$ parameters.\\
There are still some technical assumptions to be considered- they will be fully explained in Section 5. We now know how to derive residuals from the Arbitrage Pricing model, analyse their mean-reversion behaviour and transfer it to trading signals. The only, but crucial part missing is how to tackle explanatory variables for our regression model. In other words- how to create portfolios which will take every common trend out of given stock and leave it with pure mispricings component. As mentioned- $3$ approaches will be evaluated- one with already existing indices was already introduced throughout previous subsections. We will therefore provide some additional details to this one but keep the focus on other, more sophisticated two.
\chapter{Paired portfolios generation approaches}
Based on the Arbitrage Pricing model, for a given stock we want to approximate its returns using a linear combination of so called \textit{systematic factors}. Even if systematic components are strongly correlated, each one of them should have some additional explainability of our target variable. The match between model and actual returns should be strong enough that the residuals can be assumed as Gaussian and of mean $0$. Systematic factors have to be tradable: they need to represent returns of real market's equities such as exchange traded funds or unique stocks' portfolios. Although first option does not require any additional calculations beside getting appropriate $\beta$s, its main drawback is that polish ETFs spectrum is very limited (f.e. sector indices' ETFs are not available). On the other hand, creating unique portfolios suited to imitate given returns is far more flexible but requires determining daily share of each component in a quantitative manner. Additionally, transaction costs of re-balancing such portfolios may turn out too big. We are then dealing with a simplicity- flexibility trade-off.\\
The following section aims to present both existing and artificial indices approaches. The focus will be on the latter ones since they require a strong technical background to begin with. Let us start with first artificial indices' technique of Principal Components Analysis.
\section{Principal Components Analysis (PCA)}
For $d$ stocks with daily close prices $S^1_t,\ldots S^d_t$ and corresponding returns: $R^1_t,\ldots R^d_t$ given portfolio returns $F^0$ can be seen as:
$$F^0_t = \Sigma_{i=1}^d w_i R^i_t,$$
where weight $w_i,i = 1,\ldots d$ represents the (money-wise) proportion of asset $i$ in the portfolio (can be negative). For that reason we can assume $\Sigma_{i=1}^d w_i = 1$. Considering time interval $[t-(n-1)dt;t],n>d$ we can think of $R^i,i=1\ldots d$ and $F^0$ as of length-$n$ vectors. Let us consider space $\mathcal{F}$- length-$n$ returns' vectors of all possible portfolios made with $S^1_t,\ldots S^d_t$. Then matrix $\mathbf{R} = (R^1,\ldots R^d)^T$ of size $(d,n)$ spans $\mathcal{F}$- its every element is a linear combination of $\mathbf{R}$. Coming back to initial portfolio $F^0$, in a vector form we can re-write:
$$F^0 = W^T \mathbf{R}, W=(w_1,\ldots w_d).$$
Assuming that $n$ is long enough that no $R^i,i=1,\ldots d$ is a linear combination of the others, $\mathbf{R}$ is an algebraic basis of $\mathcal{F}$. Portfolios can be then represented by their weight vectors $W$, where $W_{R^i}=(0,\ldots 1, \ldots 0)$ with $1$ on the $i$-th place. There are actually many different bases that can be used instead of $\mathbf{R}$: f.e. $R^1$ can be substituted with $F^0$ if $w_1 \neq 0$. This gives us a different perspective: instead of looking at portfolios through stocks' linear combinations we can look at stocks themselves decomposed into different portfolios. In other words:
$$R^i = \Sigma_{i=1}^d w_i F^i,$$
where $F^i,i=1\ldots d$ are linearly independent. This is fairly similar to the Arbitrage Pricing model (Equation \ref{eq:apm}) but here no additional randomness is involved. Obviously, since $R_i,i=1\ldots d$ were also linearly independent, weights would be constructed in a way that all $F^i$ components offset each other leaving only $R^i$ itself. It is then not very helpful to represent stock by itself- to avoid such situation one may need to decrease the number of involved portfolios. Then, a full match will not be possible- luckily that was exactly our goal: to only approximate returns by some portfolios and profit from the residuals. Now, that we are on the edge between probability and algebra, $2$ questions come up: what is the best possible basis to begin with and how to crop it? They are answered with the theory of dimensionality reduction and one of its most common techniques: Principal Components Analysis (PCA).
\subsection{What is PCA?}
Consider $d$-dimensional normal variable $X=(X_1,\ldots X_d)$ with mean vector $\mathbb{E}[X]=0$ and covariance matrix $\mathbb{E}[XX^T]=\mathbf{\Sigma}$. We want to find such $\mathbf{W}=(W_1,\ldots W_d)$ of size $(d,d)$ with $\Sigma_{i=1}^d w_{ki} = 1,k=1\ldots d$ that
$$X'=(X'_1,\ldots X'_d) = \mathbf{W}^T X = (W_1^T X,\ldots W_d^T X) = (\Sigma_{i=1}^d w_{1i}X_i,\ldots \Sigma_{i=1}^d w_{di}X_i)$$
has uncorrelated components with maximum possible variances. Since
\begin{align*}
\text{Var}(X'_i)&=\text{Var}(w_i^T X)=\mathbb{E}[(w_i^T X)^2] - (\mathbb{E}[w_i^T X])^2 =\\
&= \mathbb{E}[w_i^T(X X^T) w_i] - (w_i^T\mathbb{E}[X])^2 = w_i^T\mathbb{E}[X^T X]w_i - 0 = w_i^T\Sigma w_i.
\end{align*}
and for $i\neq j$:
\begin{align*}
\text{Cov}(X'_i,X'_j)&=\text{Cov}(w_i^T X, w_j^T X)=\mathbb{E}[(w_i^T X)(w_j^T X)^T] - \mathbb{E}[w_i^T X] \mathbb{E}[w_j^T X]^T =\\
&= \mathbb{E}[w_i^T X X^T w_j] - w_i^T\mathbb{E}[X] \mathbb{E}[X]^T w_j = w_i^T\mathbb{E}[X^T X]w_j - 0 = w_i^T\Sigma w_j.
\end{align*}
we seek for $\mathbf{W}$ which is the solution of the recursive optimization problem:
\begin{equation}
\begin{aligned}
\text{max }&w_1^T \Sigma w_1^T\\
&\text{where }\Sigma_{i=1}^d w_{1i} = 1,\\
\text{max }&w_2^T \Sigma w_2^T\\
&\text{where }\Sigma_{i=1}^d w_{2i} = 1\text{ and }w_2^T\Sigma w_1=0,\\
\vdots\\
\text{max }&w_k^T \Sigma w_k^T\\
&\text{where }\Sigma_{i=1}^d w_{ki} = 1\text{ and }w_k^T\Sigma w_j=0\text{ for }j=1,\ldots k-1,\\
\vdots\\
\text{max }&w_d^T \Sigma w_d^T\\
&\text{where }\Sigma_{i=1}^d w_{di} = 1\text{ and }w_d^T\Sigma w_j=0\text{ for }j=1,\ldots d-1.\\
\end{aligned}
\label{eq:pca1}
\end{equation}
Recall that for a square size $(d,d)$ matrix $\mathbf{A}$ with rank $d$ (with $d$ linearly independent columns) we can re-write it in a decomposed form:
$$\mathbf{A} = \mathbf{V} \mathbf{\Lambda} \mathbf{V}^{-1},$$
where $\mathbf{\Lambda}$ is a diagonal matrix of $\lambda_1,\ldots \lambda_d \in \mathbb{C}$ and $\mathbf{V}=(v_1,\ldots v^T)$ such that $\mathbf{A}V_i = \lambda_i v_i,i=1\ldots d$. For $i\leq d$ $\lambda_i,V_i$ are called eigenvalues and eigenvectors consecutively with the latter ones being linearly independent of each other. If $\mathbf{A}$ is symmetric, eigenvalues are real and $\mathbf{V}$ can be chosen so that columns are orthogonal and of length $1$. Then $\mathbf{V}^{-1} = \mathbf{V}^T$ and therefore:
$$\mathbf{V} = \mathbf{F} \mathbf{\Lambda} \mathbf{F}^T.$$
Decomposition of symmetric matrix is usually called \textit{spectral decomposition}. Let us consider the following Lemma that will help finding the solution of Equation \ref{eq:pca1}:
\begin{lemma}
Let $\mathbf{A}$ be a positive definite matrix of size $(d,d)$ with eigenvalues $\lambda_1 \geq \lambda_2 \geq \ldots \geq \lambda_d \geq 0$ and corresponding normalized (of length $1$) eigenvectors $f_1,f_2,\ldots f_d$. Then:
\begin{align*}
&\text{max }_{x \neq 0}\frac{x^T A x}{x^T x} = \lambda_1,\text{ achieved for }x=f_1,\\
&\text{max }_{x \neq 0,x \perp f_1,\ldots  f_{k-1}}\frac{x^T A x}{x^T x} = \lambda_k,\text{ achieved for }x=f_k\text{; } k = 2,\ldots d
\end{align*}
\label{lemma:1}
\end{lemma}
\noindent Since covariance matrix $\mathbf{\Sigma}$ is from its definition a positive definite matrix we can consider its descending eigenvalues $\lambda_1 \geq \lambda_2 \geq \ldots \geq \lambda_d \geq 0$ and associated normalized eigenvectors $\mathbf{F}=(f_1,f_2,\ldots f_d)$. Then, from Lemma \ref{lemma:1} it is easy to notice that $\mathbf{W}=\mathbf{F}$ maximizes variances of $X'_i$ with $\text{Var}(X'_i)=\lambda_i$. Additionally, since eigenvectors are orthonormal, for $i \neq j$:
$$\text{Cov}(X'_i,X'_j) = f_i^T\Sigma f_j = f_i^T \mathbf{F} \mathbf{\Lambda} \mathbf{F}^T f_j = (0,\ldots \underset{\text{i-th position}}{1},\ldots 0) \mathbf{\Lambda} (0,\ldots \underset{\text{j-th position}}{1},\ldots 0)^T = 0.$$
As seen, $\mathbf{W}=\mathbf{F}$ satisfies all Equation \ref{eq:pca1}'s conditions. Note that total variance of $X$ is the same as of $X'$ since $$\Sigma_{i=1}^d \text{Var}(X_i) = \text{tr}(\Sigma) = \Sigma_{i=1}^d \lambda_i = \Sigma_{i=1}^d \text{Var}(X'_i).$$
Vectors $X'=(X'_1,\ldots X'_d)=(f_1^T X, \ldots f_d^T X)$ are commonly named \textit{Principal Components}- if one gathers samples from $X$, presenting them in terms of such components decomposes data's volatility into separate dimensions. Formally, let us consider $n$ observations from vector $X$ gathered in matrix $\mathbf{X}$ (of size $(d,n)$). Each point is mapped on a standard $\mathbb{R}^d$ basis of identity matrix columns. It turns out that the theory explained above can be directly transferred to real data- to find axes that maximize sample variances of points projected on them, one needs to use eigenvectors of sample covariance matrix $\mathbf{\widehat{\Sigma}}=\frac{1}{n-1}\mathbf{X}\mathbf{X}^T$. Normalized eigenvectors form a new, orthonormal basis $\mathbf{F}$ with transformed points $\mathbf{X}'=\mathbf{F}^T \mathbf{X}$. Notice that since $\mathbf{F}$ is orthonormal, the transition is reversible with:
\begin{equation}
\mathbf{F} \mathbf{X}' = \mathbf{F} \mathbf{F}^T \mathbf{X} = \mathbf{I} \mathbf{X} = \mathbf{X}.
\label{eq:reverse}
\end{equation}
Each sample of $\mathbf{X}$ can also be mapped to a particular $f_i$ without changing its length by using \textit{orthographic projection} matrix $\mathbf{P}_1=f_i f_i^T$ of size $(d,d)$. Note that this is analogous to Equation \ref{eq:reverse} but with only one $\mathbf{F}$'s column picked.
\begin{figure}[H]
\centering
\includegraphics[scale=0.4]{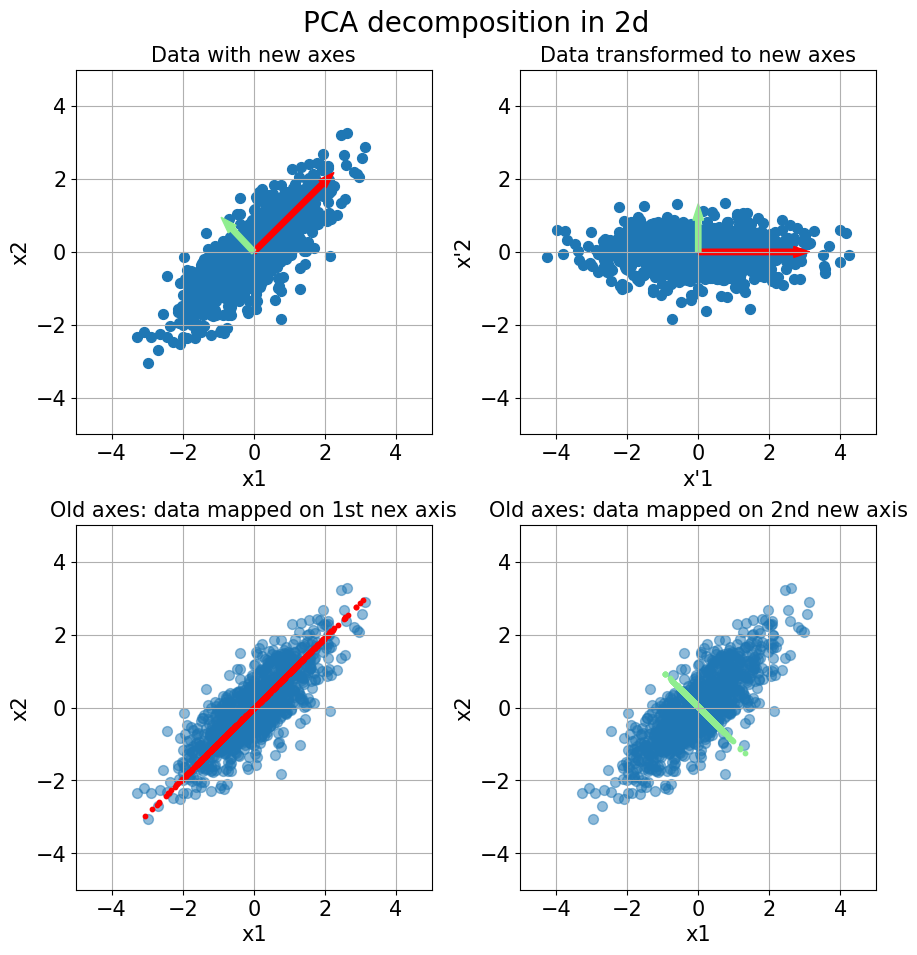}
\caption{$2$-dimensional points transformed to eigenvectors basis and mapped to singular axes}
\label{fig:pca1}
\end{figure}
\noindent Figure \ref{fig:pca1} shows PCA of $n=1000$ samples taken from $2$-dimensional centered normal distribution with $\mathbf{\Sigma}=\left[\begin{matrix}
1 & 0.8\\
0.8 & 1
\end{matrix}
\right]$. Upper left plot presents original data with new axes: 
$$f_1=(\frac{\sqrt{2}}{2},\frac{\sqrt{2}}{2})^T,f_2=(\frac{-\sqrt{2}}{2},\frac{\sqrt{2}}{2})^T$$
scaled by $\lambda_1 = 1.8348, \lambda_2=0.1997$ consecutively. It is clear that eigenvectors follow the spread of data since they are suppose to maximize sample variance represented by eigenvalues. On the upper right picture new basis was introduced with linear transformation $\mathbf{X}'=\mathbf{F}^T\mathbf{X}$. Lower images come back to the original basis and project samples on both eigenvectors separately (with the use of orthographic projection described earlier). From a $1$-dimensional perspective these linearly-dependent points can be seen as realization of principal components (eigencomponents).\\
The concept of mapping initial points to only part of the eigenvectors is called \textit{PCA dimensionality reduction}. Imagine that for given $20$-dimensional data over $75\%$ of the total variance is \quotes{stored} in the $5$ biggest-variance principal components. It might be then easier to look at samples transformed to first $5$ eigenvectors. One needs to take first $5$ components of $\mathbf{F}$ forming $\mathbf{F}_5$ and then multiplying its transposition by $\mathbf{X}$: we will end up with $\mathbf{X}'_5$- a lower dimension sample covering most of the initial volatility. If we then want to come back to the initial dimension but not retrieving back lost variance another multiplication by $\mathbf{F}_5$ needs to be applied. Note that such double transformation is basically $\mathbf{P}_5=\mathbf{F}_5\mathbf{F}_5^T$- an orthographic projection of the original data to a $5$-dimensional hyperplane. In this case dimension stays the same but rank (number of linearly independent rows of $\mathbf{X}$) decreases. As mentioned, transformed data is easier to interpret in terms of common patterns- we may lose some useful information in the process but since so much variance is already covered in $5$ dimensions- we mainly reduce the \quotes{noise}. To formalize, for given $\alpha \in [0;1]$ representing the fraction of total variance we are satisfied with $r$ is defined as:
$$r = \min_{k \leq d}{\lbrace \frac{\Sigma_{i=1}^k \lambda_i}{\Sigma_{i=1}^d \lambda_i} \geq \alpha \rbrace}.$$
Then, first $r$ eigenvectors of $\mathbf{F}$ are taken and transformed into $\mathbf{X}'_r=\mathbf{F}_r^T \mathbf{X}$.
In Figure \ref{fig:pca1}'s example reducing dimension from $2$ to $1$ achieved $\alpha=\frac{\lambda_1}{\lambda_1+\lambda_2}=0.902\approx 90\%$ of total variance. Recall that in Equation \ref{eq:reverse} reverse transformation from $\mathbf{X}'$ to $\mathbf{X}$ was possible. It is not the case with $\mathbf{X}'_{r,r<d}$ but for $\alpha = \frac{\Sigma_{i=1}^r \lambda_i}{\Sigma_{i=1}^d \lambda_i}$ we can expect analogous explainability level of $\mathbf{X}$. In other words, general linear model of initial data can be expressed as:
$$\mathbf{X} = \mathbf{F}_r \mathbf{X}'_r + U = \Sigma_{i=1}^r f_i (X'_i)^T + U,$$
where $U=\Sigma_{i=r+1}^d f_i (X'_i)^T$ and comes from a $d$-dimensional normal distribution with mean $0$. Notice that $\text{tr}(\mathbf{\Sigma}_{U})=\Sigma_{i=r+1}^d \lambda_i$ and as we already know $\text{tr}(\mathbf{\Sigma}_X)=\Sigma_{i=1}^d \lambda_i$ - on a total variance level this confirms our assumptions about model's explainability.
\begin{figure}[H]
\centering
\includegraphics[scale=0.4]{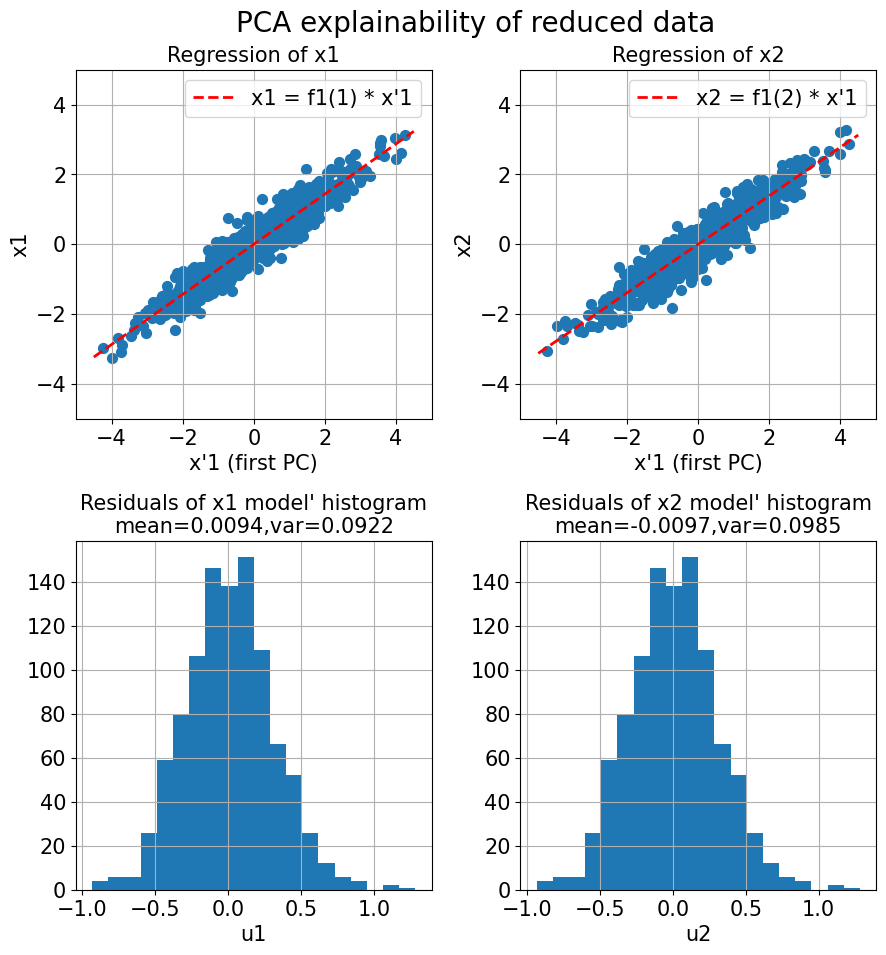}
\caption{Original data coordinates explained by first principal component}
\label{fig:pca2}
\end{figure}
\noindent On Figure \ref{fig:pca2} we come back to the previous example and show how much data is lost during the process of dimensionality reduction. We can see that on both coordinates first principal component explains data patters significantly well. Note that the sum of residuals' sample variances is equal to $\lambda_2$ described earlier.
\subsection{Considered approach}
Let us now come back to the original problem. Recall that we wanted to construct an algebraic basis of portfolios to explain singular stocks' returns in such a way that decreasing the number of components will reduce as little informations as possible. Now, we know that for data from multidimensional normal distribution the optimal approach is to use Principal Components Analysis. Throughout the paper, for example when considering Arbitrage Pricing model we assumed that actual returns can be in fact seen as samples from Gaussian distribution. It is a common conclusion in literature of the subject (in models previously considered but also famously in the original Black-Scholes model\cite{bs}) since stock prices are assumed to follow a \textit{Geometric Brownian Motion} process $S_t$. Such stochastic process, for parameters $\mu$ and $\sigma>0$, initial value $S_0$ and given $t_0\geq dt$ could be written as:
$$S_{t_0} = S_0\exp{((\mu-\frac{1}{2}\sigma^2)t_0 + \sigma \mathcal{N}(0,t))},$$
which follows a lognormal distribution. Then,
$$R_{t_0}=\frac{S_{t_0} - S_{t_0-dt}}{S_{t_0-dt}}=\exp{((\mu - \frac{1}{2}\sigma^2)dt + \sigma\mathcal{N}(0,dt))}-1.$$
From its decomposition into Taylor's series we know that for small $x$
$$\exp{(x)}-1 \approx x,$$
which makes the conclusion reasonable. With such assumption, for a time interval of length $n$ daily returns' of $d$ stocks are going to pose as $n$-dimensional vectors $R^1,\ldots R^d$. Since PCA algorithm worked on centred data, let us consider normalized returns
$$Y^i = \frac{R^i - \overline{R_i}}{\overline{\sigma_i}}, i=1,\ldots d$$
where $\overline{R_i}=\frac{1}{n}\Sigma_{j=1}^n R_j^i$ and $\overline{\sigma_i}^2=\frac{1}{n-1}\Sigma_{j=1}^n (R_j^i - \overline{R_i})^2$. These can be combined into matrix $\mathbf{Y}$ of size $(d,n)$ with $\mathbf{\widehat{\Sigma}}=\frac{1}{n-1}\mathbf{Y}\mathbf{Y}^T$. Then, normalized eigenvectors $\mathbf{F}=(f_1,\ldots f_d)$ of $\mathbf{\widehat{\Sigma}}$'s together with corresponding eigenvalues $\lambda_1 \geq \ldots \lambda_d \geq 0$ are computed and, in an initial approach, arbitrary number $r<d$ of leading components is picked for transformation matrix $\mathbf{F}_r$. In the final step, we define principal components of $\mathbf{Y}$:
$$F^i=\Sigma_{k=1}^d f_i^{(k)} \cdot Y^k, i = 1,\ldots r$$
which will be referred to as returns of \textit{eigenportfolios} and place them in orthographic projection to $r$-dimensional hyperplane formula (with $n$ samples from $i\leq d$ coordinate considered):
$$Y^i = \Sigma_{j=1}^r f_j^{(i)} \cdot F^j + U^i,i=1,\ldots d$$
where both $F^j,j=1,\ldots r$ and $U^i$ are vectors of length $n$. Since we want to trade on the actual returns and not their normalized versions, let us reformulate the formula above in the following way:
\begin{align*}
Y^i &= \Sigma_{j=1}^r f_j^{(i)} \cdot F^j + U^i / \cdot \overline{\sigma_i}\\
Y^i \cdot \overline{\sigma_i} &= \overline{\sigma_i}\Sigma_{j=1}^r f_j^{(i)} \cdot F^j + \overline{\sigma_i}U^i / + \overline{R_i}\\
Y^i \cdot \overline{\sigma_i} + \overline{R_i} &= \overline{\sigma_i}\Sigma_{j=1}^r f_j^{(i)} \cdot F^j + \overline{\sigma_i}U^i + \overline{R_i}\\
R^i &= c^i + \Sigma_{j=1}^r \overline{\sigma_i}f_j^{(i)} \cdot F^j + (U')^i\\
R^i &= c^i + \Sigma_{j=1}^r \overline{\sigma_i}f_j^{(i)} \cdot \left( \Sigma_{k=1}^d f_j^{(k)} \cdot Y^k \right) + (U')^i\\
R^i &= c^i + \Sigma_{j=1}^r \overline{\sigma_i}f_j^{(i)} \cdot \left( \Sigma_{k=1}^d f_j^{(k)} \cdot \frac{R^k - \overline{R_k}}{\overline{\sigma_k}} \right) + (U')^i\\
R^i &= (c')^i + \Sigma_{j=1}^r \overline{\sigma_i}f_j^{(i)} \cdot \left( \Sigma_{k=1}^d \frac{f_j^{(k)}}{\overline{\sigma_k}} \cdot R^k \right) + (U')^i
\end{align*}
Then, for constant $(c')^i = \alpha_i dt$, parameter $\beta_{ij}=\overline{\sigma_i}f_j^{(i)}$, $n$-dimensional normal variable $(U')^i = dI^i$ and eigenportfolios rewritten as
\begin{equation}
F^i = \Sigma_{k=1}^d \frac{f_i^{(k)}}{\overline{\sigma_k}} \cdot R^k = \Sigma_{k=1}^d Q^i_k \cdot R^k, i = 1,\ldots r,
\label{eq:eigenportfolio}
\end{equation}
we end up with Arbitrage Pricing model from Equation \ref{eq:apm}.\\
Approach of using PCA technique for extract meaningful information from financial data was already tackled in many papers. We can f.e. refer the works of Laloux\cite{pca1} and Plerou\cite{pca2}. They analysed $d=500$ biggest American companies (capitalization-wise) gathered in \textit{SP500} index. Same index basis was taken in M. Avellaneda and J-H. Lee's paper that we mainly reference to throughout this work. Due to the fact that polish stock exchange is far less developed in comparison to NYSE (there are not even $500$ companies participating in GPW) one needs to significantly decrease $d$ in a way that preserves further eigenportfolios market's explainability. Smaller companies do not influence the overall market significantly (and therefore their corresponding dimensions would have near-zero eigenvectors' coordinates) so keeping them as parts of artificial indices is not needed. Additionally, trading with their stocks may distort further results due to higher prices' instability. Since Poland's domestic stock market has around $6$ times less members, to only consider the most liquid and influential polish companies and have them gathered in indices we decided on combining \textit{WIG20} with \textit{mWIG40} making $d=60$.
Additionally, for calculating covariance matrix $\widehat{\mathbf{\Sigma}}$, $n=252$ and PCA parameters $n=252$ samples will be used- this should incorporate similarities between stocks' returns during the whole year's circle.
\begin{figure}[H]
\centering
\includegraphics[scale=0.4]{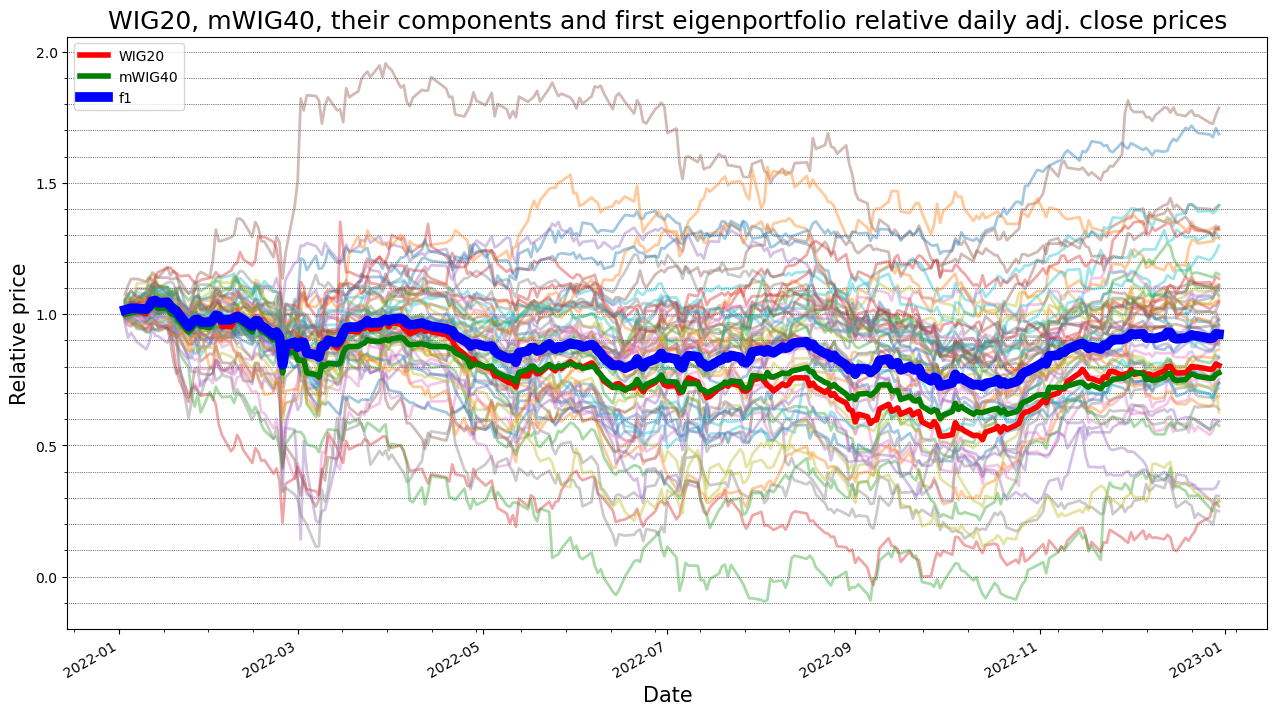}
\caption{Relative eigenportfolio price in comparison to corresponding components and overall indices}
\label{fig:pca_years}
\end{figure}
\noindent Figure \ref{fig:pca_years} shows first PCA decomposition of biggest $d=60$ polish companies. Following previously explained methodology relative adjusted daily prices of all members are shown (with higher transparency) together with corresponding indices and most importantly- first eigenportfolio. Eigenvectors used in the process were calculated based on $252$-days window before 2022. It is clear that first eigenportfolio already possesses main \quotes{flow} of the market tracking both real indices almost perfectly. Highlighting that we did not use capitalization nor total turnover data to create $F^1$ but only actual returns makes the results even more impressive.
\begin{figure}[H]
\centering
\includegraphics[scale=0.4]{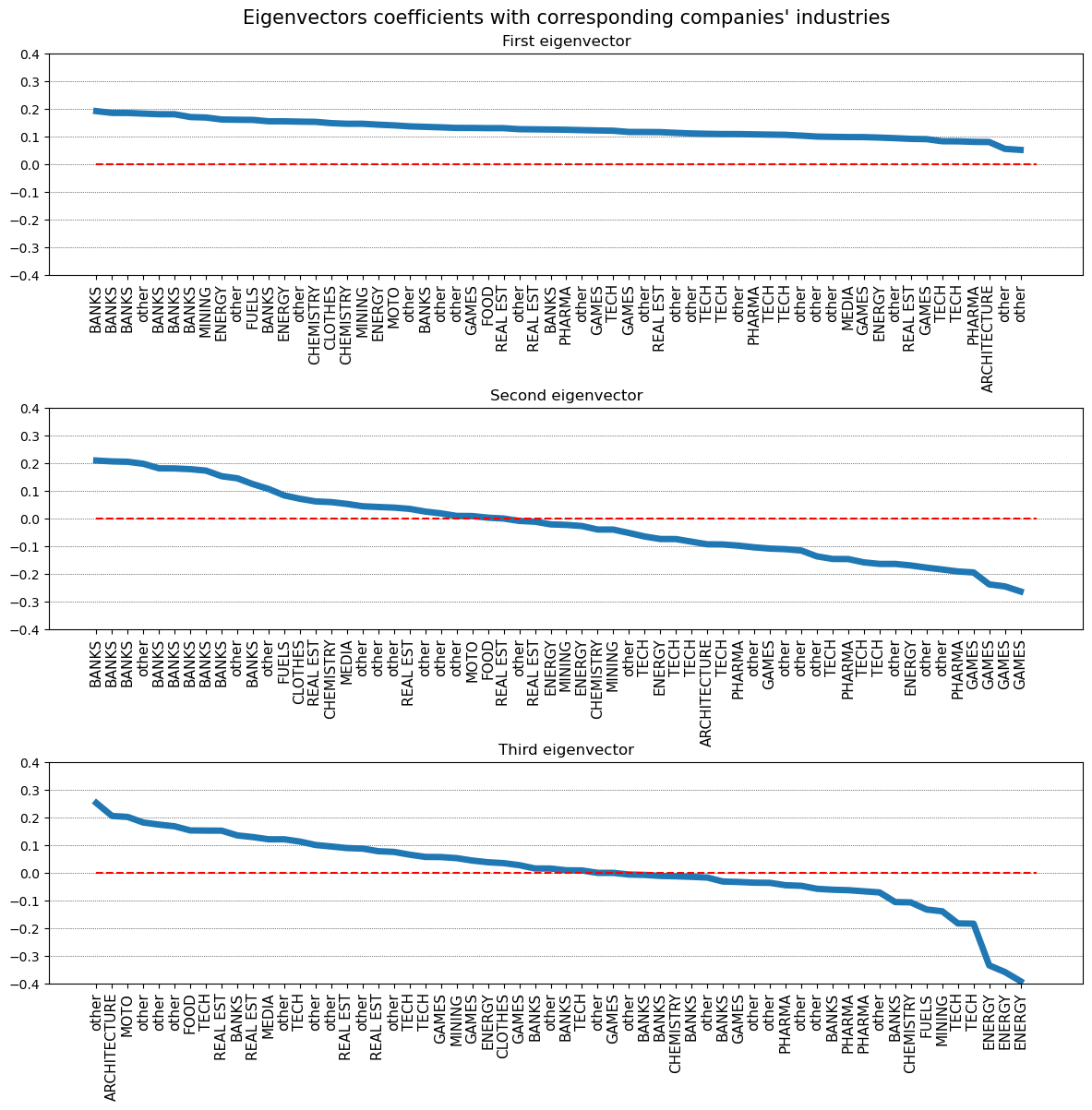}
\caption{Coefficients sizes of first three eigenvectors with assigned industries of corresponding stocks}
\label{fig:pca_eigenvectors}
\end{figure}
\noindent Figure \ref{fig:pca_eigenvectors} shows magnitudes of the first three eigenvectors' coefficients. To give a better insight for readers not so familiar with the polish equities market, companies' tickers were substituted with sector indices' names based on affiliations. Due to low number of indices in Poland some of the companies do not belong to any sector one and were named as \quotes{others}. In the literature of subject first eigenvector shown on the upper picture is referred to as the \quotes{market} one\cite{pca1}. Its coordinates are all positive if the initial correlations between companies' returns were positive- this follows directly from Perron–Frobenius theorem. It was almost the case for our $\widehat{\mathbf{\Sigma}}$ with a few marginal negative values that, as can be seen, did not influence the signs of $f_1^{(k)}, k=1, \ldots 60$. It is clear that the impact of banks in the first eigenportfolio is the highest- this was also true for \textit{WIG20} which we analysed earlier. These are one of the biggest companies on GPW and by carrying loans and deposit influence the entire economy. Note that actual weights $Q^1_k,k=1\ldots60$ of first eigenportfolio members additionally carry the inverse of returns' standard deviations. Higher capitalization companies tend to be more stable (and therefore have lower price volatility) which makes their leading eigenportfolio's share even higher. Switching to second and third eigenvector we can see that some of the coordinates are negative. It is obvious since $f_2$ and $f_3$ have to be orthogonal to $f_1$. In their work, Avellaneda and Lee\cite{main_paper} noticed that stocks of common industries form \quotes{groups} with similar coordinate sizes- on a plot like the one considered they are next to each other. Authors interpreted it as pairs trading on the level of industries. Bigger common factors are offset to achieve the lower ones. We can observe this especially with second eigenvector (middle picture)- banks and some of tech, real estate or games companies are in fact quotes{grouping} together. High impact of banks in the first eigenvector is \quotes{offset} here to achieve independent eigenportfolio capturing lower level trends. For $f_3$ and corresponding $F^3$ we are capturing trends in what was already left by the first two components- such \quotes{pairing} behaviour is also visible although it is not trivial what is offsetting what. With lower components coefficients would become harder and harder to justify since at some point PCA is looking for trends in pure noise.\\
As already seen on Figure \ref{fig:pca_years} using just the first eigeportfolio to decompose unique behaviours of stocks is almost equivalent to considering real indices (in a form of corresponding ETFs). For that reason to fully benefit from PCA technique we need to take more eigenportfolios into account. Obviously using all of them is not optimal since that would offset our entire pairs portfolio to $0$. Crucial question arises then: how many is enough? We can start with looking at how much of total variance we seek to explain.
\begin{figure}[H]
\centering
\includegraphics[scale=0.48]{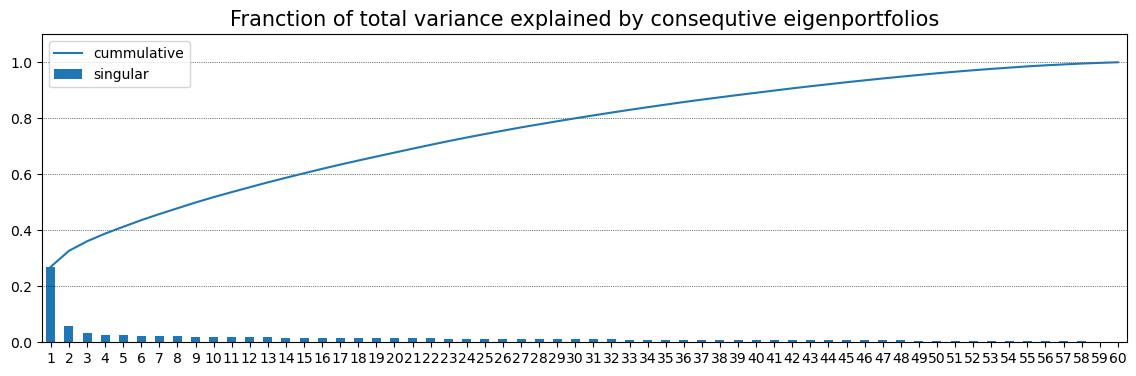}
\caption{Fraction of total variance explained by eigenportfolios}
\label{fig:pca_eigenvalues}
\end{figure}
\noindent As already stated total variance of given returns' data $\mathbf{X}$ with empiric covariance matrix $\widehat{\mathbf{\Sigma}}$ is defined as the sum of all eigenvalues. Each eigenportfolio $F^i$ explains some fraction of the total variance directly proportional to $\lambda_i$. Figure \ref{fig:pca_eigenvalues} shows these fractions for all $d=60$ components we can derive from $\mathbf{X}$. They are presented separately and cumulatively such that all of them explain $100\%$ of volatility. It can be seen that first, \quotes{market} eigencomponent describes more than $15\%$ of total variance with the second one already dropping to around $5\%$. More than half of the volatility explained is achieved by combining $14-15$ components together. To posses explainability higher than $90\%$ one would need to consider more than $40$ components. In our strategy we want to explain so much volatility that what the \quotes{leftovers} can be seen as unique, idiosyncratic movement of each, individual stock. At the same time too much variance explained may lead to no signals for arbitrage since $2$ assets- main and the offsetting one are going to be too close. Additional and practical reason for not using too many eigenportfolios is the existence of transaction costs- higher number of portfolios leads to more transactions and more taxes zeroing any potential profits.
\begin{figure}[H]
\centering
\includegraphics[scale=0.55]{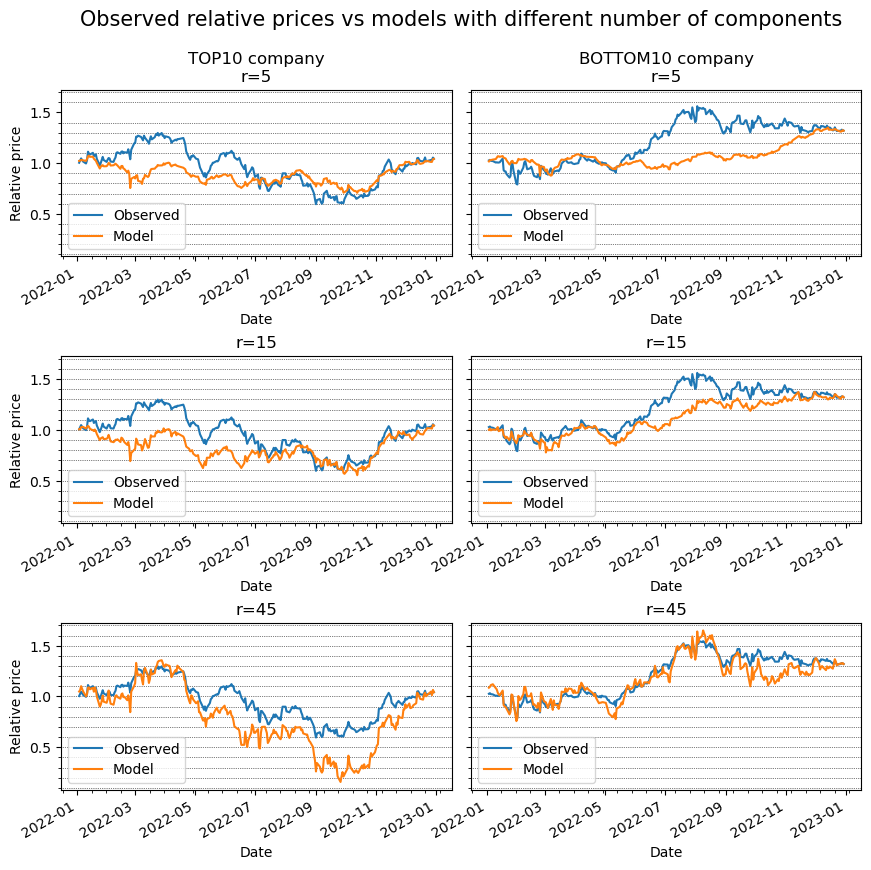}
\caption{Comparison between observed and model's relative prices depending on number of eigencomponents used}
\label{fig:diff_rs}
\end{figure}
\noindent For the same time period as above we picked $2$ companies from our spectrum: KGHM Polska Miedź  and Polenergia - first is one of the leading companies of \textit{WIG20} while the latter one occupies one of the bottom places of \textit{mWIG40}. Figure \ref{fig:diff_rs} presents relative prices of the actual stocks together with their pairing portfolios. Mentioned companies were named according to their rankings and in each row different number of eigenportfolios were put together matching the Arbitrage Pricing model (Equation \ref{eq:apm}). It can be seen that $r=5$ components are enough to decently track KGHM Polska Miedź  which is not the case for Polenergia. On the other hand, $r=45$ seems to add too much \quotes{noise} to the TOP10 company although it gives good predictions for the BOTTOM10 one. A satisfying equilibrium can be observed with $r=15$- main movement of both prices is imitated with still enough space for possible arbitrage opportunities.\\
It turns out that choosing $15$ components was seen as the optimal choice by Avellaneda and Lee\cite{main_paper}. They used the argument that for the biggest economies in the world (members of \textbf{G8}) around $10-20$ components are most relevant for the Arbitrage Pricing Model of stocks' returns. Relying on previous authors' (Laloux (2000)\cite{pca1}, Plerou (2002)\cite{pca2}) findings they also pointed out that such number should explain around $50\%$ of the total variance. As the final argument, $15$ components were also relevant due to similar number of main sectors in country's economy. We shall stick to such approach and also consider $r=15$ as the optimal choice. One can note that Poland is certainly not a part of \textbf{G8} and the number of stocks our predecessors considered was $6$ times larger. At the same time, we are explaining a similar part of national stock market as they did and Poland is highly influenced by the EU which has a special place in \textbf{G8}. 
\begin{table}[H]
\centering
\begin{tabular}{lr}
\toprule
{} &  Explained \\
\midrule
2015 &        0.571 \\
2016 &        0.564 \\
2017 &        0.527 \\
2018 &        0.561 \\
2019 &        0.528 \\
2020 &        0.669 \\
2021 &        0.557 \\
2022 &        0.633 \\
\bottomrule
\end{tabular}
\caption{Fraction of total variance explained by $15$ out of $60$ components throughout years}
\label{tab:3}
\end{table}
\noindent Another, more quantitative argument behind using $r=15$ in our circumstances is that as can been seen in Table \ref{tab:3}, approximately more than $50\%$ of the total variance was in fact explained by $15$ eigenportfolios in last years. All values are based on one-year window PCA fit- combined \textit{WIG20} and \textit{mWIG40}' members were annually adjusted based on historical rankings. COVID-19 recession and 2022 economical crisis resulted in lower volatility of the market- every company was following a bearish movement. It is reflected in higher explainability of $15$ components in the corresponding years. As a second approach (besides considering constant $r=15$) we will also analyse a reverse approach where the fraction of total variance is set to $55\%$ each recalculation we consider a variable number of eigenportfolios trying to achieve it. This may slightly increase the transaction costs in $<55\%$ total variance fraction coming from $r=15$' periods.
\section{Long short-term memory (LSTM) networks}
The PCA approach demonstrated how to represent the market with so called \textit{eigenportfolios}. Each portfolio consisted of all $60$ trading stocks with different amounts depending on eigenvectors of historical returns' correlation matrix. If you then not separate by eigenportfolios, we are basically replicating given company returns by all companies including the main one- \quotes{dummy} \textit{1-1} match is avoided due to the universality of each $F^i$. Such perspective may be an obvious one, but it shows that instead of picking compact portfolios and adjusting coefficients on their level, we can pick weights for all companies in the portfolio directly. This offers us more tailored portfolios with $59$ variables (this time the main company replicated needs to be directly skipped) fitted instead of $r \approx 15$ ones. All coefficients should be based on the current and historical relations between explained and explanatory ones. To find them and capture those $2$ dimensions (time and companies' characteristics) we will consider Recurrent Neural Networks.\\
\subsection{What is a Neural Network?}
Human nerve system consists of special cells called \quotes{neurons} which are connected to each other and transmit electric signals throughout the body. Together with the brain it is a communication and translation system between our concious/unconscious mind's desires and parts capable of fulfilling them. If we, for example,  see a good trading opportunity on the internet, the signal goes from our vision system to the brain and then a message is send to the hand muscles: move the cursor and click on that \quotes{Buy} button immediately. This is of course a simplification but it offers a good basis for understanding the purpose of machine learning technique called \textit{Artificial Neural Network}- possessed information gets translated to desired outcomes after multiple transformations.\\
First artificial neural network was developed before we even considered biological neurons (which can be dated to the end of 19th century)- Gauss (1795) wanted to predict planetary movements with numerical inputs multiplied by a vector of weights and summed up together with additional bias term. To achieve the most appropriate weights and bias he tried to minimize the mean square error between predictions and actual movements. This is obviously the origin of standard OLS regression model. From a neural network perspective, we have $d$ dimensional inputs that are combined together with linear transformation to achieve a scalar output in a single \textit{neuron} with transformation parameters aimed to minimize certain quality determining function called \textit{loss function}. Imagining that the outputs may for example represent probabilities (i.e. be between $0$ and $1$) one could add an additional transformation (\textit{activation function}) to the linear output making it fit the desired range- with \textit{sigmoid function} $\sigma(x)=\frac{1}{1+\exp{(-x)}}$. That would probably also require the change of loss function from MSE to a more appropriate one like the \textit{cross entropy}- then there is no exact formula for weights and bias minimizing it and numerical optimization methods are required. In the end more and more transformation layers combined of linear mapping and activations can be added to the initial input not only shrinking the dimensions' number to $1$ neuron but also increasing it to several ones.
\begin{figure}[H]
\centering
\includegraphics[scale=0.4]{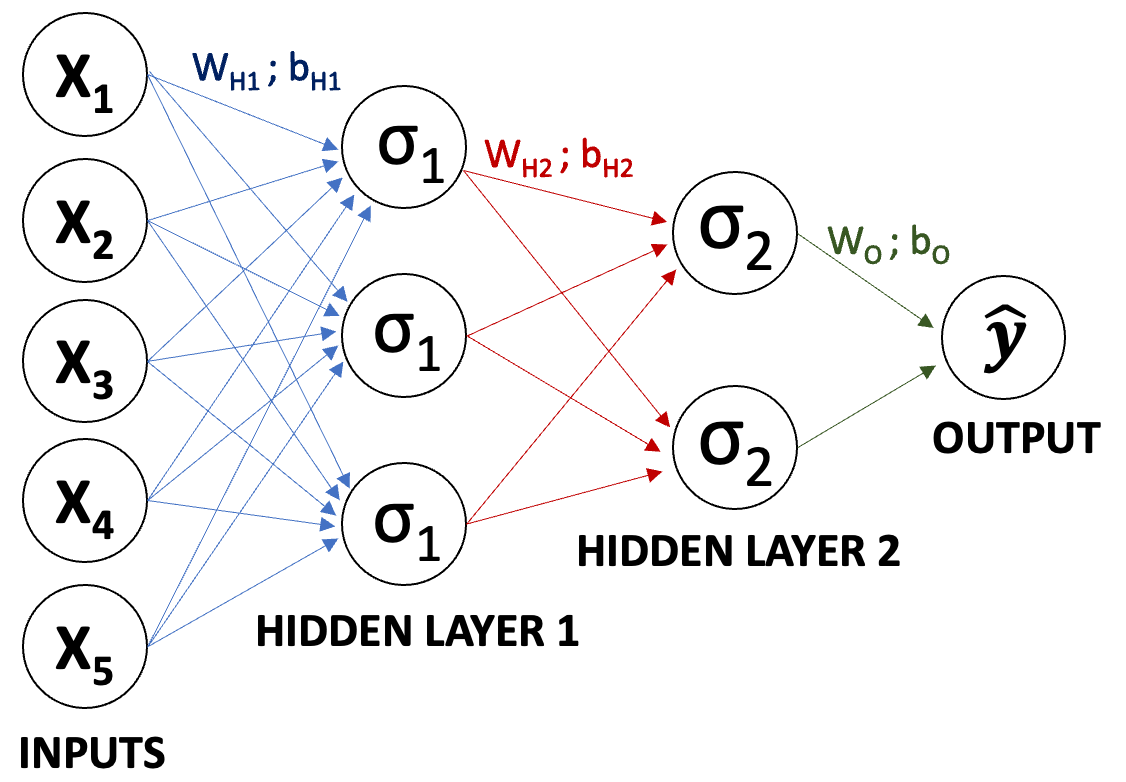}
\caption{Basic $2$ layers neural network schema}
\label{fig:nn_simple}
\end{figure}
\noindent Figure \ref{fig:nn_simple} shows a simple neural networks schema. Inputs are linearly transformed with appropriate weights and biases to $3$ neurons of the first hidden (between inputs and output) layer where activation functions $\sigma$ are applied. Then the results go to the second hidden layer combined of $2$ neurons- both the linear transformation and activation function are again used. The final outcome is a linear regression of $H_2$ outputs- no additional activation function is applied (although it would also be allowed). Formally, for inputs $X=(x_1, \ldots x_5) \in \mathbb{R}^5$ final output of running the network is:
$$\widehat{y} = W_O\sigma_2\left(W_{H2}\sigma_1\left(W_{H1} X + b_{H1}\right) + b_{H2}\right) + b_O,$$
with $W_{H1} \in \mathbb{R}^{3 \times 5}, b_{H1} \in \mathbb{R}^3$; $W_{H2} \in \mathbb{R}^{2 \times 3}, b_{H2} \in \mathbb{R}^2$ and $W_O \in \mathbb{R}^{1 \times 2}, b_O \in \mathbb{R}$. For a set of samples $X_1,\ldots X_N$ with known observed results $y_1,\ldots y_N$ loss function $L:\mathbb{R}^N \times \mathbb{R}^N \longrightarrow \mathbf{R}$ describes the accumulated dissimilarity between predictions $\widehat{y}_1,\ldots \widehat{y}_N$ and actual outcomes. As mentioned, popular choices of $\mathcal{L}$ are mean squared error or cross entropy but $\mathcal{L}$ can be directly adjusted to the task we are dealing with. Since both inputs and actual results are given, loss can be seen as the function of all used weights/biases used in the process of predicting. Each of these parameters ($W_{H1}, b_{H1}$; $W_{H2}, b_{H2}$ and $W_O,b_O$ in presented example) should then be optimized to make $\mathcal{L}$ the smallest possible.  As mentioned, for even the simplest neural network addition of at least one non-linear transformation makes it impossible to find direct parameters' formulas that minimize loss. Numerical techniques are used to approximate arguments of $\mathcal{L}$'s global (or local if $\mathcal{L}$ is not convex) minimum. Due to its simplicity and effectiveness one of the most popular algorithms is \textit{Gradient Descent}.\\
For a function $F:\mathbb{R}^d \longrightarrow \mathbb{R}$ that is differentiable around point $x_0 \in \mathbb{R}^d$, it turns out that starting from $x_0$ $F$ decreases the most if one goes in the direction of negative gradient $\nabla F = (\frac{\partial F}{\partial x_1},\ldots \frac{\partial F}{\partial x_d})$ calculated at $x_0$. This leaves us with an iterative algorithm of:
$$x_{k+1} = x_k - \alpha \nabla F(x_k), k>0$$
with $x_0$ being the initial guess of optimal $x$ and $\alpha \in \mathbb{R}_+$ being the \textit{learning rate} responsible for shrinking the gradient term (to not \quotes{jump over} the local minimum while iterating) and can depend on step number. For small enough $\alpha$, $F(x_{k+1})<F(x_k)$ till the optimal argument $x_{\text{min}}$ is potentially reached with $\nabla F(x_{\text{min}}) = 0$. Usually, for more convoluted functions achieving the $x_{\text{min}}$ is almost impossible- after some iteration steps we usually start oscillating around the optimum not being able to completely zero-down the gradient. It is then important to set a stopping point of the algorithm based f.e. on number of steps, desired $F$ level or small enough gradient magnitude. Gradient descent can be compared to going down the mountain in low-visibility conditions- to arrive at the bottom one should aim against the steepest way upward and frequently re-measure the direction to avoid starting on a different hill (do not try it- the steepest way down usually involves a fall).\\
Within the neural networks' framework differentiable losses and activation functions are usually use to make gradient descent algorithm possible. Partial derivatives can be achieved throughout the \textit{chain rule} i.e. the formula for calculating the derivative of $2$ differentiable functions' composition $h = f \circ g = f(g(x))$. According to the rule:
$$h'(x) = \frac{dh}{dx} = \frac{df}{dg}\frac{dg}{dx} = f'(g(x))g'(x).$$
It can be also extended to multiple compositions- coming back to our example for a single prediction $\widehat{y}$ we can write partial derivative of $\mathcal{L}$ in accordance to $b_{H1}^{(1)}$ in the following way:
$$\frac{\partial L}{\partial b_{H1}^{(1)}} = \frac{\partial L}{\partial \widehat{y}}\frac{\partial \widehat{y}}{\partial b_{H1}^{(1)}} = \frac{\partial L}{\partial \widehat{y}}\frac{\partial \widehat{y}}{\partial \sigma_2}\frac{\partial \sigma_2}{\partial b_{H1}^{(1)}} = \frac{\partial L}{\partial \widehat{y}}\frac{\partial \widehat{y}}{\partial \sigma_2}\frac{\partial \sigma_2}{\partial \sigma_1}\frac{\partial \sigma_1}{\partial b_{H1}^{(1)}}.$$
Same can be showed for all the other biases and weight matrices' coefficients. Notice how using the chain rule we are recreating the steps of neural network in the opposite order- starting from the final prediction up to the initial activation function. For that reason, process of using current coefficients' values to calculate gradients and then improving them by singular step of gradient descent algorithm is called \textit{backpropagation}. Then the training of neural network involves:
\begin{enumerate}
\item Setting up initial parameters.
\item Making predictions based on current parameters' values (propagation)
\item Calculating accumulated loss for all predictions.
\item Deriving gradients and updating parameters (backpropagation).
\item Repeating step $2$.\\
$\vdots$
\item[T.] Terminating algorithm based on gradient descent stopping condition.
\end{enumerate}
Since we are dealing with multiple samples loss function can be calculated using the entire training dataset for each gradient descent step. This is the classic version of the algorithm where the optimized function $F$ stays the same for the entire process. But our actual interest lies on the so called \textit{out-of-sample} data that is not involved in the training- model is suppose to predict the unknown future. Obviously, the assumption is that upcoming data should be at least similar to samples that we trained on (and even that may be a stretch) but an exact match is highly unlikely. How to prepare weights and biases to samples that are not available yet? Method that can help is a stochastic variant of gradient descent algorithm (SGD). Instead of calculating $\mathcal{L}$ on all training samples each time before backpropagation, a random set of $n<N$ samples (a \textit{batch}) is selected to define loss. Batches can also be determined before the algorithm, then each time we are picking a random one and backpropagating based purely on its predictions. One can notice that the optimized function changes slightly throughout steps- it may result in slower convergence since parameters suitable for one batch are not exactly as good for another one. We are trying to minimize according to the common part of each batch- although accumulated loss on the entire data set is going to be larger than in the classic approach, stochastic gradient descent offers more universality to unknown samples especially for smaller batches. It does not completely avoid \textit{overfitting} i.e. trying to reflect the unique noise of the training data (not relevant in testing) but is certainly less sensitive to such error. Also, from a practical perspective- it is performing way faster on large sets due to smaller loss inputs' number in each iteration. There is much more to the topic of gradient descent- further analyses on the performance and additional updates can be found throughout literature~\cite{gd}. Version that we will ultimately use is going ton rely on \textit{Adam} algorithm where SGD is combined with gradient stabilization and scaling.\\
But why are neural networks so powerful? How are they outperforming classic statistical methods such as generalized linear models? It is all about their flexibility. Due to a large number of free parameters (weights and biases) they are able to approximate any non-linear relations between inputs and outputs whereas standard techniques usually require you to select the type of relation before training. Networks try to \quotes{successively refine} and compress the input's signal information to fit the output. Activation functions such as sigmoid or \textit{ReLu}\footnote{ReLU: $\text{ReLu}(x) = \begin{cases} x, x>0 \\ 0, \textit{elsewhere}\end{cases}$} decide whether a certain neuron's input is significant or not for future predictions. Flexibility of neural networks is formally explained by their fundamental theorem from 1989~\cite{nn}:
\begin{theorem}[\textit{Universal approximation theorem}]
Let $C(X,\mathbb{R}^m)$ denote a set of continuous functions from $X \subseteq \mathbb{R}^n$ to $\mathbb{R}^m$. Let $\sigma \in C(\mathbb{R},\mathbb{R})$ be the activation function.\\
Then $\sigma$ is not a polynomial if and only if for every $n, m \in \mathbb{N}$, compact $K \subseteq \mathbb{R}^n$, $f \in C(K, \mathbb{R}^m)$ and $\epsilon>0$ there exists $k \in \mathbb{N}, W_{H1} \in \mathbb{R}^{k \times n}, b_{H1} \in \mathbb{R}^k, W_O \in \mathbb{R}^{m \times k}$ such that:
$$\sup_{x \in K}{||f(x) - W_O \sigma(W_{H1}x + b_{H1})||} < \epsilon.$$
\end{theorem}
\noindent According to the theorem every continuous function can be approximated to a predetermined level by a one layer neural network with enough neurons as long as activation function is not a polynomial. Dual version of the theorem where the number of neurons is predefined and there are unlimited layers to use was also proved for any $f \in L^1$~\cite{nn2}. Both versions of the theorem are not fully practical since they do not give any insight on the size of required network. Nevertheless, they show the power behind using NNs.\\
Neural network that was described on Figure \ref{fig:nn_simple} is a so called \textit{feedforward neural network}. The flow of data is unidirectional- there are no cycles or loops. Also, the length of inputs needs to be predefined. In opposition to these, we are going to use \textit{recurrent neural networks} (RNN) that were designed to re-use transformed inputs multiple times.
\subsection{Recurrent neural networks (RRNs) and Long short-term memory (LSTM)}
Classic, feedforward neural networks don't have a notion for directionality of the inputs. When we are predicting time series, the input is usually a historical window of values- it is then intuitive that the latest values are more important to the prediction than the ones from earlier. Also, feedforward neural networks require a predetermined input size- with time series more flexibility would be desired. For example, although $60$ days' samples were used to speed up the training, we would like to make valid predictions on longer test sets. Recurrent neural networks solve these problems by treating each coordinate of the input vector as separate entry and keeping already processed ones to predict the future.\\
Let us consider a concrete example to better understand how RNNs work. Our task is to predict next day's return of company A based on its historical values. Let's say that a relevant window of previous returns to consider is $X=(x_1, \ldots x_W)$ and the goal is to come up with $\widehat{Y}=\widehat{x}_{W+1}$. 
\begin{figure}[H]
\centering
\includegraphics[scale=0.4]{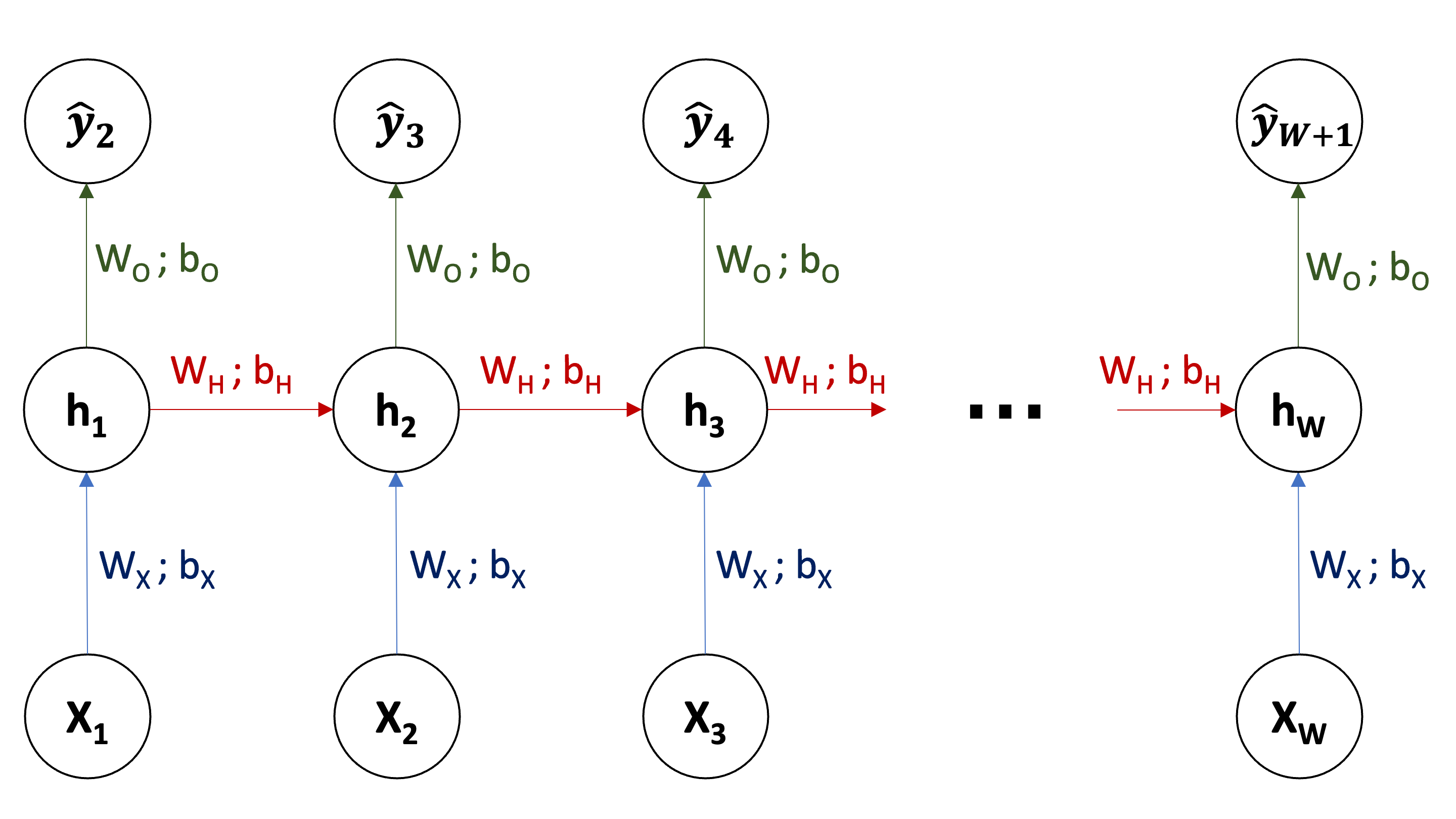}
\caption{Basic $1$ layer unfolded recurrent neural network schema}
\label{fig:rnn_simple}
\end{figure}
\noindent Figure \ref{fig:rnn_simple} shows a $1$ layer recurrent neural network that can be used for the task. Process starts from linearly transforming the first historical value captured $x_1$ by $W_X,b_X$. Although it is not specified on the schema, to achieve $h_1$- hidden state value at time $1$ activation function $\sigma_H$ is used on the input. From $h_1$ prediction $\widehat{x}_2$ can already be made using $W_O, b_O$ and activation function $\sigma_O$- this is an attempt to already approximate the next value- $x_2$. Since we are only interested in $\widehat{x}_{W+1}$ here, it can be skipped. Note how $h_1$ possesses transformed information about $x_1$. It is then adjusted by $W_H, b_H$ and added to $W_X x_2 + b_X$ before applying $\sigma_H$. Expression $\sigma_H(W_X x_2 + b_X + W_H h_1 + b_H)$ represents $h_2$- hidden layer value at $t=2$ capturing information about both $x_1$ (in $h_1$) and $x_2$. Another prediction can then be made from $h_2$. The process continues for the next inputs which are always paired with previous value of the hidden state. In other words, for $k>1$:
\begin{align*}
&h_k = \sigma_H (W_X x_k + b_X + W_H h_{k-1} + b_H)\\
&\widehat{x}_{k+1} = \sigma_O (W_O h_k + b_O)
\end{align*}
The final output $\widehat{y}_{W+1}$ is the one we care about. By handling hidden state values for previous time steps we make sure that not only the latest input is used for prediction but also a transformed version of all previous ones. The role of $\sigma_H$ is then to extract only the features that may be relevant later. Although, we differentiated between both activation functions used- they may be of the same type. All $3$ weights and biases are consistent throughout the process. Presented network is \textit{unfolded}- we assumed a certain length of input vector. This does not need to be the case and for that reason RNNs are often presented in a folded way where only the loop is shown as a recursive step. To train recurrent neural networks a \textit{backpropagation through time} is used. Sticking to considered example, let us assume that $N$ historical returns are used both as inputs and actual $y$-s (using $1$-day forward shift). We can split the training data into windows of size $W$ (i.e. $X_k = (x_{k},\ldots x_{k+W}, Y_k = x_{k+W+1}$ for $k<N-W$) creating $N-W-1$ samples. For each sample, network is unfolded and all coefficients flow through it (like on Figure \ref{fig:rnn_simple}) providing one or multiple predictions which are then used to derive the loss function. Gradients for all parameters are accumulated through backpropagation since each weight matrix and bias vector shows up multiple times in the unfolded network. Finally, through one of gradient descent variants, all coefficients are updated. Although it is a common error in all neural networks, recurrent ones more frequently suffer from the \quotes{vanishing/exploding gradient} problem. When calculating gradients for the impacts of initial state values with the chain rule we have to do a lot of multiplications backropagating throughout the entire unfolded network. If the size of factors is smaller than $1$, we may converge to $0$ and in the other case- diverge to infinity (at least from computer's limited perspective). For that reason, we would like to pass at least some of the information in an unchanged manner- to do that special recurrent neural networks called \textit{Long short-term memory networks} (LSTM) were designed.\\
Long short-term network, as the name suggests, uses $2$ types of information storage: short- similar to one in classic RNNs, and long where data passes without linear transformations. The latter one is then not sensitive to the vanishing/exploding gradient problem thus long-term relations between inputs can be \quotes{remembered}. Typical LSTM unit (one step of the recursion) consists of \textit{cell}, \textit{forget gate}, \textit{input gate} and an \textit{output gate}. Cell stores long-term data in an unchanged manner while all $3$ gates are used to regulate the flow of information to and from the cell.
\begin{figure}[H]
\centering
\includegraphics[scale=0.4]{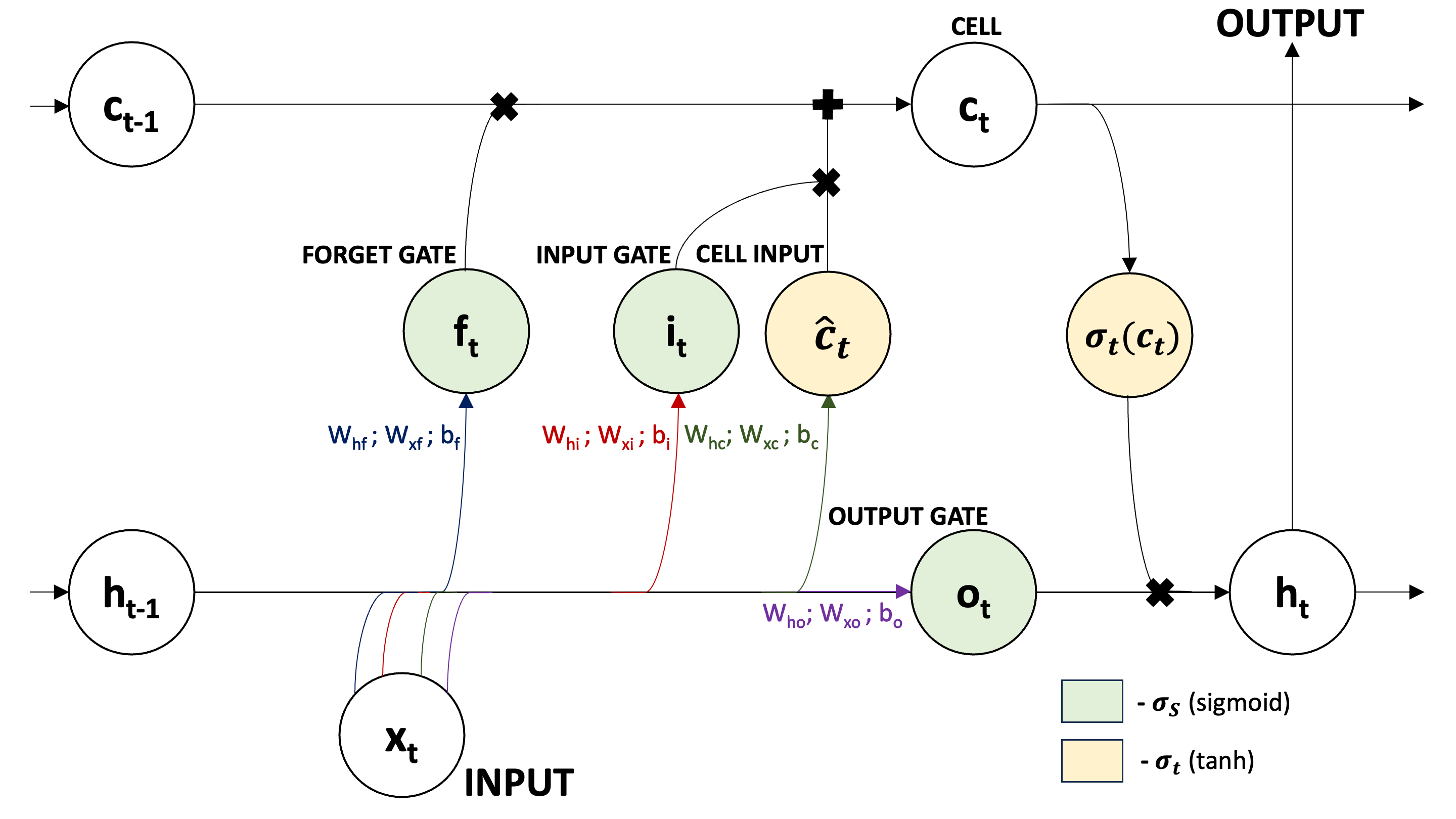}
\caption{One unit of LSTM schema}
\label{fig:lstm_simple}
\end{figure}
\noindent Figure \ref{fig:lstm_simple} presents an unit of a LSTM. There are $2$ main data flows from left to right- of hidden state $h_t$ that ends up being the output and of $c_t$ representing the cell. Given previous hidden state value $h_{t-1}$ and a new input $x_t$, weights $W_{hf},W_{xf}$ and bias $b_f$ are applied together with sigmoid activation function $\sigma_S$ to achieve the forget gate $f_t$. Notice that $f_t$ is between $0$ and $1$- it regulates what percentage of long-term information will be kept. For that reason we multiply (using $\odot$ i.e. element-wise) previous cell $c_{t-1}$ by $f_t$. Both input and previous hidden state value are again used to derive cell input $\widehat{c}_t$- this time instead of sigmoid function hyperbolic tangent $\sigma_t$ \footnote{hyperbolic tangent: $\sigma_t = \frac{\exp{(x)} - \exp{(-x)}}{\exp{(x)}+\exp{(-x)}}$} with values within $[-1;1]$ is used. Cell input is the new data flow coming to $c_t$ as an update of long-term memory. It does not go there entirely, another regulating percentage variable in form of $i_t$ is calculated from linearly transformed $h_{t-1}$ and $x_t$ with $\sigma_S$ additionally applied. Product of $i_t$ and $\widehat{c}_t$ is added to $c_{t-1} \odot f_t$ creating new cell: $c_t$. Meanwhile, in the bottom flow $o_t$ is calculated as the output gate- it regulates how much of the updated long-term memory is going to be used in deriving the output (and for that reason $\sigma_S$ is applied as the activation function). As mentioned, the final output and a new hidden state $h_t$ is a element-wise product of $o_t$ and $c_t$. We can re-write the exact formulas behind the process in the following way:
\begin{align*}
f_t &= \sigma_S (W_{hf}h_{t-1} + W_{xf}x_t + b_f)\text{ (forget gate)}\\
\widehat{c}_t &= \sigma_t (W_{hc}h_{t-1} + W_{xc}x_t + b_c)\text{ (cell input)}\\
i_t &= \sigma_S (W_{hi}h_{t-1} + W_{xi}x_t + b_i)\text{ (input gate)}\\
c_t &= c_{t-1}\odot f_t + i_t \odot \widehat{c}_t\text{ (cell)}\\
o_t &= \sigma_S (W_{ho}h_{t-1} + W_{xo}x_t + b_o)\text{ (output gate)}\\
h_t &= o_t \odot \sigma_t (c_t)\text{ output}
\end{align*}
Notice that as already mentioned- there are no weights applied in the upper stream. Information possessed several steps before is therefore not vanishing on a gradient level. Training not only adjusts the transformation of the new input together with previous ones stored in $h_{t-1}$ to minimize the loss function, but also regulates how relevant are previous inputs to the output and what should be kept for the future. An intuitive comparison to LSTM is one that we started this whole section with: the way our mind functions. When making any decision (from not taking an umbrella for a walk to buying shares of a prospering company) you must assess the current situation but also recall your previous experience. Not all of the previous knowledge is relevant- last time you were just going to the store and now you'll be out for few hours, or last time you lost a lot of money but COVID-19 recession was the main driver behind it. Also the current state may not be so informative- you must find a perfect mix-up between experience and proper current circumstances assessment to decide. And the next time you will be in the same position, today's experience might help.\\
Similarly to the gradient descent algorithm, feedforward neural networks and the classic recurrent ones- there is much more to be said about LSTMs. Nevertheless, to not lose too much focus on the main topic of the thesis, and since we feel like the main idea behind such networks was captured in the short description above, no further theoretical details are going to be explained- with no particular paper mentioned we refer curious readers to the growing literature of the subject. Let us now proceed to how we are actually going to use the \quotes{magic} of Long short-term memory networks to perform statistical arbitrage.
\subsection{Considered approach}
As mentioned at the beginning of this section the aim is to uniquely construct pairs portfolios at the level of single stocks and not entire indices. In other words, for given company's returns we want to adjust weights for all $59$ remaining ones (of \textit{WIG20} and \textit{mWIG40} combined) to explain it. This should allow for higher explainability of stocks' unique behaviours since we are not limited to artificial/real market indices common for all participants. Note that we still remain in scope of the Arbitrage Pricing Theory- each stock can be seen as a separate market indicator (with much lower explainability than a typical index). The use of LSTM instead of intuitive OLS approach is meant to incorporate time-relations between main variable (today's return of given company) and explanatory ones (other companies' historical returns). Training phase should \quotes{teach} the network to also rely on some long-term dependencies and not treat each historical input as equally important. This relativity of predictions is especially important due to the nature of our trading strategy- parameters triggering signal for closing are then calculated \quotes{remembering} inputs when the position was opened. It is important to highlight that the use of LSTMs (or RNNs in general) to derive market portfolios' coefficients seems not to be covered in literature of the subject. Our approach is then the first step in such direction and an initial check whether it has the potential to be developed further. This has many implications, one of them being that some of the technical assumptions will be made on a basis of expert judgements (by the \quotes{rule of thumb})- a full review of different LSTM settings' sensitivity may pose as a natural consequence for future works assuming that this paper's results are going to be promising enough. Let us now consider the technicalities of considered technique in greater detail.\\
Two stacked LSTMs are going to be considered instead of one- they can be seen as $2$ layers of a neural network. As with typical feedforward NN, the increase of depth makes the model more flexible allowing for higher-level, more abstract representations of the inputs. Stacking is understood as using outputs of the first LSTM as inputs to the second one. Predicting particular stock's $R_t$, a single step starts with $59$-elements input vector $X_t$ with other stocks returns at time $t$. First LSTM layer $(1)$ works as follows:
\begin{align*}
f^{(1)}_t &= \sigma_S (W^{(1)}_{hf}h^{(1)}_{t-1} + W^{(1)}_{xf}X_t + b^{(1)}_f)\\
\widehat{c}^{(1)}_t &= \sigma_t (W^{(1)}_{hc}h^{(1)}_{t-1} + W^{(1)}_{xc}X_t + b^{(1)}_c)\\
i^{(1)}_t &= \sigma_S (W^{(1)}_{hi}h^{(1)}_{t-1} + W^{(1)}_{xi}X_t + b^{(1)}_i)\\
c^{(1)}_t &= c^{(1)}_{t-1}\odot f^{(1)}_t + i^{(1)}_t \odot \widehat{c}^{(1)}_t\\
o^{(1)}_t &= \sigma_S (W^{(1)}_{ho}h^{(1)}_{t-1} + W^{(1)}_{xo}X_t + b^{(1)}_o)\\
h^{(1)}_t &= o^{(1)}_t \odot \sigma_t (c^{(1)}_t).
\end{align*}
Then, the second one $(2)$ takes $h^{(1)}_t$ as input and:
\begin{align*}
f^{(2)}_t &= \sigma_S (W^{(2)}_{hf}h^{(2)}_{t-1} + W^{(2)}_{xf}h^{(1)}_t + b^{(2)}_f)\\
\widehat{c}^{(2)}_t &= \sigma_t (W^{(2)}_{hc}h^{(2)}_{t-1} + W^{(2)}_{xc}h^{(1)}_t + b^{(2)}_c)\\
i^{(2)}_t &= \sigma_S (W^{(2)}_{hi}h^{(2)}_{t-1} + W^{(2)}_{xi}h^{(1)}_t + b^{(2)}_i)\\
c^{(2)}_t &= c^{(2)}_{t-1}\odot f^{(2)}_t + i^{(2)}_t \odot \widehat{c}^{(2)}_t\\
o^{(2)}_t &= \sigma_S (W^{(2)}_{ho}h^{(2)}_{t-1} + W^{(2)}_{xo}h^{(1)}_t + b^{(2)}_o)\\
h^{(2)}_t &= o^{(2)}_t \odot \sigma_t (c^{(2)}_t)\\
\beta_t & = \sigma_t(W_{h\beta} h^{(2)}_t + b_{\beta}) = (\beta_{t,1}, \ldots \beta_{t,59}).
\end{align*}
The final output $\beta_t$ will be a vector of the same size as $X_t$, describing coefficients to be used on other returns to describe $R_t$. For the training of weights and biases cumulative loss is defined in the following way:
$$\mathcal{L}(\mathbf{W};\mathbf{b}) = \frac{1}{W}\left(\Sigma_{t=1}^W (R_t - X_t^T\beta_t )^2 + p\frac{1}{59}\Sigma_{i=1}^{59} |\beta_{t,i}|\right),$$
which can be seen as the standard mean squared error of all predictions with additional $L^1$ norm penalty. Such penalty with predetermined intensity $p$ to has the property of zeroing-down the unnecessary coefficients which should decrease the number of transactions needed (since squared differences of returns are going to be small, we will set $p=10^{-5}$ not to overshadow them in the overall loss). As before, $W$ represents window size i.e. the number of recursions of the stacked network- it should indicate for how long time-dependencies should the network be prepared in the testing phase. Singular training sample $\mathbf{X}=(X_1,\ldots X_W)$ has then the size of $59 \times W$ providing a historical period of explaining stocks' returns. After the initial training to tune the network we end up with a tool that can be \quotes{fed} with new sequential data and provide time-dependent coefficients' sets to model $R_t$ not knowing its actual value. One can see this process as \quotes{blind}, recursive OLS where instead of passing the entire training set at once, historical values are given one-per-step and the predicted $\beta$ vector is updated without any knowledge of response variables. As mentioned, OLS would treat each new input as equally important- LSTM aims to rate $X_t$'s explainability first and then combine it with important informations from previous steps.\\
We decided to use $64$ as both the first and second layer's hidden sizes. It is a power of $2$ which is the usual \quotes{rule of thumb} requirement and just above both the input and output sizes. A choice of $128$, at least for one of the layers would also be reasonable but more computationally requiring, especially since single stacked Long short-term memory model is trained to provide parameters for just one predetermined stock. Training of all weights and biases is done with the use of Adam algorithm- an extension of the stochastic gradient descent. Through backpropagation, instead of using just calculated gradient it relies on a moving weighted average with the previously computed one (usually with $1:99$ proportion between them). This technique called \textit{momentum} stabilizes noisy gradient avoiding getting attracted by insufficient local minimas- its disadvantage is that it oscillates vertically resulting in slower convergence. To shrink them making the search more direct Adam combines momentum with \textit{RMSProp} that additionally scales the step size with gradient's quadratic norm weighted average (with $1:999$ proportion between current norm and previous average). The stochastic part of the algorithm relies on batches- $\mathcal{L}$ is averaged over randomly selected samples before gradients calculation. For every step (epoch) of the overall training we will use samples $\mathbf{X}^1,\ldots \mathbf{X}^B$ with $B=16$ as not necessary adjacent $W$-days periods of the training historical data. Training window $W$ (i.e. the first dimension of each sample and the number of recursions throughout LSTM units) is going to be set to $120$ days reflecting half-year circle dependencies. We did not decide to use an entire year for highlighting long-term dependencies due to limited relevant training data (some of the companies are relatively new on the polish market or joined the top $60$ recently so we cannot use $<$2015 historical records) and since trading opportunities come from potential mispricings which should not hold for longer than half a year anyway. Note that model will still be able to work with longer periods' predictions, it may just not hold wider time-relations in its long-term memory.\\
Let us now analyse how the stacked LSTM is working on real data- in the first example ING Bank Polski (bank sector) holding a place in \textit{WIG20} was selected as a potentially easy replicable one due to common movements across most banks. Network was then trained on 2019-2021 data to give appropriate weightings for the portfolio in 2022. In the testing phase we allow LSTM to first \quotes{speed-up} on the second half of 2021 before jumping to the actual testing dataset. This way at the beginning of 2022 we will not have predictions based on one or few historical values only.
\begin{figure}[H]
\centering
\includegraphics[scale=0.31]{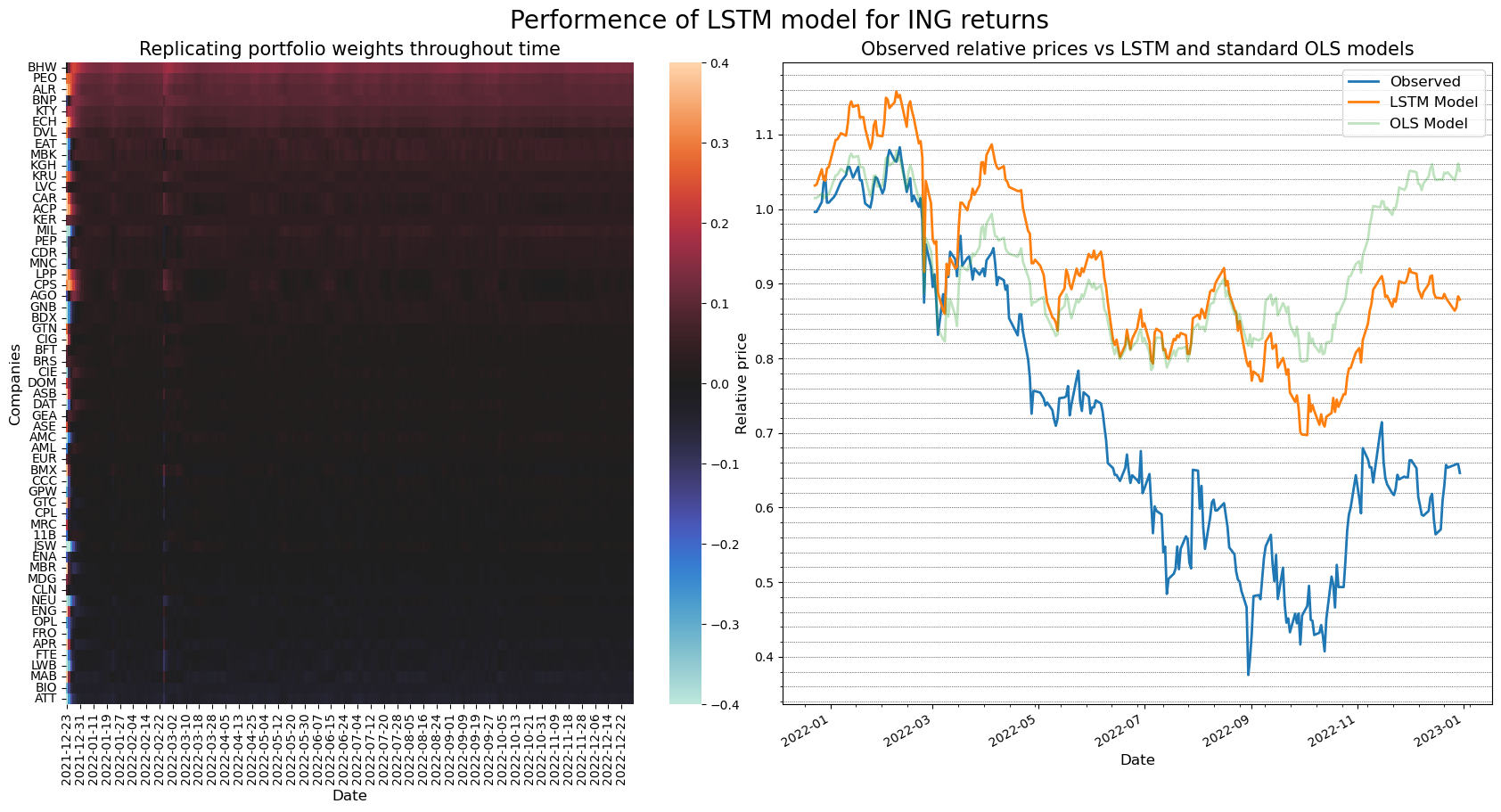}
\caption{Performance of the LSTM model for replicating ING Bank Śląski returns}
\label{fig:lstm_diffs1}
\end{figure}
\noindent Figure \ref{fig:lstm_diffs1} consists of $2$ plots: left one presents the development of $\beta_t$ vector for replicating ING Bank Śląski returns throughout 2022. Companies participating in portfolio were ordered by their average weight. As can be seen, the weightings vary from $-0.5$ to $0.5$ \quotes{vertically} (between components) with small \quotes{horizontal} (time-wise) adjustments. It is not surprising that the most influential companies in the portfolio (BHW, PEO, ALR and BNP) are also banks. Right plot, similarly to previous ones in PCA approach presents the actual relative prices versus ones of the LSTM model. Additionally, standard OLS regression model with constant $\beta$ derived on the 2019-2021 set was also included. Although both prediction lines are quite similar, LSTM seems to replicate actual relative price in a slightly more adjusted manner. 
\begin{figure}[H]
\centering
\includegraphics[scale=0.31]{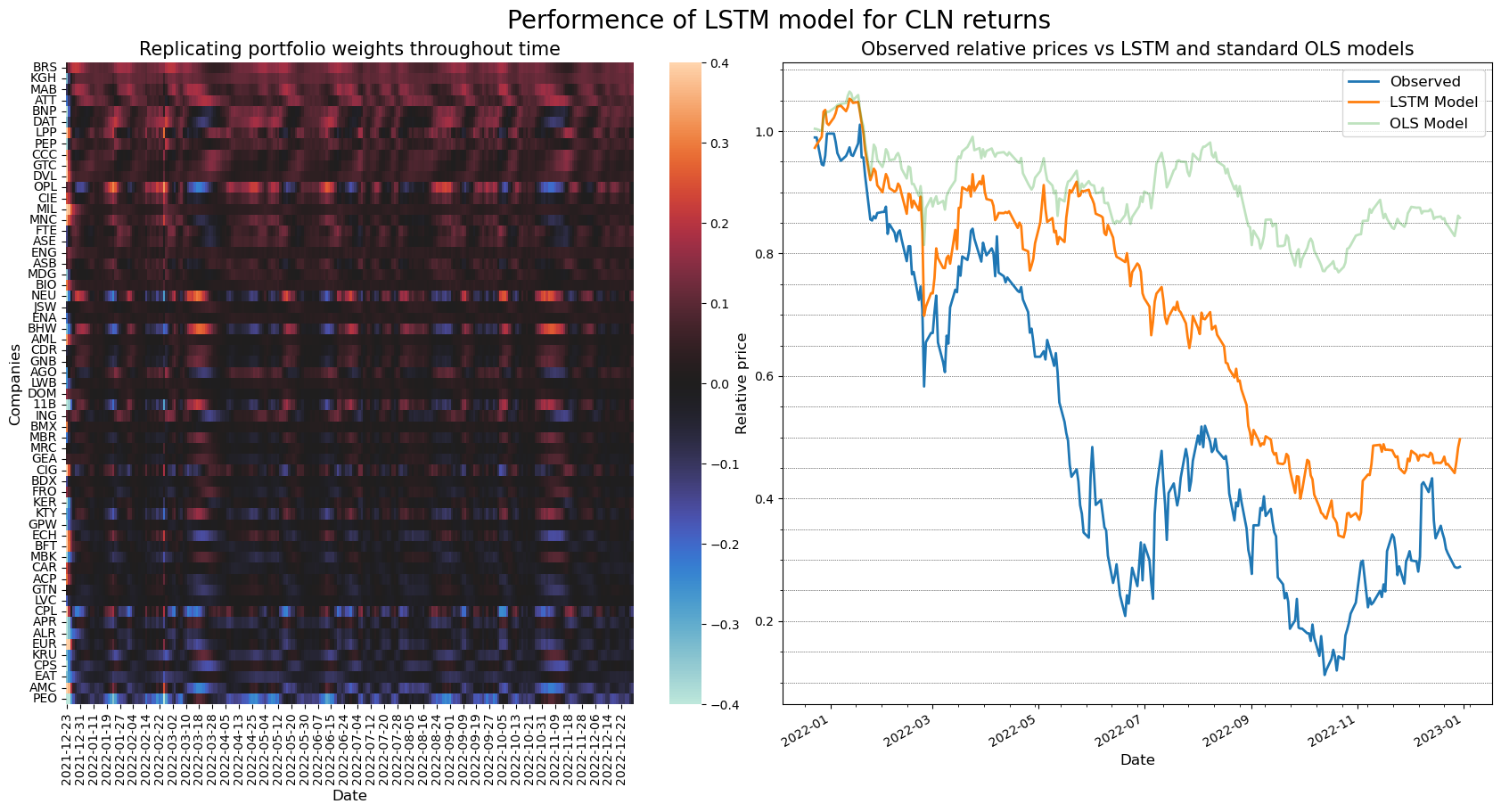}
\caption{Performance of the LSTM model for replicating Celon Pharma returns}
\label{fig:lstm_diffs2}
\end{figure}
\noindent Using the exact same methodology, plots were also generated (Figure \ref{fig:lstm_diffs2}) for Celon Pharma- a pharmacological company from \textit{mWIG40} whose behaviour should be harder to track by most of remaining components. As can be seen, weights are far more spread between all components than with ING. For that reason it is harder to see clear, intuitive connections between most influencing companies and the main one- perhaps except Mabion SA (MAB) occupying the third spot since it is also a pharmaceutical concern. Weights are also more unstable throughout the year with some of them varying between $0$. The predictions themselves are again better than the ones made by standard OLS. It is important to remember that trained (tuned through gradient descent variant) LSTM is just a tool for calculating $\beta_t$ coefficients- the testing phase i.e. providing output vectors for 2022 can be seen as  additional \quotes{blind} training. The actual replicated returns are not used but adjustments to $\beta$ vector are constantly made based on explanatory variables' previously input values. This may be beneficial for the neural network, especially if potential long-term dependencies are captured by adjusted weights and biases.\\
Cases where predictions of LSTM are worse than ones of standard regression can also be found- if no relevant long-term relations are identified network's replicating portfolio is just a more unstable copy of the OLS's one. Nevertheless, as already mentioned, desired potential of LSTM does not fully lie in replication capabilities- we want to incorporate relativeness of proposed weightings such that ones indicating that position should be closed are also based on the opening date's values. This may lead to better identification of actual mispricings  between two porftolios and of their potential corrections. Potential proofs of that are going to be determined by backtesting in relation to different approaches- last one being the use of existing indices' linear combinations instead of artificially adjusted portfolios is going to be further explained in the following section.
\section{Exchange traded funds of market indices}
The use of existing indices as systematic components was already discussed in previous examples (see Figures \ref{fig:regression}, \ref{fig:regression3d} and \ref{fig:signals}). The approach was to use \textit{WIG20}'s returns as the one and only explanatory variable. Although the results were promising, sections concerning artificial indices (especially with the use of PCA) showed us that an all-market factor may not be enough to explain smaller companies stocks' behaviours. Natural approach would be to use sector indices that are more specific than the overall market ones. Such approach was in fact used by Avellaneda and Lee\cite{main_paper}- they considered $15$ exchange traded funds dedicated to various sectors of \textit{SP500} stocks and plugged them into the Arbitrage Pricing model. As already noted, polish equities market does not have any sector ETFs\footnote{This one is not exactly true because there is a sector index with exchange traded fund tracking it- \textit{WIGTECH} which is calculated since 2019 and cover polish tech companies. Due to its short existence in comparison to our backtesting period (2017-2022) we decided not co consider it.}. For that reason a direct transition of our predecessors' approach is not possible. Two substitute solutions will be considered. In a \quotes{sparse} approach, we are going to explain stocks' returns by existing ETFs of $3$ ranking indices. As a \quotes{dense} alternative artificial ones imitating replication of sector indices' portfolios are going to be used. While the second approach (which can be compared to using larger $r$ in the PCA technique shown on Figure \ref{fig:diff_rs}) should give more appropriate residuals than the first, more general one- it is not fully realistic due to ommitment of tracking errors. Staying in the \quotes{number of eigenportfolios} analogy, we expect the \quotes{sparse} technique to perform better with leading companies of considered ranking indices.
\subsection{Existing ETFs approach}
In accordance with previous assumptions we will trade on $60$ components of \textit{WIG20} and \textit{mWIG40}. For that reason, in the \quotes{sparse} approach, the following existing ETFs will pose as model's basis:
\begin{itemize}
\item \textit{BETA ETF WIG20TR},
\item \textit{BETA ETF mWIG40TR},
\item \textit{BETA ETF sWIG80TR}.
\end{itemize}
One can note that the third one is not fully relevant to our stocks' portfolio- nevertheless we decided to add it since it may be more suitable to lower-place companies of \textit{mWIG40} than the actual corresponding fund. All $3$ funds aim to track daily increments of indices' portfolios including all additional payments such as dividends- for \textit{WIG20} it was already shown on Figure \ref{fig:wig20etf-years}.
\begin{figure}[H]
\centering
\includegraphics[scale=0.4]{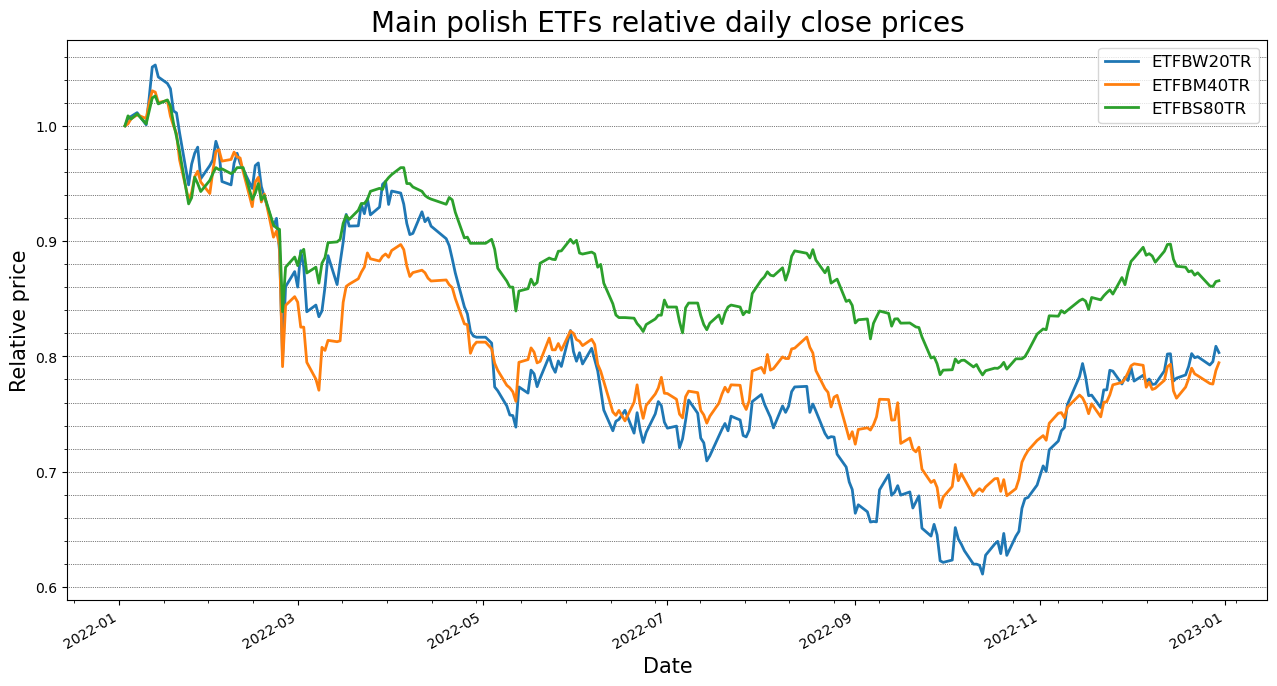}
\caption{Relative daily close prices of \textit{BETA ETF WIG20TR},\textit{BETA ETF mWIG40TR} and \textit{BETA ETF sWIG80TR}}
\label{fig:etfs}
\end{figure}
\noindent Considering ETFs daily returns $F_{\mathit{W20TR}}, F_{\mathit{mW40TR}}, F_{\mathit{sW80TR}}$ and following the Arbitrage Pricing theory (Equation \ref{eq:apm}), for a given stock $i$ we can write:
$$R^i_t = \alpha_i dt + \beta_{i,\mathit{W20TR}} F_t^{\mathit{W20TR}} + \beta_{i,\mathit{mW40TR}} F_t^{\mathit{mW40TR}} + \beta_{i,\mathit{sW80TR}} F_t^{\mathit{sW80TR}} + dI^i_t.$$
One, significant issue with the regression model which was not the case in the PCA approach is that explanatory variables are strongly correlated (as can be seen on Figure \ref{fig:etfs} where one year's relative prices were considered). For that reason, it may not be fully sufficient to keep all $3$ factors in the portfolio cause they provide similar informations. We are fully aware of the issue but since the number of components is already very low, no actions are going to be taken to reduce it. The idea is to use unique behaviours of all funds as an additional insight, no matter how small it is. Furthermore, from a practical perspective owning multiple exchange traded funds does not require that much transaction costs since we do not need to trade with their components. All $\beta$-coefficients expressing how much of funds should be purchased are going to be derived using a standard regression approach of minimizing the mean squared errors.
\begin{figure}[H]
\centering
\includegraphics[scale=0.55]{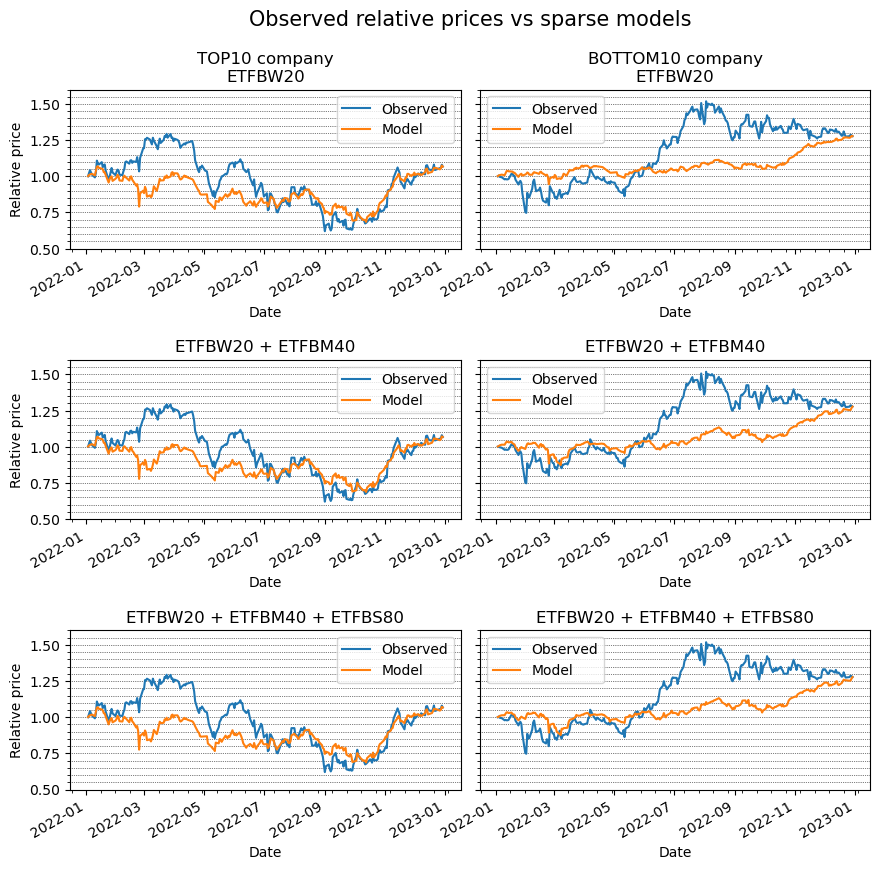}
\caption{Comparison between observed and model's relative prices in existing ETFs' approach}
\label{fig:diffs_etfs}
\end{figure}
\noindent In analogy to Figure \ref{fig:diff_rs}, we consider the same $2$ companies from top and bottom of our scope. Consecutively, $1$, $2$ nd $3$ components in form of exchange traded funds were used to describe stocks' returns (as shown in equation above). Figure \ref{fig:diffs_etfs} shows how observed relative prices from 2022 compare to the predicted ones. By just looking at the plots we can see that both lines are diverging and coming back together with overall trends remaining similar. For KGH Polska Miedź - a TOP10 company, there was practically no gain from adding $2$ additional components besides \textit{WIG20}'s ETF. With Polenergia  again posing as the bottom-place one a small improvement (especially in the first half of 2022) can be seen with second and third factors added.\\
There is one additional problem that was omitted throughout this subsection. Exchange traded funds considered above were introduced after 2018 while our backtesting is supposed to start in 2017. For that reason we will \quotes{extend} their existence period based on corresponding Total Return indices. This simplification assumes that already existing funds got introduced earlier and therefore should not have any significant distorting impacts on final conclusions of strategies' usefulness in the future.
\subsection{Artificial ETFs approach}
As a counter to the existing ETFs' approach let us consider using artificial funds of sector indices. As we all know it is impossible to directly trade on indices- they only pose as benchmarks for analysis and theoretical underlyings of derivative instruments. At the same time, for a given day replicating the exact squad of an index so that the total cost is equal to $1 PLN$, then holding it for up to a quarter and finally closing the trade should give very similar profit (including dividends etc.) as theoretical return coming from the index (assuming it is a Total Return one) on the same time-interval. Obviously, the latter one is more intricate since there exist frequent adjustments of index's portfolio such as changing distribution of shares or introduction of new participants- this results in what is called a \textit{tracking error} between an static replicating portfolio and actual index. If the aim is to profit from market's trends then such error could make it impossible. Nevertheless, strategy of pairs trading benefits from under- and overpricings of stocks in comparison to the market. Gain is then not dependent on the actual trends thus assuming that we are able to replicate the index perfectly should not lead to any additional benefits for the approach itself.\\
Recall that we have considered $14$ sector indices. For each calendar year companies in scope of our interest can be mapped to these sectors- it was already presented on Figure \ref{fig:pca_eigenvectors} for 2022. Not all $60$ participants belonged to one of the indices and some of them cover a great majority of the index they are in (for example PKN Orlen in \textit{WIG-PALIWA}). For that reason we will use all $14$ artificial ETFs for each company returns' model- this should compensate for missing sectors and provide some additional informations from correlated sectors in case of index-determining stocks. Ranking ETFs were skipped in this approach to further differentiate $2$ considered approaches- having $14$ components should already be sufficient enough. Analogous to previous example, Arbitrage Pricing models were trained using 2021's data using $14$ sector indices as explanatory variables and then tested on 2022. This time we are also interested in $\beta$-coefficients' sizes to validate how much information is provided by each factor.
\begin{figure}[H]
\centering
\includegraphics[scale=0.37]{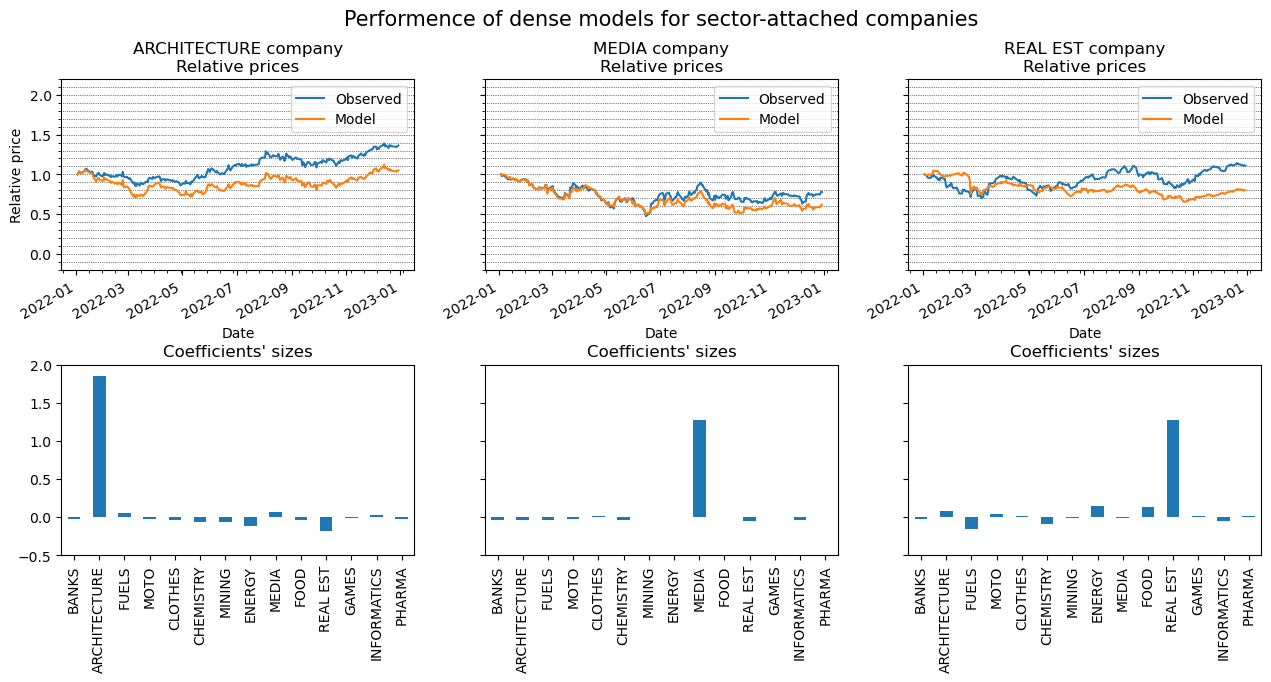}
\caption{Performence of artificial ETFs' approach for sector indices' participating companies}
\label{fig:diffs_etfs_art1}
\end{figure}
\noindent On Figure \ref{fig:diffs_etfs_art1} we can see models' performance for $3$ companies assigned to specific sectors. Sectors were picked to have as wide overview as possible- architecture, media and real estate indices are represented by Budimex, Wirtualna Polska  and Dom Development. In all cases, movement of relative prices is well-represented by the predictions although a non-converging overperformance of the actual values is seen for Budimex  in comparison to the model. Note that in the OU process responsible for fitting the mean-reversion behaviour we also consider such situations due to the presence of non-zero mean $\mu$- we would then profit from any significant deviations from the usual spread between the $2$ assets. In regression models coefficients' sizes indicate how strong is the impact of particular explanatory variables. It is then not surprising that they are the biggest ones are of the same exact indices we picked companies from.
\begin{figure}[H]
\centering
\includegraphics[scale=0.37]{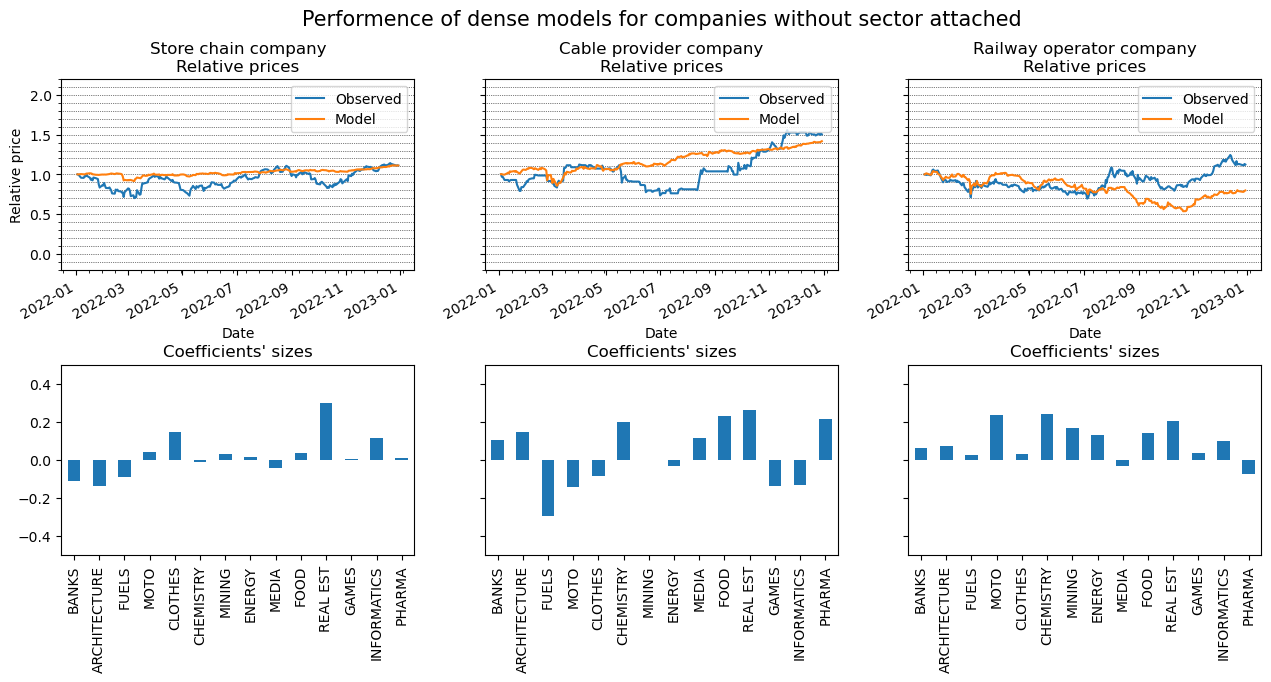}
\caption{Performence of artificial ETFs' approach for companies not participating in any sector index}
\label{fig:diffs_etfs_art2}
\end{figure}
\noindent Companies that do not belong to any sector index were also considered in the same manner- during 2021 they accounted for around $30\%$ of all $60$ companies considered. We again picked representatives from uncorrelated industries- results for Dino Polska (stores' owner), Cyfrowy Polsat (cable provider) and PKP (railway operator) are shown in each column consecutively. Predictions are not as good as in the analysis above but since there was no direct match between stocks' and artificial ETFs they can be seen as sufficient. Explanatory power is far more spread between all indices, some negative impacts are also introduced. We leave detailed analysis on the $\beta$-s magnitudes in relation to companies' characteristics to eager readers- from our perspective more trades are going to be necessary to replicate presented models which will introduce more transaction costs than with sector-attached companies.
\chapter{Backtesting}
The ultimate goal of the following paper is to validate whether statistical arbitrage strategy of pairs trading can be successfully merged with the Arbitrage Pricing Theory using different approaches for portfolio replication. Many of the assumptions, such as the way we transit from residuals of replicated portfolios to actual, transaction determining signals, directly follow an influential paper of Avellaneda and Lee\cite{main_paper}. Our contribution to the discipline is then an addition of a new, deep learning oriented approach and adjustment of real-life conditions to a smaller, far less fluid market of Poland. For the backtesting, a mix-up between mentioned authors' testing rules and our adjustments is also going to be suggested. We will start with setting up the scope of backtesting and some market, quantitative assumptions that needs to be set and later proceed to picked day-to-day trading method description. Next subsection is going to cover individual set-ups for $3$ considered approaches that were not fully determined earlier. We will then proceed with choosing metrics for validation and finally present the results divided into \quotes{standard} and \quotes{recessive} market conditions' ones.\\
Let us begin with setting up necessary background for trading activities.
\section{Principles}
\subsection{Scope of backtesting and market assumptions}
For backtesting purposes it is important to determine what historical data is going to be used period- and recentness-wise. As we already saw (f.e. on Figure \ref{fig:idio-years}) residuals based on daily candles usually fluctuate within months or even quarters. For that reason trading period should be measured in years rather than in days or months. Avellaneda and Lee\cite{main_paper} performed day-to-day trading simulation for $11$ or $6$ years depending on the approach (between 1997/2002 and 2007 stopping before 2008's global crisis). We decided to pick a shorter period of $3$ trading years between 2017 and 2019 due to the fact that squad of \textit{SP500} considered by mentioned authors is far more stable participants-wise than top $60$ \textit{WIG}'s companies. Using $10$ years would require annual re-builds of the entire portfolio adding additional costs not connected to the strategy itself. Within the $3$-years interval $60$ companies that were in \textit{WIG20} or \textit{mWIG40} with its reserves list during the period will be used. Starting year of 2017 was selected to keep the backtesting results relevant and to separate COVID-19 pandemic economic effects out of the scope (similarly to how 2008 was avoided in referenced paper). Such bearish period (2020) is going to be used separately as an additional check whether strategy can profit even in case of strong recessive trend of the market.\\
To incorporate dividends and additional payments in transaction profits, we will buy and short-sell stocks for adjusted close prices with no limitation on shorting. Transaction cost of $c=0.1\%$ per buy/sell is going to be added. Even though the GPW (polish stock market) taxation tables suggest higher percentages of around $0.29\%$, pages with lower fees are available among traders. It is important to note that Avellaneda and Lee' set the only transaction costs to $0.05\%$  which is $2$ times smaller than our assumption. This will be an important factor in validating our results against theirs. Last but not least, so-called \textit{risk-free rate} reflecting time-value of money needs to be determined. The usual approach is to consider rates of zero-coupon government bonds with appropriate duration as their risk is practically negligible. For $<5$ years of constant trading, $52$-weeks bond' yield to maturity is the most appropriate one- based on historical data from GPW $r_f = 1.5\%$ in the main period of 2017-2019 and $r_f=0.1\%$ in 2020 are going to be used as constant rates. For simplicity $r_f$ is going to apply for both owned and owed money time-development.
\subsection{Trading rules}
Let us recall that the strategy for each trading-day and given company goes as follows:
\begin{enumerate}
\item Construct APT model based on eigenportfolios, all remaining companies' stocks combined or existing indices on a theoretical basis (i.e. determine all necessary weights with one of $3$ approaches).
\item Calculate residuals of the model on last $W$ days.
\item Determine Orstein-Uhlenbeck process's parameters from the residuals.
\item Calculate value of $G_t$ (normalized idiosyncratic component) for current day.
\item Based on thresholds decide on the type of potential transaction.
\item[6a.] If you hold a position and $G_t$'s value indicates closing it- do so.
\item[6b.] If you do not hold a position and $G_t$'s value indicates opening it- do so based on weights determined in point 1.
\item[6c.] Else: do nothing.
\end{enumerate}
First point is mainly dependent on the approach we took and for that reason was already explained in individual sections- any remaining technicalities are going to be specified later. Other parts of presented algorithm are universal except for the actual values of signals' thresholds which are going to be optimized on 2014-2016 interval (using the same validation metrics and risk-free rate as for the main backtesting) for each technique separately. We decided to unify window used for residuals' calculation (not to mistake with individually set windows used to train both PCA and LSTM models) to $W=120$- for that reason stocks mean-reverting within $60$ days (with $\kappa>4$) should be considered. 
\begin{table}[H]
\tiny
\centering
\begin{tabular}{ccccccccccc}
\toprule
& \multicolumn{2}{c}{Constant PCA} & \multicolumn{2}{c}{Variable PCA} & \multicolumn{2}{c}{LSTM} & \multicolumn{2}{c}{Real ETFs} & \multicolumn{2}{c}{Artificial ETFs}\\
\midrule
{} &         $\kappa$ &   $\tau$ (days) &         $\kappa$ &   $\tau$ (days)&         $\kappa$ &   $\tau$ (days)&         $\kappa$ &   $\tau$ (days) &         $\kappa$ &   $\tau$ (days)\\
\midrule
ARCHT &	 $17.05$  	&	$ 14.78$  	&	$ 21.61$  	&	$ 11.66$  	&	$21.07$	&	$11.96$	&	 $24.13$  	&	$10.44$	&	 $17.84$  	&	$ 14.12$  	\\
BANKS	&	 $22.56$  	&	$ 11.17$  	&	$ 21.72$  	&	$ 11.60$  	&	$21.01$	&	$11.99$	&	 $26.77$  	&	$9.41$	&	 $26.26$  	&	$ 9.60$  	\\
CHEM	&	 $15.80$  	&	$ 15.95$  	&	$ 14.93$  	&	$ 16.88$  	&	$20.69$	&	$12.18$	&	 $19.30$  	&	$13.05$	&	 $21.43$  	&	$ 11.76$  	\\
CLOTHES	&	 $13.44$  	&	$ 18.76$  	&	$ 12.49$  	&	$ 20.17$  	&	$17.56$	&	$14.35$	&	 $19.28$  	&	$13.07$	&	 $27.05$  	&	$ 9.31$  	\\
ENERGY	&	 $18.43$  	&	$ 13.67$  	&	$ 17.20$  	&	$ 14.65$  	&	$18.87$	&	$13.36$	&	 $14.92$  	&	$16.89$	&	 $16.09$  	&	$ 15.66$  	\\
FOOD	&	 $15.47$  	&	$ 16.29$  	&	$ 14.44$  	&	$ 17.45$  	&	$20.05$	&	$12.57$	&	 $16.76$  	&	$15.04$	&	$ 22.83$  	&	$ 11.04$  	\\
FUELS	&	 $24.79$  	&	$ 10.16$  	&	$ 25.19$  	&    $10.00$  	&	$19.12$	&	$13.18$	&	 $20.00$  	&	$12.6$	&	 $20.32$  	&	$ 12.40$  	\\
GAMES	&	 $22.25$  	&	$ 11.32$  	&	$ 23.73$  	&	$ 10.62$  	&	$27.05$	&	$9.32$	&	$ 23.91$  	&	$10.54$	&	 $27.87$  	&	$ 9.04$  	\\
INFRMTCS	&	 $22.40$  	&	$ 11.25$  	&	$ 21.72$  	&	$ 11.60$  	&	$21.45$	&	$11.75$	&	 $23.81$  	&	$10.58$	&	$ 21.39$  	&	$ 11.78$  	\\
MEDIA	&	 $21.04$  	&	$ 11.98$  	&	$ 23.54$  	&	$10.71$  	&	$17.28$	&	$14.58$	&	 $19.20$  	&	$13.13$	&	$ 20.58$  	&	$ 12.24$  	\\
MINING	&	 $22.33$  	&	$ 11.29$  	&	$ 25.49$  	&	$9.89$  	&	$21.55$	&	$11.69$	&	 $19.41$  	&	$12.99$	&	$ 21.08$  	&	$ 11.95$  	\\
MOTO	&	$ 20.22$  	&	$ 12.47$  	&	$ 22.39$  	&	$11.26$  	&	$21.85$	&	$11.54$	&	 $22.30$  	&	$11.3$	&	$ 18.87$  	&	$ 13.36$  	\\
PHARMA	&	 $16.82$  	&	$ 14.99$  	&	$ 15.05$  	&	$ 16.75$  	&	$20.37$	&	$12.37$	&	$ 20.73$  	&	$12.16$	&	 $20.16$  	&	$ 12.50$  	\\
REAL EST	&	 $20.85$  	&	$ 12.09$  	&	$ 21.69$  	&	$ 11.62$  	&	$23.55$	&	$10.7$	&	 $29.60$  	&	$8.51$	&	 $31.38$  	&	$ 8.03$  	\\
other	&	 $22.78$  	&	$ 11.06$  	&	$ 20.86$  	&	$ 12.08$  	&	$22.95$	&	$10.98$	&	 $25.17$  	&	$10.01$	&	 $23.69$  	&	$ 10.64$  	\\
\bottomrule
\end{tabular}
\caption{Average mean-reversion speeds and their reverses across 2019 and industries based on moving $120$-days windows}
\label{tab:4}
\end{table}
\noindent Table \ref{tab:4} presents averaged mean-reversion speeds $\kappa$ calculated for each day of 2019 and for each participating company together with $\tau$- their inverse scaled with $\frac{1}{dt}=252$ representing how many days, on average, are needed for technical arbitrage opportunity to be corrected. Ornstein-Uhlenbeck's parameters were calculated based on rolling $W=120$ days window. As can be seen,  energy and food industries' portfolios are mean-reverting the slowest with games and real estate sectors' ones being the fastest. This is obviously strictly correlated to individual behaviour of stocks within sectors throughout 2019- nevertheless average $\kappa$s larger than $4$ are a promising sign for all approaches. As a \quotes{sanity check} it is worth noticing that even though considered approaches have different residuals- their final OU parameters are comparable.\\
All $60$ traded stocks are going to be used for replication purposes together with real or artificial ETFs. Trading on every stock's residuals portfolio is going to be completely independent from other ones thus final results can be seen as sum of $60$ \quotes{traders}' incomes with each one focusing on one company only. Since $2$ out of $3$ main approaches use stocks as both explanatory and explained variables, transfer of already possessed assets between \quotes{traders} would decrease some transaction costs connected to buying shares to open position and simultaneously short-selling some to close another one (for different main company). For simplicity and to keep track of every opened position (to successfully close it when appropriate signal comes) we will not implement such common trading account. Even though some weights of individual replicating portfolios (f.e. $\beta$ coefficients in PCA approach or all weightings in the LSTM one) are calculated daily to produce new set of OU process' coefficients and thus generate signals, amounts defined by initial weightings for already opened position are not going to be adjusted throughout time. Using a specific example that can be translated to all techniques: if signal for \textit{open long} appears in the real ETFs approach for a given day and stock $i$ we are going to buy $1$ unit of stock $i$ and short-sell relevant $\beta_{i,\textit{W20TR}},\beta_{i,\textit{mW40TR}}$ and $\beta_{i,\textit{W20TR}}$ amounts of appropriate funds. Then, till a new signal comes for the same stock indicating a optimal moment to close long position, created portfolio is going to remain untouched. This is not ideal theoretically since closing signal is generated on a basis of potentially different weightings than ones we actually own but, from a more practical perspective this skips multiple cost-generating transactions. To further justify- within a maximum of few months we do not expect significant changes in generated weightings since they are based on a long enough historical periods. Although it was already mentioned in signals generation' section, let us also recall that only one position  can be opened at a given time moment for concrete main company. Therefore all \quotes{traders}' states can always be described with $-1,1$ and $0$ indicating long position opened, short position opened and no position (empty portfolio) consecutively. Near the end of the $3$ years trading period ($\frac{W}{2}=60$ days before) we will stop any potential openings to optimally close most of owned positions before the final day. Any position still opened at $t=3$ (years) is going to be sold-out to completely empty the overall portfolio.\\
There are many takes to how much capital shall be used for trading purposes. Invested money can be dependent on already-made profits or adjusted to signals magnitudes. We have decided to stick with amounts proportional to current equity state. Additionally a $2:1$ leverage ratio is assumed. Leverage is about magnifying potential profits by borrowing money for the investments, $2$ to $1$ means that for each $1$ PLN own $2$ PLNs can be spend on both types of transactions (long/short)- it is a fairly common and not significantly risky assumption on broker accounts. Each transaction i.e. opening position for given stock $i$ at $t_0$ and closing it at $t_1>t_0$ generates a profit of:
\begin{align*}
P^i_{[t_0,t_1]} &= \Lambda_t[(-1)^{(\text{transaction}=\text{\textit{short}})}\cdot (R^i_{[t_0,t_1]}-Q^M_tR^M_{[t_0,t_1]}) -\\
 &-(-1)^{(\text{transaction}=\text{\textit{short}})}\exp{(r_f(t_1-t_0))}\cdot(1 - Q^M_t) - \\
 &- c\cdot\left(\exp{(r_f(t_1-t_0))}|1 +  Q^M_t| + |(1+R^i_{[t_0,t_1]}) + Q^M_t(1+R^M_{[t_0,t_1]})|\right)],
\end{align*}
where $R^i_{[t_0,t_1]},R^M_{[t_0,t_1]}$ are returns from the main stock and its replicating portfolio consecutively, $Q^M_t$ is a total PLN amount that was invested in all participants of the replicating portfolio (taken directly from the Arbitrage Pricing model of returns), $\Lambda_t$ is the scaling factor and $r_f, c$ are as already defined. Note that if the transaction is a short one, we sell scaled $1$ PLN of the main entity and buy the replicating portfolio- for that reason $(-1)^{(\text{transaction}=\text{\textit{short}})}$ was added. The last part of the formula is the transaction fee- it indicates that cost has to be accounted for both opening and closing the transaction. For initial amount of money $E_0$ let us define summed, accumulated equity $E_t$ as:
\begin{equation}
E_t = E_0\exp{(r_f t)}+ \Sigma_{i=1}^{60} \Sigma_{t_0,t_1 \leq t} P^i_{[t_0,t_1]}\exp{(r_f(t-t_1))}.
\label{eq:e_t}
\end{equation}
It sumps up profits that happened till $t$ for all $N=60$ companies. Then, with the assumption of $2:1$ average leverage level, previously used $\Lambda_t$ can be formally written as:
$$\Lambda_t = \frac{2}{60} P_t.$$
Factor of $\frac{2}{60}$ is determined by the fact that maximum of $60$ companies' positions can be opened at given moment and we since are allowed to trade with twice as much money as we own. Parameter $E_0$ is going to be set at $100$ PLN making initial individual trades scaled by $\Lambda_{dt}\approx 3.33$ PLN. As already mentioned, we are going to trade on adjusted closing prices- this will take into account any dividends coming from owning stocks automatically. As can be seen all trading assumptions are fairly simple and basic- it is because while validating the results we aim to put an emphasis on the mathematical theory behind the strategy. The following thesis is therefore more of a check whether our take on statistical arbitrage has the necessary potential to generate any repetitive and scalable profits.
\section{Individual set-ups}
Some of the choices that we had to make in the process of constructing the entire framework are native to individual characteristics of taken approaches- especially in case of PCA and LSTM. Additionally, optimal signal thresholds are also assumed to be uniquely selected for each approach. Let us then go through all techniques and fill in the blank spots. Like when explaining the paired portfolio generation approaches, we will start with the Principal Components Analysis technique.
\subsection{Principal Components Analysis (PCA)}
Recall that in the PCA approach we will use a constant $r=15$ number of eigenportfolios to explain stocks or, as a separate sub-technique: variable $r$ such that at least $55\%$ of the total variance is explained. In both approaches eingenvectors will be recalculated once every $252$ days making a total of $3$ changes in our main trading interval of 2017-2019. The reasoning behind not doing it more frequently is as we already saw in PCA dedicated section, eigenportfolios are capable of producing weights appropriate also for next year's data not participating in training phase. To justify further, squad of real indices is usually fairly stable with only slight adjustments within a given year. When the time for recalculation comes correlation matrix and thus eigenvectors are going to be derived on a basis of $252$ preceding days. For example, in the entire 2017 we will use eigenportfolios with weightings from 2016 training data and switch to 2017-trained ones at the beginning of 2018- in a variable components approach, these are the only opportunities to change $r$. During additional analysis of 2020, the only difference between two PCA methods will be the initially selected $r$. Note that even though eigenportfolios' weights stay the same within days- $\beta$ parameters corresponding to $F^k, k=1,\ldots r$ are daily recalculated for signals' generation purposes.\\
A basic search for optimal thresholds $\overline{g}_{ol},\overline{g}_{os},\overline{g}_{cl}$ and $\overline{g}_{cs}$ was performed on 2015-2016 data assuming same trading technique and market parameters as in the main backtesting. For simplicity and symmetry opening and closing signals are assumed to be the same pair-wise with $\overline{g}_{ol},\overline{g}_{os} \in [1.1;2.1]$ and $\overline{g}_{cl},\overline{g}_{cs} \in [-2.0;-1.0]$. To validate the best combination we will consider with the final accumulated profit i.e. $E_T - E_0$. All runs are identical except for the thresholds so there seems to be no need for more sophisticated metric.
\begin{figure}[H]
\centering
\includegraphics[scale=0.4]{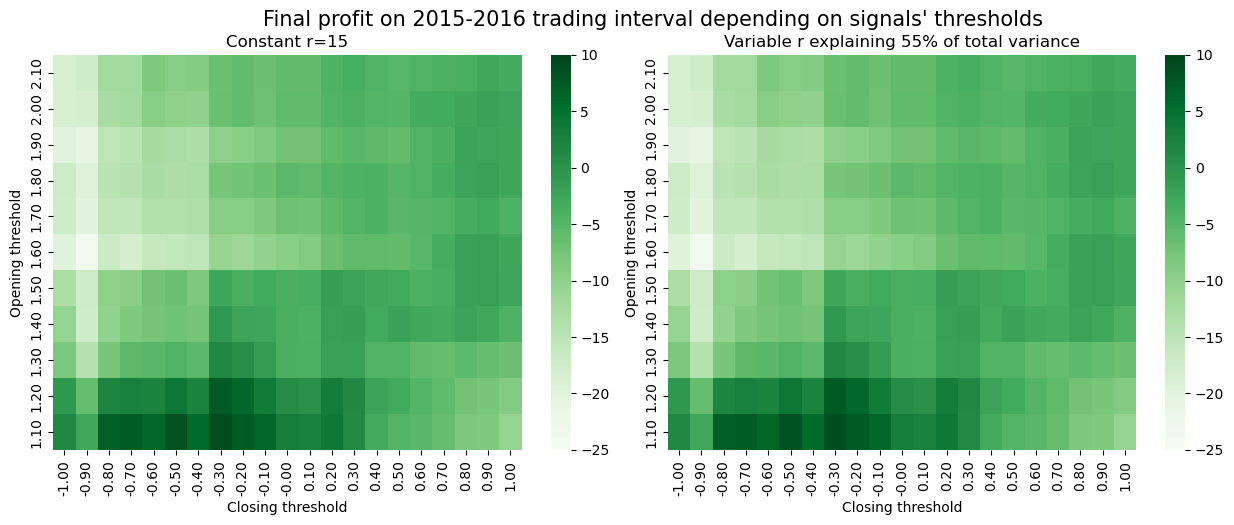}
\caption{Final accumulated profit on the optimization interval for PCA approaches depending on signal cut-offs values}
\label{fig:pca_opt}
\end{figure}
\noindent Figure \ref{fig:pca_opt} presents heatmaps of the final $E_t-E_0$ with given combinations of opening and closing thresholds for both versions of the PCA approach. As can be seen only some of the runs ended up profiting with the best ones having the opening signal $\overline{g}_{ol}=\overline{g}_{os} \in [1.1;1.3]$ and closing one $\overline{g}_{cl}=\overline{g}_{cs} \in [-0.8;-0.3]$. The latter one suggests that the best way to close potential arbitrage opportunities is waiting till $G_t$ not only hits $0$ but also starts destabilizing in the other direction. Note that both sub-approaches gave practically identical results- we may conclude that $r\approx15$ was selected twice to match $55\%$ desired total variance explained ratio. Let us now consider a narrower search grid to find the optimal thresholds.
\begin{figure}[H]
\centering
\includegraphics[scale=0.4]{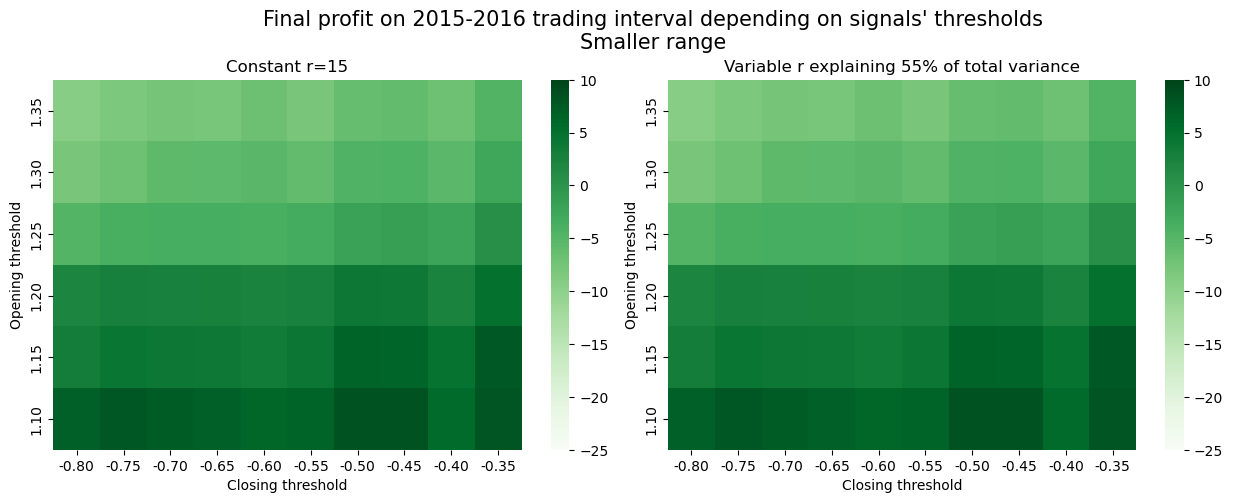}
\caption{Final accumulated profit on the optimization interval depending on signal cut-offs values- second search}
\label{fig:pca_opt2}
\end{figure}
\noindent As before, Figure \ref{fig:pca_opt2} presents final accumulated profits for both sub-strategies- this time within a narrower range of thresholds. It is clear that for both sub-approaches $\overline{g}_{ol}=\overline{g}_{os}=1.10$ and $\overline{g}_{cl}=\overline{g}_{cs}=-0.50$ give the best results oscilating around $10$ and for that reason they will be used for backtesting.
\subsection{Long short-term memory (LSTM) networks}
In the LSTM approach we need to first tune network's weights and biases so that it can correctly use historical inputs to predict appropriate set of $\beta$ coefficients. This part is going to occur once a year based on $3$-years historical window. In other words, LSTM tuned on 2014-2016 will be run $W=120$ days before the start of the next year to start giving valid predictions when 2017 comes and later stopped as 2018 arrives. Then a new tuning, this time relying on 2015-2017 data is going to be performed and a new network will be run in a similar manner. Note that we are not stopping the trading, only the LSTM gets interrupted after a year and $120$ days of running. As in the PCA approach we will then have $3$ full trainings throughout the main testing period. During each tuning, samples of length $120$ are going to be used to make the network focus mainly on this long windows in the derivation of weightings vectors. For 2020 analysis, network is going to be trained only once based on 2017-2019 data.\\
Like before we constructed $2$ grid-searches for optimal thresholds- with a wider and narrowed range of possible pairs. Since the stacked LSTM network had to be first tuned on years 2012-2014 for 2015's run it was necessary to adjust the scope of considered stocks such that they all existed back in 2012. For that reason optimization is performed on slightly different set of companies than in the PCA sub-approaches.
\begin{figure}[H]
\centering
\includegraphics[scale=0.4]{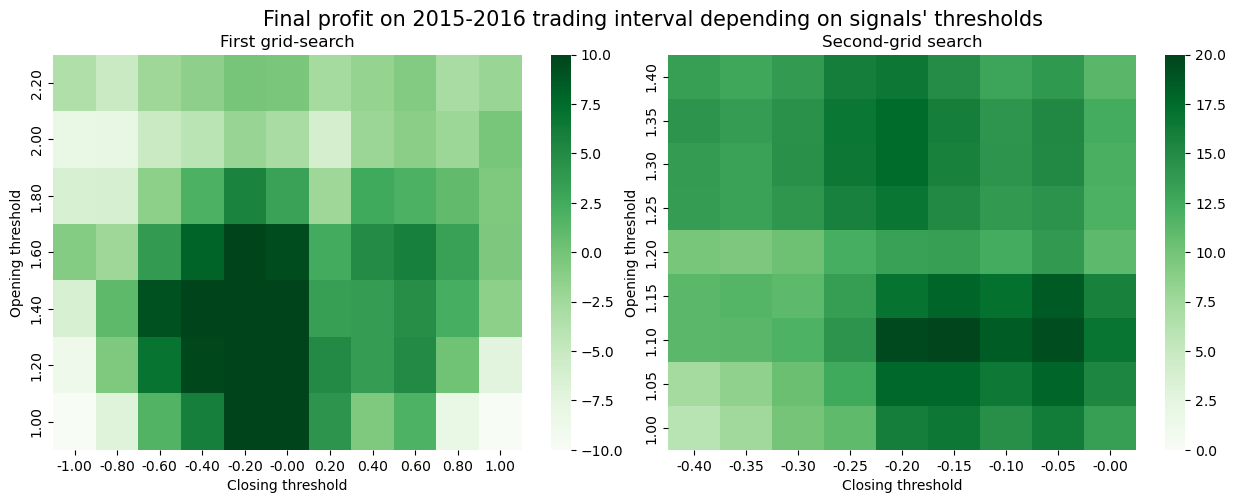}
\caption{Final accumulated profit on the optimization interval for LSTM approach depending on signal cut-offs values}
\label{fig:lstm_opt}
\end{figure}
\noindent Since there is only one LSTM technique we decided to put both grid-searches on one plot (Figure \ref{fig:lstm_opt}). First observation is that the final values of $E_T-E_0$ have similar magnitude to PCA results. Thresholds $\overline{g}_{ol}=\overline{g}_{os}=1.10$ and $\overline{g}_{cl}=\overline{g}_{cs}=-0.15$ are our final choice of thresholds- they are also fairly comparable to chosen PCA cut-offs.
\subsection{Exchange traded funds of market indices}
in the ETFs' approaches, we already have all components of the Arbitrage Pricing model therefore no training besides calculating the $\beta$ coefficients is necessary. Like with other techniques, these coefficients are going to be derived daily based on a moving $W=120$ days window.\\
Continuing with previous methodologies 2015-2016 data is used to identify potentially optimal signals' thresholds for ETFs sub-approaches. In case of artificial ETFs' technique we had to resign from $3$ out of $14$ sector indices: \textit{WIG-MOTO} (moto), \textit{WIG-ODZIEŻ} (clothes) and \textit{WIG-LEKI} (pharma) since they did not exist back in the optimization period. For the other sub-approach all $3$ ranking indices are used in form of their Total Return equivalents imitating actual funds.\\
 \begin{figure}[H]
\centering
\includegraphics[scale=0.4]{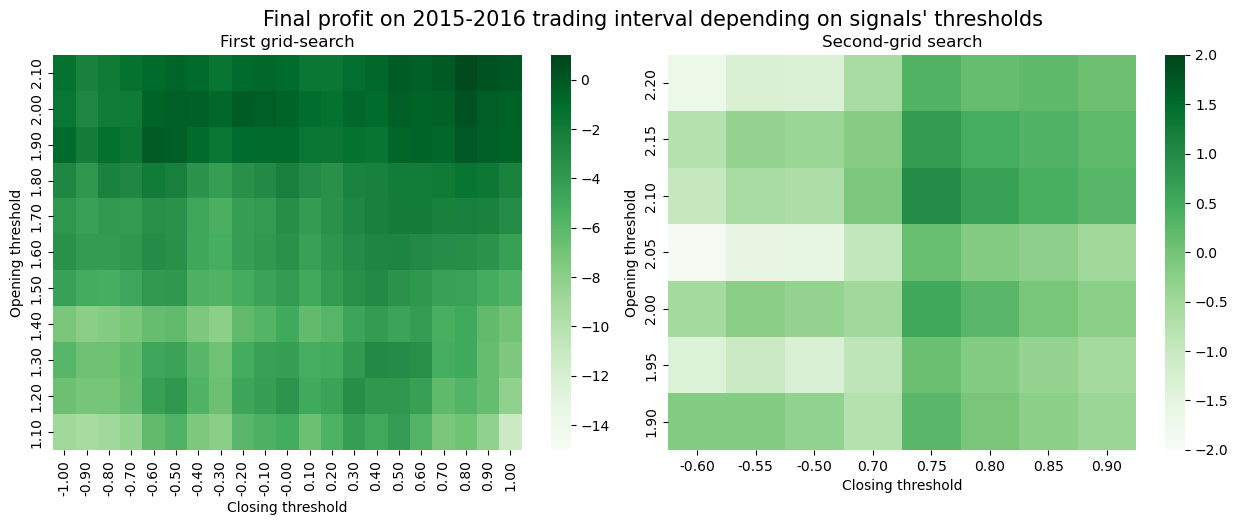}
\caption{Final accumulated profit on the optimization interval for existing ETFs approach depending on signal cut-offs values}
\label{fig:etf_opt1}
\end{figure}
\noindent Note that the final results are way smaller than in previous approaches- most of thresholds' pairs were not profitable. The optimal pair for real funds' technique is $\overline{g}_{ol}=\overline{g}_{os}=2.10$ and $\overline{g}_{cl}=\overline{g}_{cs}=0.75$ with around $1$ PLN profit achieved in the second round of grid-search (Figure \ref{fig:etf_opt1}).
 \begin{figure}[H]
\centering
\includegraphics[scale=0.4]{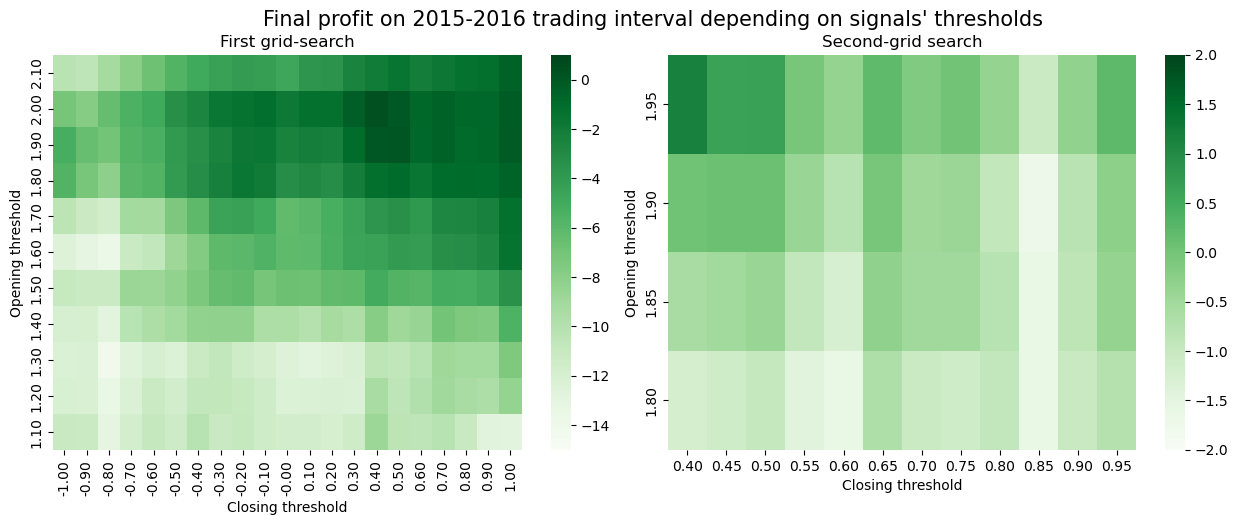}
\caption{Final accumulated profit on the optimization interval for artificial (sector) ETFs approach depending on signal cut-offs values}
\label{fig:etf_opt2}
\end{figure}
\noindent With the second sub-approach, magnitude of results is even smaller (Figure \ref{fig:etf_opt2}). Second grid-search suggests $\overline{g}_{ol}=\overline{g}_{os}=1.95$ and $\overline{g}_{cl}=\overline{g}_{cs}=0.40$ as the potentially optimal choices for backtesting purposes.\\
Exchange traded funds' approaches optimal thresholds are more similar to ones of Avellaneda and Lee. Both opening and closing singnals of given type lay within one side of $0$.
\section{Validation metrics}
The main goal of backtesting is to validate whether considered strategy is working i.e. providing repetitive profits throughout trading years. Since we are dealing with $3$ main approaches for replicating portfolios: PCA, LSTM and the use of actual market indices- it is also important to compare their outcomes between each other. For both of these tasks we require a quantitative measure of performance. One was already presented in a form of accumulated equity $E_t$ (Equation \ref{eq:e_t}). Its final level $E_T$ shows whether the initial capital of $100$ was increased by daily trades of $60$ components based on generated signals. Since in the main backtesting period $3$ years are considered, it is also important to see how relevant are our earnings- if each year we generate similar profits strategy can be seen as more applicable in the future. For that reason annual returns should be considered instead of an entire trading period one. Another desire is to outperform the overall market growth which can be represented by the risk-free rate $r_f$. If we are not able to beat it there may be no point in taking the risk. Finally, equity level should not fluctuate much by add-ups of new transaction closings. If a positive annualized return comes from a lot of accumulated losses and one massive profit we may not want to count on such a rare event the future.  To incorporate all of those demands, a very basic yet commonly used metric called \textit{Sharpe ratio} of annualized returns between $t$ and $t+1$ can be used. For trading year starting at $t=j$ its formula goes as follows:
\begin{equation}
\mathcal{S}_y = \frac{252\overline{R_j} - r_f}{\sqrt{252}\sigma_j},
\label{eq:sharpe}
\end{equation}
where $R_j$ is an average of daily returns throughout year $j$ and $\sigma_j$ describes the estimate of daily returns' standard deviation. Factors of $252$ are then used for annualizing purposes. We seek to maximize Sharpe ratio- risk-free rate is subtracted to see the potential outperformence of proposed strategy and any potential instabilities make the metric smaller. Note how $E_t$ and its sub-components $P^i_{[t_0,t_1]};t_0,t_1 < t$ only describe the jumps connected to positions' closings. Assuming that we only trade with one company as the main one, nothing changes in $E_t$ value when a position is opened and until it is closed. Possibility exists that even though we set an average leverage level, scaled weightings indicate opening with almost all of currently owned capital. Even though in the end we may end up making a small profit from closing the position using such a large amount to trade was a major risk-driver. To give a different perspective on the trades incorporating all changes of owned cash an additional performance metric $C_t$ will be introduced. It aims to track every-day state of an additional account where the initial $E_0=100$ and all already made profits (or losses) are stored and used for following trades. Each pair of open/close will lead to a rectangular \quotes{spike} on $C_t$'s plot where ultimate profit can be seen as difference in the initial and final level (before and after the \quotes{spike}).
We will show all $3$ metrics (via table in case of Sharpe ratio and plots for the other two) for the overall portfolio of $60$ stocks representing the entire market. Additionally, results are going to be separated according to participants' sectors seeking for any over- or under-performing ones. Note that since the list of companies used is relatively small, we may see practically no movement for the less represented ones.
\section{Results of backtesting}
Throughout the entire paper with watchman's precision we collected different pieces of Pairs Trading puzzle and placed them in correct places. Now, to complete the picture one final piece is needed. The \textit{coup de grâce}, the \textit{icing on a cake}, the \textit{bull's eye}- results.
\subsection{Main historical period of 2017-2019}
\subsubsection{Principal Components Analysis (PCA) approach}
Since this analysis is the first one out of three we will provide additional insight results' validation methodology. First, annual Sharpe ratios defined as above (Equation \ref{eq:sharpe}) are going to be shown for sub-portfolios gathering stocks' of specific sectors and for the entire portfolio consisting $60$ components. Since two sub-approaches were introduced (with constant and variable number of eigenportfolios)- two results' tables are going to be merged together.
\begin{table}[H]
\small
\centering
\begin{tabular}{llrrr}
\toprule
           &           &  2017 &  2018 &  2019 \\
\midrule
$r=15$ & ARCHITECTURE &  1.39 &  0.49 &  0.75 \\
           & BANKS &  0.33 &  0.36 & -1.00 \\
           & CHEMISTRY &  1.70 &  0.54 & -0.67 \\
           & CLOTHES &  1.46 &  1.77 & -0.27 \\
           & ENERGY & -0.47 &  2.22 & -0.02 \\
           & FOOD & -1.78 &  0.59 &  0.15 \\
           & FUELS &  1.79 &  1.10 &  0.36 \\
           & GAMES & -0.27 &  0.75 &  1.11 \\
           & INFORMATICS &  1.16 &  0.96 &  0.15 \\
           & MEDIA &  0.30 & -0.27 & -0.10 \\
           & MINING &  0.43 &  2.18 & -0.51 \\
           & MOTO & -0.66 &  0.13 & -0.96 \\
           & PHARMA &  1.70 & -1.02 & -1.31 \\
           & REAL EST &  1.66 & -0.49 & -0.32 \\
           & \textit{other} &  2.80 & -0.51 & -0.68 \\
           & \textbf{PORTFOLIO} &  2.63 &  1.01 & -1.16 \\
\hline
Variable $r$ & ARCHITECTURE &  1.51 &  0.08 &  0.47 \\
           & BANKS & -0.12 &  0.81 & -1.15 \\
           & CHEMISTRY &  1.56 &  0.76 &  0.02 \\
           & CLOTHES &  1.67 &  1.88 & -0.86 \\
           & ENERGY & -0.57 &  1.96 &  0.22 \\
           & FOOD &  0.08 &  0.43 & -0.37 \\
           & FUELS &  1.65 &  0.99 &  0.51 \\
           & GAMES & -0.20 & -0.91 &  1.97 \\
           & INFORMATICS &  1.09 & -0.67 & -0.10 \\
           & MEDIA &  0.08 &  1.20 & -1.10 \\
           & MINING &  0.61 &  2.34 & -0.39 \\
           & MOTO & -1.11 &  0.12 & -0.54 \\
           & PHARMA &  1.49 & -1.46 & -1.53 \\
           & REAL EST &  1.71 & -1.27 &  0.10 \\
           & \textit{other} &  2.66 & -0.34 &  0.01 \\
           & \textbf{PORTFOLIO} &  2.51 &  0.44 & -0.91 \\
\bottomrule
\end{tabular}
\caption{Sharpe ratios for PCA approaches based on 2017-2019 trading interval}
\label{tab:5}
\end{table}
\noindent Analysing Table \ref{tab:5} let us start with ratios of the entire portfolios throughout years. In both sub-approaches 2017 had the highest metric values around $2.5$. Performance dropped in the next years with 2019 underperforming in relation to the risk free rate. Positive ratios are slightly higher in the entire portfolio of constant $r$ approach but so is the magnitude of negative one (during 2019). To give a comparison, on their entire portfolio Avellaneda and Lee\cite{main_paper} achieved annual Sharpe ratios between $-0.7$ and $3.4$ with $r=15$ and between $-0.4$ and $2.6$ with variable $r$- although there are some differences in taken assumptions and the overall trading scope, our results can be seen as fairly comparable. Interestingly, mentioned authors also achieved the lowest, negative ratios in the last trading year (in their case it was 2007). In our case it may have been determined by some potentially forced early closings made to clear off the portfolio. With sector sub-portfolios, companies that were not assigned to any specific industry are the best performing group (in both sub-approaches). It is also interesting that games' companies were practically the only well-managing group in 2019 but failed beat the market in previous years. On the other hand, fuels companies formed the most stable performing group. For both $r$ types, mean Sharpe ratios of sector sub-portfolios oscillated around $0.35$.\\
In the second validation step we will take a closer look on development of equity $E_t$ and cash $C_t$ throughout trading period. Again, overall portfolio is going to be analysed first.
\begin{figure}[H]
\centering
\includegraphics[scale=0.3]{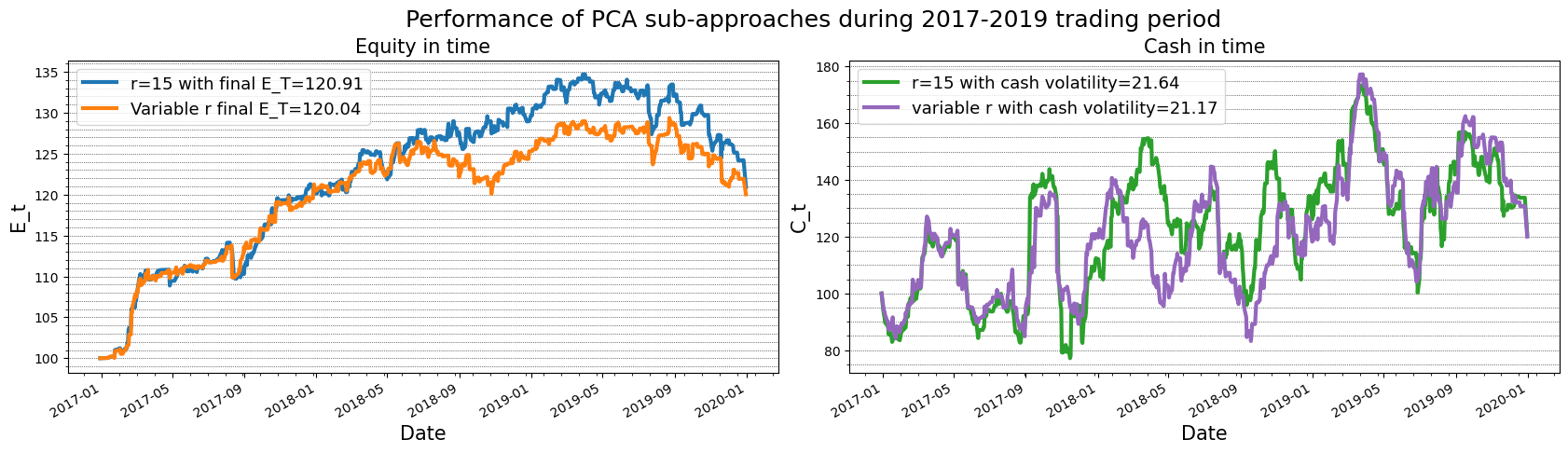}
\caption{Equity and cash levels throughout the 2017-2019 trading interval for PCA approaches}
\label{fig:pca_results_years}
\end{figure}
\noindent The left plot of Figure \ref{fig:pca_results_years} presents equity $E_t$ that accumulates all achieved profits from positions' closings. Therefore, separate costs of opening and closing transactions are not included. Right plot fills this gap gathering all money flows in and out of the portfolio as its time cash status $C_t$. Analysing $E_t$ we can see that constant $r$ technique outperformed the latter one throughout 2018 and 2019 where $r$ was set to $18$ and $17$ consecutively. In 2017, $r=16$ explained around $55\%$ of total variance- for that reason $E_t$ values were very similar. Final value of $E_t$ gives practically identical profit of around $20$ PLN. Switching back to the right plot, we can see all the fluctuations caused by spending or gaining money from either buying or short selling stocks. Since the final cash value $C_T$ is equal to $E_T$ an information about volatility was added to the sub-plot. These are also very similar and around $21$- since we did not see any other techniques it is hard to raise any quantitative conclusions based on these numbers. Intuitively, since $E_t$ was increasing, fluctuations got wider due to bigger $\Lambda_t$ scaling.
\begin{figure}[H]
\centering
\includegraphics[scale=0.3]{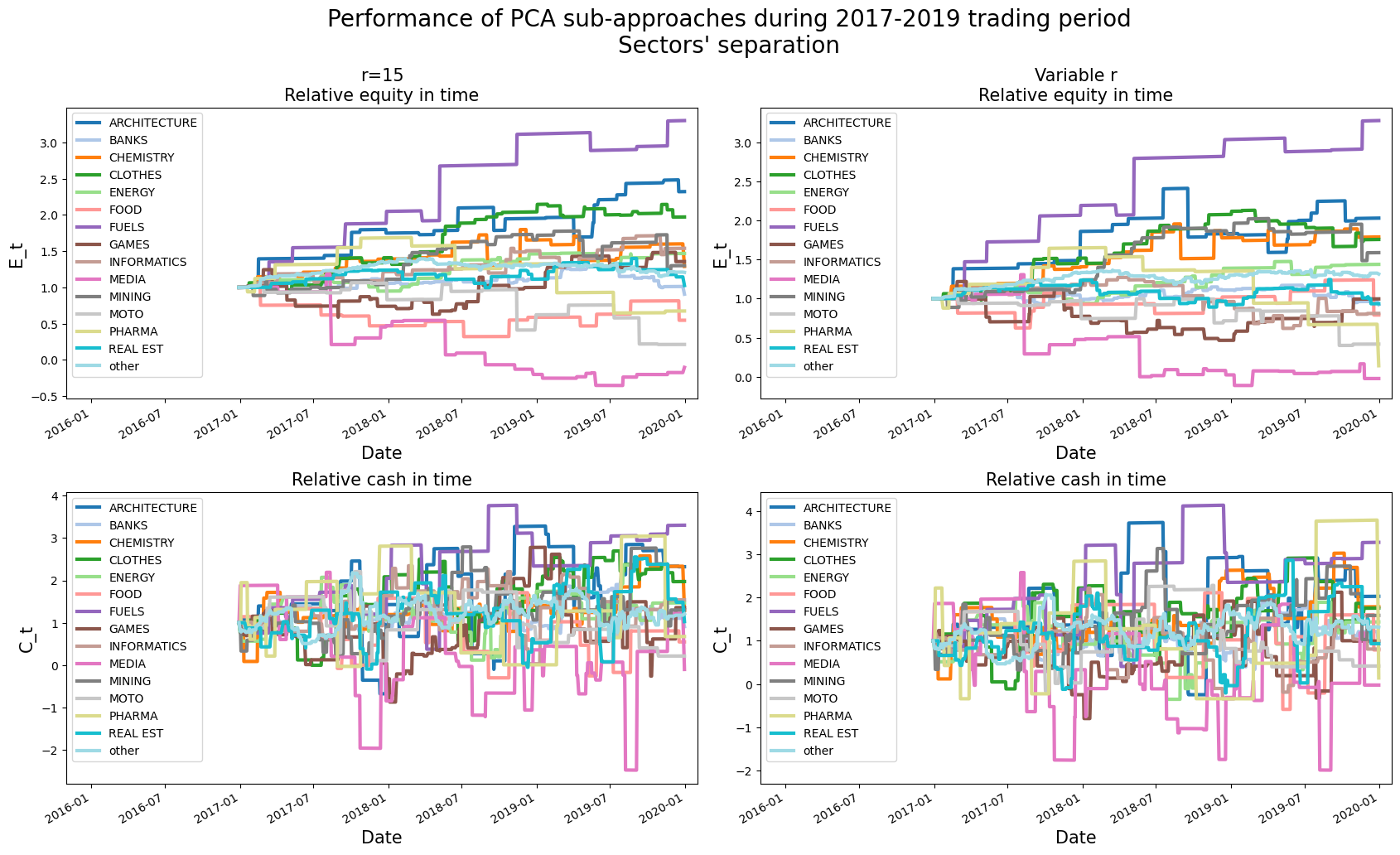}
\caption{Equity and cash relative levels throughout the 2017-2019 trading interval for PCA approaches- sectors' separation}
\label{fig:pca_results_sectors_years}
\end{figure}
\noindent Analogous time-series were presented on Figure \ref{fig:pca_results_sectors_years}- this time separating overall portfolio into stocks' coming from different sectors. Notice that since the numbers of components from given industry are different- sub-portfolios start from different initial capitals. To make them comparable we thus divided all values of both $E_t$ and $C_t$ by starting ones. It is then somehow similar to the way we compared stocks' closing prices via relative worth of $1$ PLN investment. Upper plots focus on the relative equity level with the lower ones presenting cash movement. Due to inclusion of $15$ different sub-portfolios there is a lot happening on each picture and thus not everything can be easily interpreted. One of the reasons we decided to include these plots is to again highlight nature and differences between two performance indicators. If profits are positive, $E_t$ constantly increases while $C_t$ oscillates being sensitive to every money flow. Even with so many lines it is clear that for both PCA approaches fuels sub-portfolio profited the most in relation to its initial capital with media one having the biggest relative decrease- fluctuations of these groups were also the highest (as lower plots suggest).\\
Already at this point, without comparing PCA to other techniques of deriving components' weightings, we can be satisfied with its performance. Most importantly, Sharpe ratios of the overall portfolio for both sub-approaches were positive in the first 2 years of the strategy with significantly high ones observed in 2017. Stocks within particular sectors also gave $\mathcal{S}>0$ with only few exceptions. It is not trivial to differentiate which PCA sub-technique is the superior one- $55\%$ was selected in a way that the actual number of eigenportfolios picked is usually very similar and thus so are the results.
\subsubsection{Long short-term memory networks approach}
Similarly to the PCA approaches we will validate trading with the use of stacked LSTM network aimed to produce appropriate replicating portfolios' weightings. This part is especially important since using recurrent neural networks to perform pairs trading does not seem to be yet covered in the literature of the subject. Any promising results achieved here may potentially lead to development of a new branch in Statistical Arbitrage spectrum. Let us now consider $3$ metrics to validate outcomes of LSTM based 2017-2019 trading.
\begin{table}[H]
\small
\centering
\begin{tabular}{lrrr}
\toprule
{} &  2017 &  2018 &  2019 \\
\midrule
ARCHITECTURE &  0.63 &  1.81 &  0.32 \\
BANKS        &  1.50 &  1.53 &  0.40 \\
CHEMISTRY    & -1.40 & -0.03 &  1.41 \\
CLOTHES      & -1.52 &  1.52 & -0.48 \\
ENERGY       &  0.99 &  0.90 &  0.72 \\
FOOD         &  2.09 & -0.58 & -1.58 \\
FUELS        & -1.72 &  0.79 &  0.02 \\
GAMES        & -0.58 & -1.17 & -1.15 \\
INFORMATICS  &  0.98 & -0.40 & -1.34 \\
MEDIA        & -0.34 & -0.93 & -1.90 \\
MINING       & -1.87 &  0.82 & -0.16 \\
MOTO         &  0.22 &  0.31 & -1.34 \\
PHARMA       &  0.10 &  1.17 & -1.26 \\
REAL EST     &  0.30 &  2.05 & -1.80 \\
\textit{other}        &  0.22 & -0.41 & -1.35 \\
\textbf{PORTFOLIO}    &  0.60 &  2.09 & -1.53 \\
\bottomrule
\end{tabular}
\caption{Sharpe ratios for LSTM approach based on 2017-2019 trading interval}
\label{tab:5.5}
\end{table}
\noindent Table \ref{tab:5.5} gives us an overview of annualized Sharpe ratios across sectors and for the overall portfolio. Similarly to the PCA approaches, LSTM is underperforming in 2019 for most of the groups. Highest ratios were achieved in 2018 with 2017 giving more than half of positive ratios. Some of the sector sub-portfolios are outperforming the overall one.
\begin{figure}[H]
\centering
\includegraphics[scale=0.3]{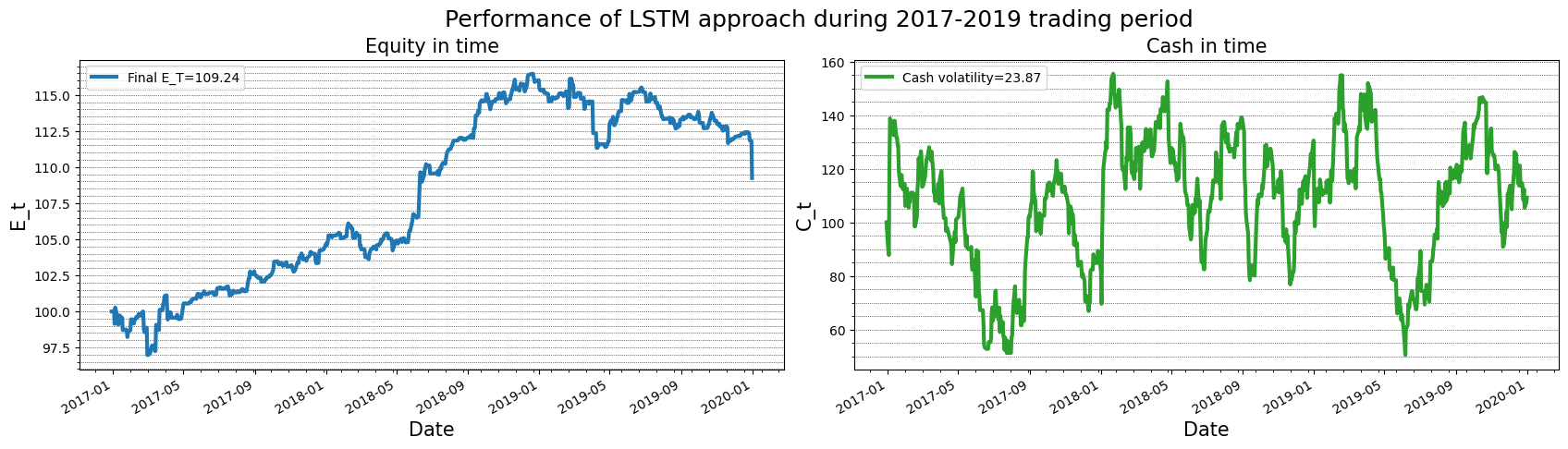}
\caption{Equity and cash levels throughout the 2017-2019 trading interval for LSTM approach}
\label{fig:lstm_results_years}
\end{figure}
\noindent Figure \ref{fig:lstm_results_years} shows further similarities between PCA and LSTM results- both equity and cash levels move similarly to ones seen for both Principal Components Analysis techniques. There is a bullish run of $E_t$ throughout the first $2$ years and a slight drop in 2019 (already confirmed by negative Sharpe ratios). The final return is around $10\%$ which is worse than with both PCA approaches.
\begin{figure}[H]
\centering
\includegraphics[scale=0.3]{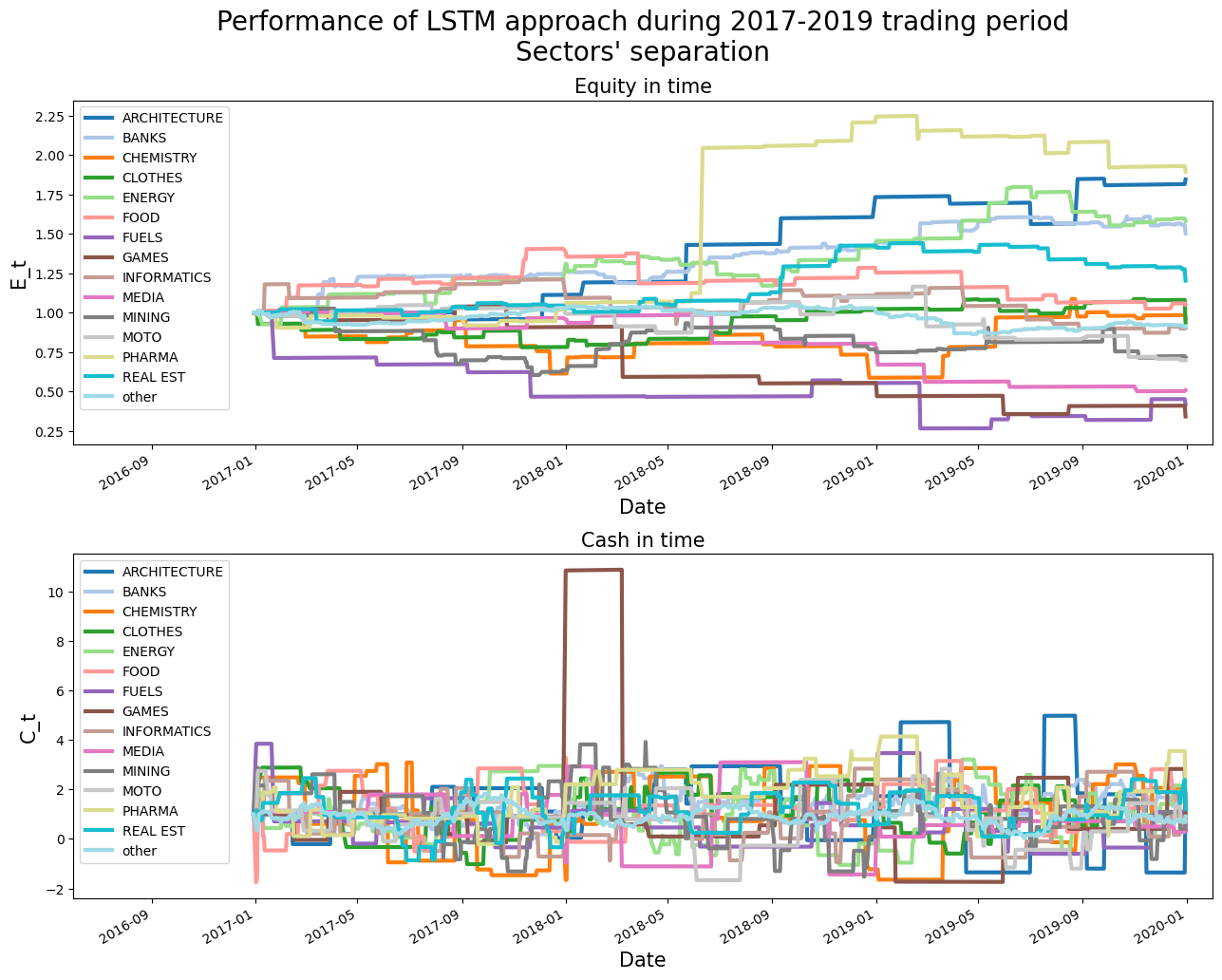}
\caption{Equity and cash relative levels throughout the 2017-2019 trading interval for LSTM approach- sectors' separation}
\label{fig:lstm_results_sectors_years}
\end{figure}
\noindent Dividing replicated stocks into sector portfolios and scaling their cash levels by the initial capital we can observe (lower plot of Figure \ref{fig:lstm_results_sectors_years}) one large transactions' pair in the gaming industry and a lot of smaller ones with maximum of $4$ PLN spend on opening long or closing short. Performing the same separation and scaling for equity level $E_t$ it is visible (upper plot of Figure \ref{fig:lstm_results_sectors_years}) that such relatively large movement of games' sector cash level ended up in a loss- same can be said about further closings in this sub-portfolio. More than half of the other sectors had $>0$ returns with pharmaceutical and energy ones being the best.\\
There are certainly some similarities between PCA and LSTM replication methodologies (especially when comparing them to ETFs approach). Even though the latter one gives more flexibility for modelling individual companies' returns, they both rely on the entire trading portfolio as a set of explanatory variables. These similarities resulted in fairly similar results achieved by both methods. Both methods produced profits in the first two years and experienced a bearish movement during 2019.
\subsubsection{Exchange Traded Funds of market indices approach}
Although we analysed PCA sub-techniques together, both ETFs' methods are going to be validated separately (with short comparison at the end). It is because they are not different by just one parameter (as it was with PCA) but by an entire explanatory variables' set. Let us then start with approach concerning a limited set of existing ETFs.\\
\textbf{Existing ETFs sub-approach}\\
As before, first and possibly the most important metric that will be analysed is Sharpe ratio $\mathcal{S}$.
\begin{table}[H]
\small
\centering
\begin{tabular}{lrrr}
\toprule
{} &  2017 &  2018 &  2019 \\
\midrule
ARCHITECTURE &  -inf & -0.98 & -0.98 \\
BANKS        & -1.89 &  0.31 &  0.40 \\
CHEMISTRY    &  1.12 & -1.17 &  1.34 \\
CLOTHES      &  -inf &  0.98 & -0.98 \\
ENERGY       & -1.38 &  0.32 &  0.98 \\
FOOD         &  1.41 &  0.23 &  -inf \\
FUELS        & -1.39 &  -inf &  0.98 \\
GAMES        &  0.99 & -0.85 & -0.98 \\
INFORMATICS  & -0.71 & -0.84 & -1.09 \\
MEDIA        &  0.14 &  0.93 & -0.30 \\
MINING       &  0.94 &  0.98 &  1.05 \\
MOTO         &  1.28 &  0.02 &  1.29 \\
PHARMA       & -0.15 & -1.05 & -0.98 \\
REAL EST     &  0.69 &  1.22 &  1.64 \\
\textit{other}        & -0.27 & -0.82 &  0.88 \\
\textbf{PORTFOLIO}    & -0.25 & -0.46 &  1.43 \\
\bottomrule
\end{tabular}
\caption{Sharpe ratios for existing ETFs' approach based on 2017-2019 trading interval}
\label{tab:6}
\end{table}
\noindent Table \ref{tab:6} presents annualized Sharpe ratios for the overall portfolio and sub-portfolios of common industry stocks'. Since we already know the PCA and LSTM results, it can be noticed that all $60$ components are performing worse during the first $2$ years but overcome previous techniques during 2019. Looking at sectors, we can see that $\mathcal{S}$ are mostly small with a mean of $0.06$. In some cases there were no positions opened throughout an entire year resulting in Sharpe ratio set to \textit{-inf}.
\begin{figure}[H]
\centering
\includegraphics[scale=0.3]{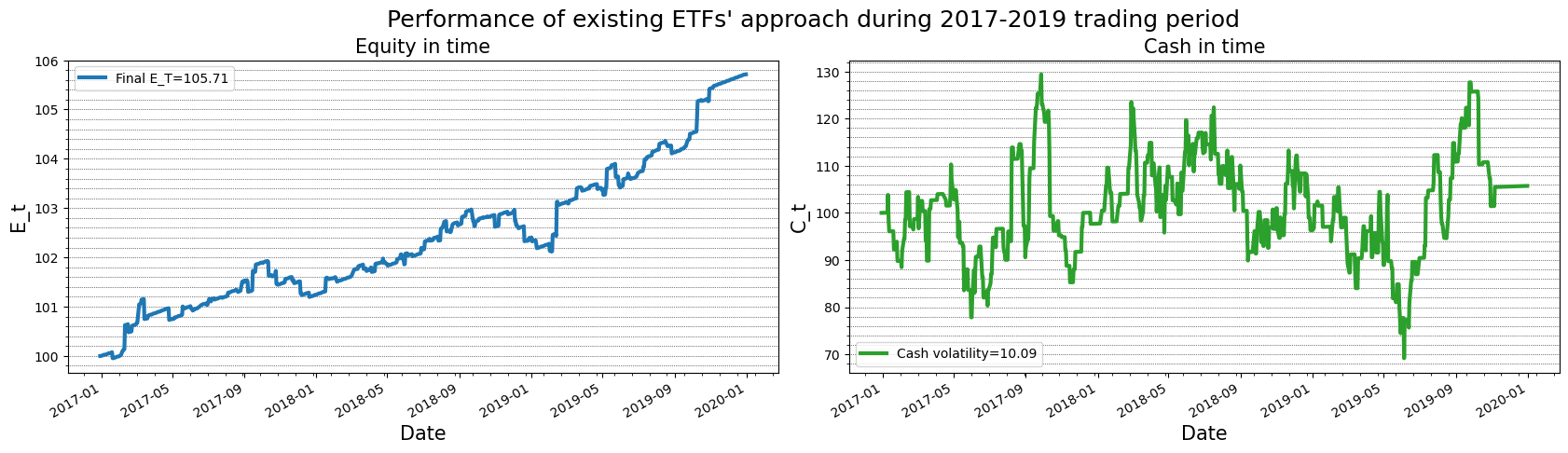}
\caption{Equity and cash levels throughout the 2017-2019 trading interval for existing ETFs approach}
\label{fig:etf1_results_years}
\end{figure}
\noindent As can be seen on the left plot of Figure \ref{fig:etf1_results_years} strategy managed to make a profit of around $5$ PLN. It is a decrease in relation to previous methods- we could have expected it judging by the performance during grid searches and since explainability of eigenportfolios and LSTM regression model is superior to just $3$ funds. Nevertheless, smaller profits are partially compensated with lower volatility of cash used for all trading activities.\\
\begin{figure}[H]
\centering
\includegraphics[scale=0.3]{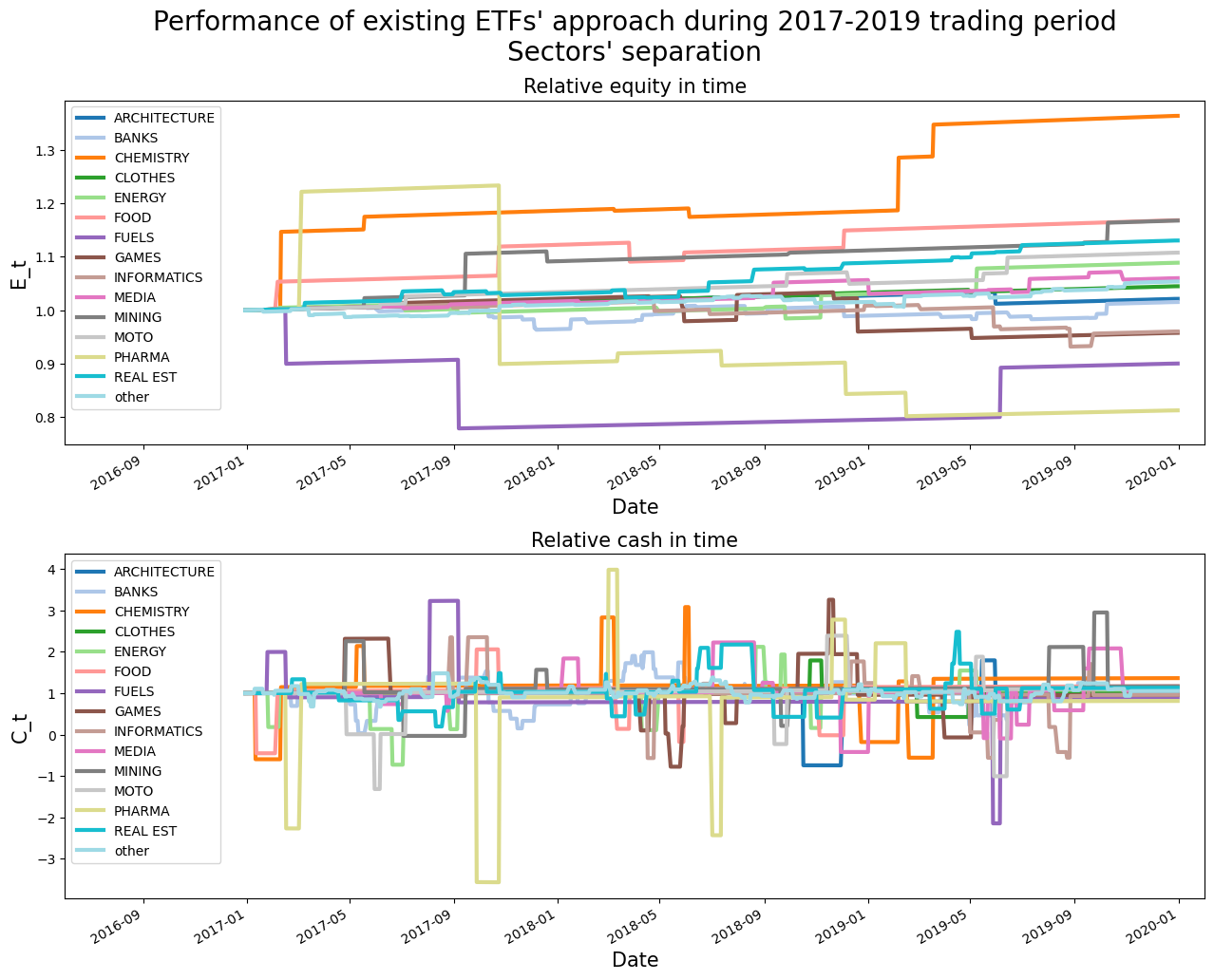}
\caption{Equity and cash relative levels throughout the 2017-2019 trading interval for existing ETFs approach- sectors' separation}
\label{fig:etf1_results_sectors_years}
\end{figure}
\noindent Figure \ref{fig:etf1_results_sectors_years}'s perspective highlights our previous assumptions about practically no movement for some of the sectors. Even the ones that changed their equity of cash value (again calculated in a relative manner) more than once do not have more than a few trades in total.\\
Notice that if we only use $3$ components, residuals are consistently larger- overall market funds are unable to model smaller trends in companies' movements. Since the residuals are less volatile, it is not that likely for their cumulative sums to deviate. This ultimately results in less signals and thus less trades. At the same time, more stability  leads to better decisions concerning transactions making the profits more consistent throughout the entire trading period.\\
\textbf{Artificial ETFs sub-approach}\\
Intuitively, artificial ETFs should perform better due to higher explainability of the systematic factors. At the same time more transactions has to be made resulting in additional taxation.
\begin{table}[H]
\small
\centering
\begin{tabular}{lrrr}
\toprule
{} &  2017 &  2018 &  2019 \\
\midrule
ARCHITECTURE &  1.01 &  0.28 & -1.61 \\
BANKS        &  0.77 &  0.37 &  1.29 \\
CHEMISTRY    & -1.04 & -0.87 & -1.18 \\
CLOTHES      &  0.38 & -1.01 & -0.55 \\
ENERGY       &  1.48 &  1.46 & -0.12 \\
FOOD         &  0.19 &  1.12 & -0.62 \\
FUELS        & -0.99 & -0.98 & -0.98 \\
GAMES        & -1.04 & -1.08 &  1.16 \\
INFORMATICS  &  0.85 &  1.57 & -1.23 \\
MEDIA        & -1.43 & -0.87 &  0.85 \\
MINING       &  1.42 &  0.58 & -1.47 \\
MOTO         & -0.83 & -0.98 &  0.54 \\
PHARMA       &  1.45 &  0.98 & -0.98 \\
REAL EST     & -0.30 & -1.17 &  1.64 \\
\textit{other}        &  1.59 & -2.36 & -1.28 \\
\textbf{PORTFOLIO}    &  1.28 & -1.63 & -0.84 \\
\bottomrule
\end{tabular}
\caption{Sharpe ratios for artificial ETFs' approach based on 2017-2019 trading interval}
\label{tab:7}
\end{table}
\noindent Sharpe ratios presented in Table \ref{tab:7} show larger magnitudes of $\mathcal{S}$s than with existing funds' approach. This is not necessary good due to more negative ratios for particular sectors. Overall portfolio is only profiting in 2017, then it suffers from significant underperformences relative to the market (represented by $r_f$).
\begin{figure}[H]
\centering
\includegraphics[scale=0.3]{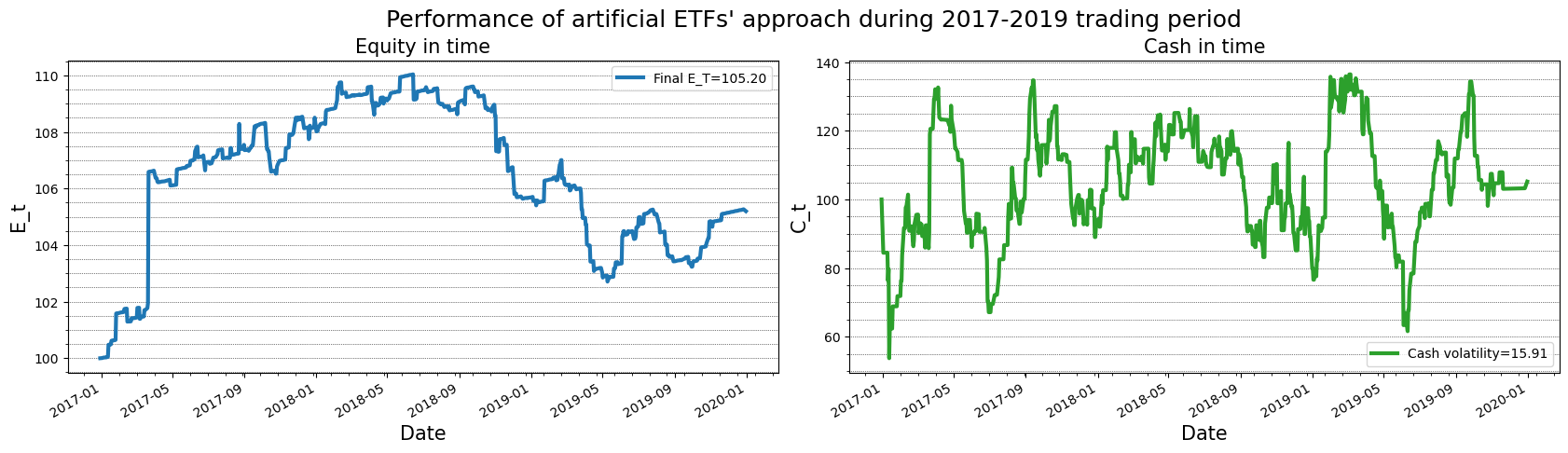}
\caption{Equity and cash levels throughout the 2017-2019 trading interval for artificial ETFs approach}
\label{fig:etf2_results_years}
\end{figure}
\noindent As a confirmation of our overall portfolio Sharpe ratio's conclusions, Figure \ref{fig:etf2_results_years} presents how equity $E_t$ significantly raised in the first year just to drop in the following ones. Such instability is also seen with wider fluctuations of owned cash. Although indicators' paths are noticeably different, the ultimate result (profit) of considered approach is practically identical to one of existing ETFs' technique.
\begin{figure}[H]
\centering
\includegraphics[scale=0.3]{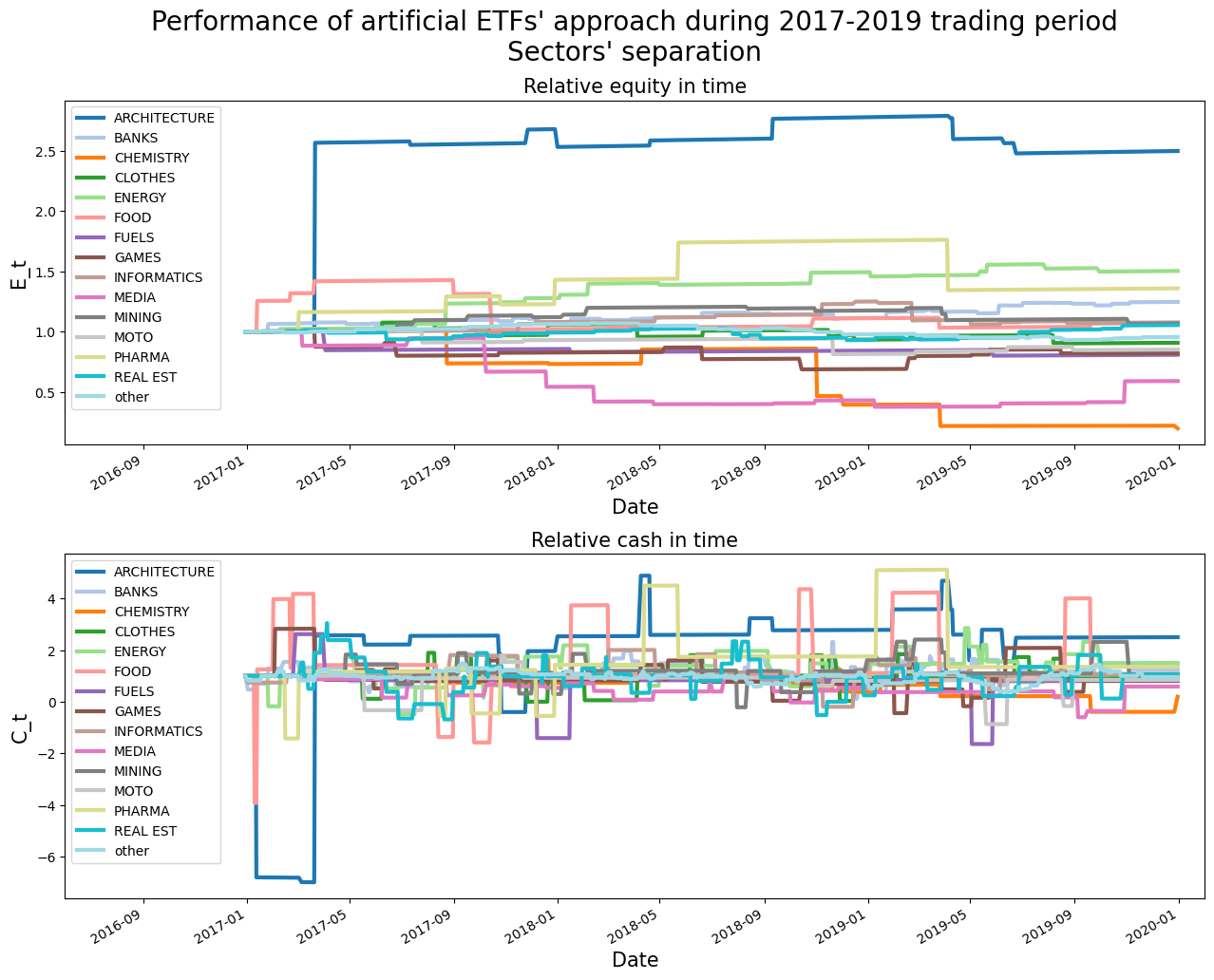}
\caption{Equity and cash relative levels throughout the 2017-2019 trading interval for artificial ETFs approach- sectors' separation}
\label{fig:etf2_results_sectors_years}
\end{figure}
\noindent Final figure (\ref{fig:etf2_results_sectors_years}) again presents the relative equity and cash values for stocks' sectors. Notice how the ultimate profit of overall portfolio is mainly determined by a positive jump in architecture's sector and later two negative ones of media and chemistry stocks.\\
Both techniques give similar profits with the latter one (with more explanatory variables) being far more unstable. Their performance is rather worse than ones of more sophisticated methods. This was also the general outcome in Avellaneda and Lee's paper (with PCA as the counter-technique).
\subsection{COVID-19 pandemic recession in 2020}
Partially imitating Avellaneda and Lee analysis of liquidity crisis in summer of 2007, we are going to take a closer look at our strategy performance in 2020 COVID-19 recession. All previous assumptions will be held (including optimal thresholds from 2016-2017 grid searches) except arisk free rate $r_f$ that is going toi be set at $0.5\%$.
\subsubsection{Principal Components Analysis approach}
\begin{table}[H]
\small
\centering
\begin{tabular}{llr}
\toprule
           &           &  2020 \\
\midrule
r=15 & ARCHITECTURE & -0.98 \\
           & BANKS & -1.83 \\
           & CHEMISTRY &  0.19 \\
           & CLOTHES &  1.03 \\
           & ENERGY & -0.86 \\
           & FOOD & -1.02 \\
           & FUELS & -0.98 \\
           & GAMES & -1.02 \\
           & INFORMATICS & -1.18 \\
           & MEDIA &  1.20 \\
           & MINING &  0.89 \\
           & MOTO &  1.33 \\
           & PHARMA &  1.30 \\
           & REAL EST &  0.56 \\
           & \textit{other} & -0.28 \\
           & \textbf{PORTFOLIO} & -1.39 \\
variable r & ARCHITECTURE & -0.82 \\
           & BANKS & -1.62 \\
           & CHEMISTRY & -0.12 \\
           & CLOTHES &  1.12 \\
           & ENERGY & -0.88 \\
           & FOOD & -0.99 \\
           & FUELS & -0.98 \\
           & GAMES &  1.40 \\
           & INFORMATICS & -0.97 \\
           & MEDIA &  1.08 \\
           & MINING &  0.73 \\
           & MOTO & -0.98 \\
           & PHARMA & -1.11 \\
           & REAL EST &  1.19 \\
           & \textit{other} &  0.97 \\
           & \textbf{PORTFOLIO} &  0.59 \\
\bottomrule
\end{tabular}
\caption{Sharpe ratios for PCA approaches based on 2020 recession trading interval}
\label{tab:8}
\end{table}
\noindent Since, based on 2019 data $r=18$ was picked in the second sub-approach both methods perform differently. Higher number of components used actually increased overall portfolio's $\mathcal{S}$ (Table \ref{tab:8}). Still, a more general observation for both methods is that the ratios are mostly smaller than during 2017-2018 and thus more similar to ones from 2019. Next plots should give us more insight on the overall and individual performances.
\begin{figure}[H]
\centering
\includegraphics[scale=0.3]{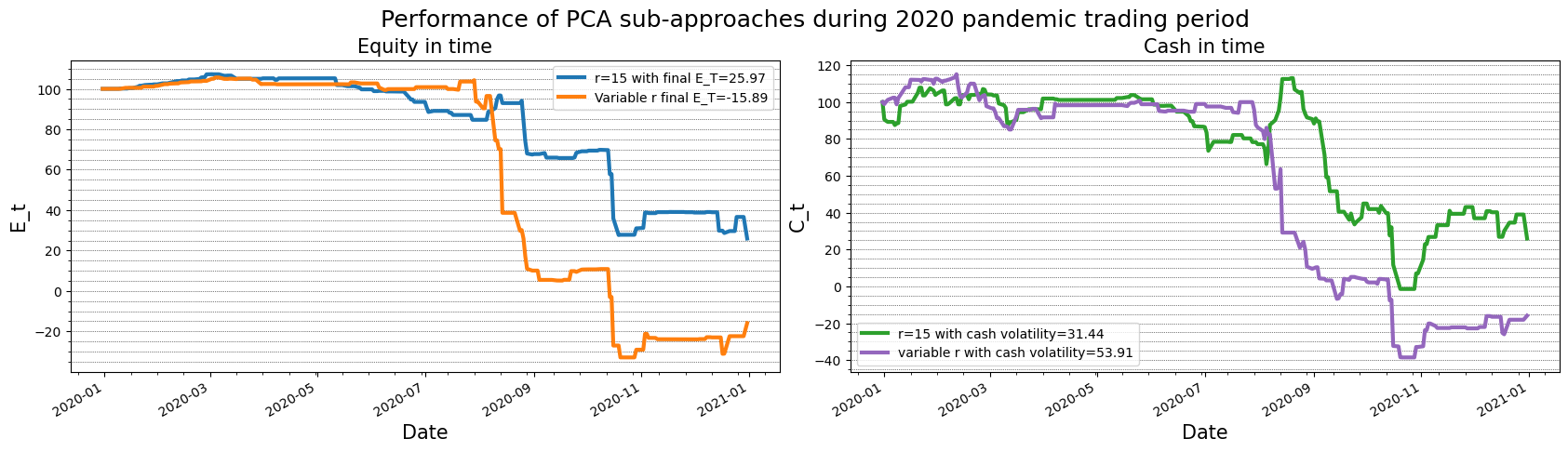}
\caption{Equity and cash levels throughout the 2020 recession trading interval for PCA approaches}
\label{fig:pca_results_years_covid}
\end{figure}
\begin{figure}[H]
\centering
\includegraphics[scale=0.3]{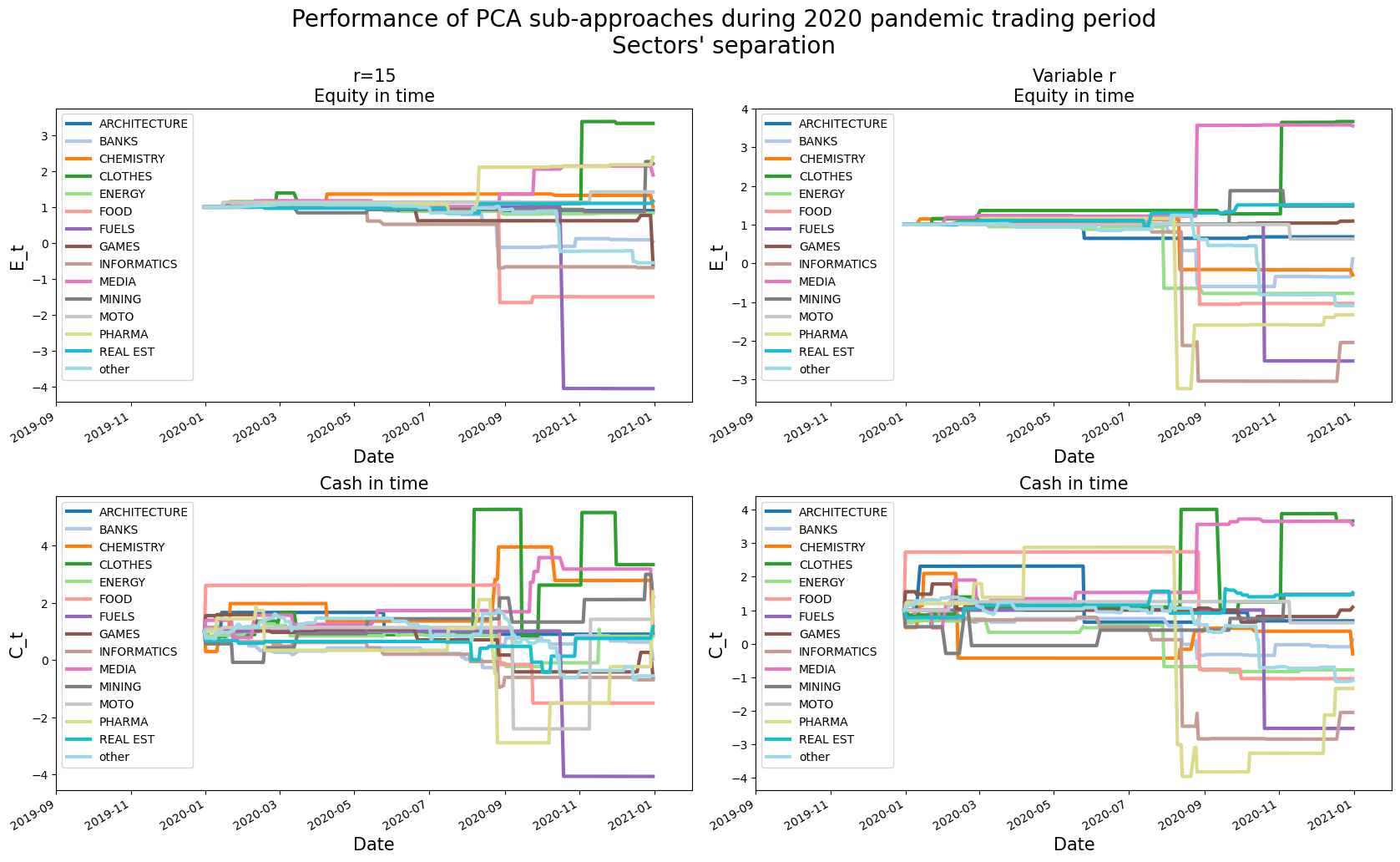}
\caption{Equity and cash relative levels throughout the 2020 recession trading interval for PCA approaches- sectors' separation}
\label{fig:pca_results_sectors_years_covid}
\end{figure}
\noindent For the sake of brevity let us consider equity and cash movement for both the overall portfolio and sub-portfolios together. Starting with the upper Figure (\ref{fig:pca_results_years_covid}) the most important conclusion is that both strategies not only failed to bring any ultimate profits but also resulted huge losses. Significant drop of $E_t$ begins in the middle of the year- it seems like the moving window used to provide $\beta$ started to be more influences by prices' falls during spring 2020 which ultimately resulted in corrupted signals and thus incorrect decisions. Notice how variable $r$' approach still manage to give a positive Sharpe ratio since mean used in the formula is not as sensitive to enormous but rather singular negative returns. This is exactly the reason why wider perspective needs to be presented. Switching to lower Figure (\ref{fig:pca_results_sectors_years_covid}) shows that the majority of large amounts' opening were done in the second half of 2020- more than half of them resulted in significant losses. This is in line with our previous conclusions about signals' being corrupted by growing influence of not typical prices' dives in the first half of the year.
\subsubsection{Long short-term memory networks approach}
\begin{table}[H]
\small
\centering
\begin{tabular}{lr}
\toprule
{} &  2020 \\
\midrule
ARCHITECTURE &  0.51 \\
BANKS        &  1.27 \\
CHEMISTRY    & -0.86 \\
CLOTHES      &  0.11 \\
ENERGY       & -2.11 \\
FOOD         &  0.79 \\
FUELS        &  1.50 \\
GAMES        & -1.25 \\
INFORMATICS  &  0.96 \\
MEDIA        &  0.73 \\
MINING       &  1.13 \\
MOTO         &  1.10 \\
PHARMA       &  0.78 \\
REAL EST     & -0.44 \\
\textit{other}        & -1.26 \\
\textbf{PORTFOLIO}    & -0.34 \\
\bottomrule
\end{tabular}
\caption{Sharpe ratios for LSTM approach based on 2020 recession trading interval}
\label{tab:9}
\end{table}
\noindent Although the overall portfolio ended up in a negative Sharpe ratio (Table \ref{tab:9}), more than half of sectors' sub-portfolios manage to overperform the risk-free rate of $0.5\%$. Nevertheless, the usual set of time-series plots needs to be included for further conclusions.
\begin{figure}[H]
\centering
\includegraphics[scale=0.3]{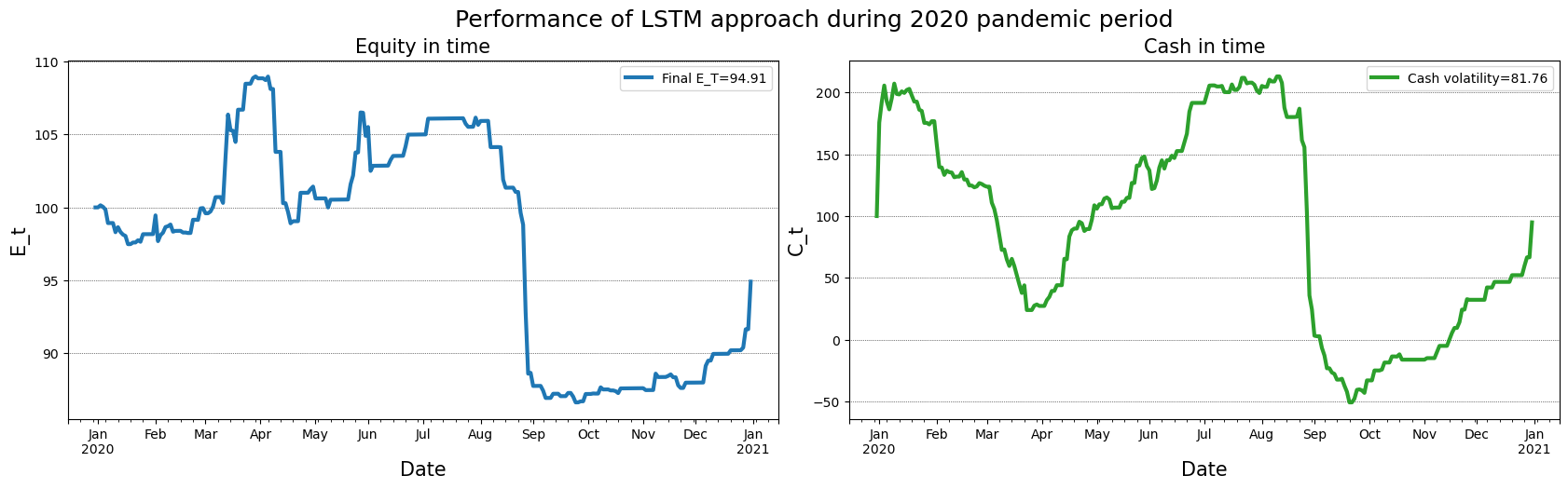}
\caption{Equity and cash levels throughout the 2020 recession trading interval for LSTM approach}
\label{fig:lstm_results_years_covid}
\end{figure}
\begin{figure}[H]
\centering
\includegraphics[scale=0.3]{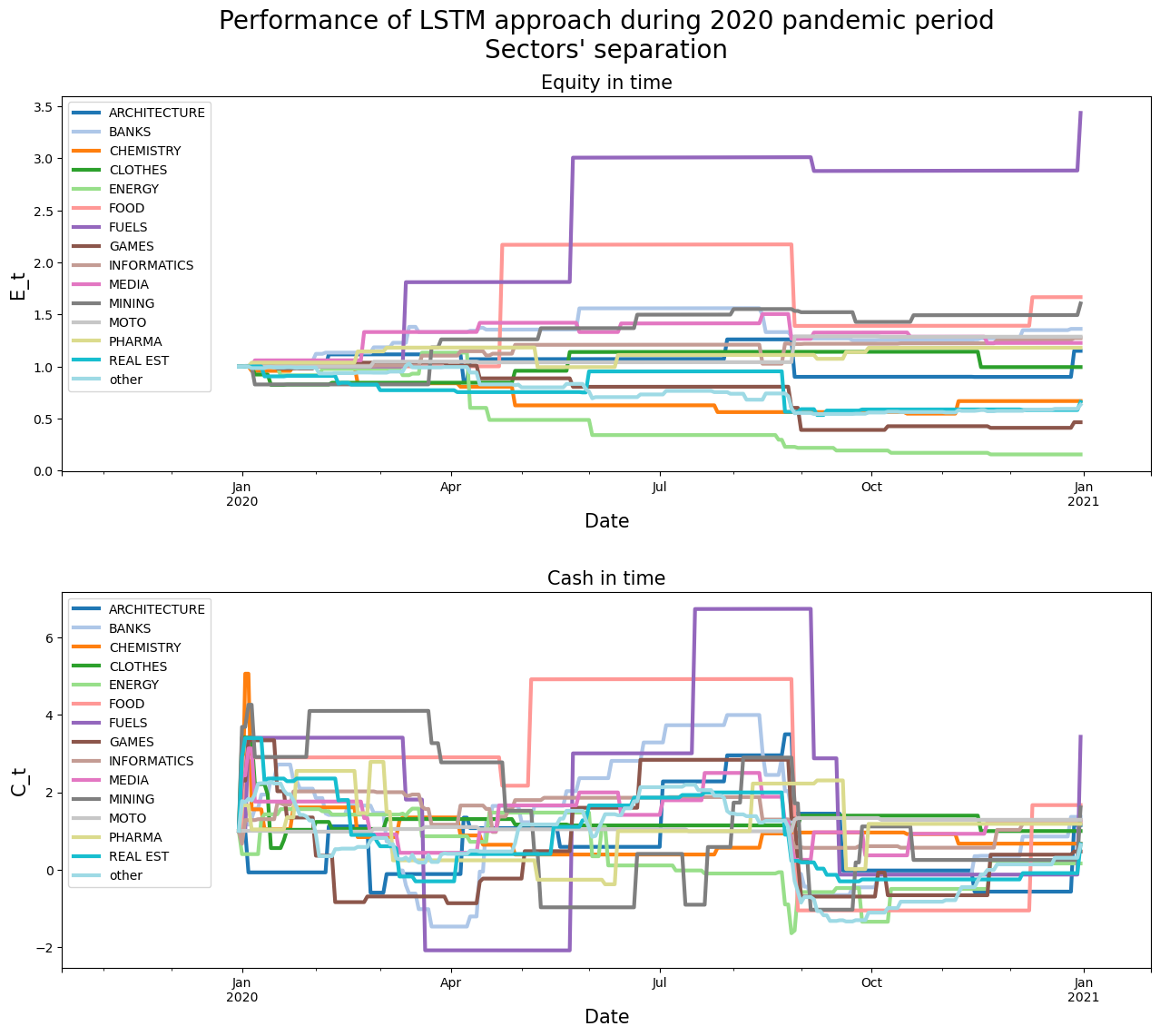}
\caption{Equity and cash relative levels throughout the 2020 recession trading interval for LSTM approach- sectors' separation}
\label{fig:lstm_results_sectors_years_covid}
\end{figure}
\noindent Like in the PCA approaches, LSTM turns out not to be profitable in 2020 even though the final results are far better than dimensionality reduction techniques (Figure \ref{fig:lstm_results_years_covid}). This time the strategy was actually profiting in the first half of the year. Notice how there is less cash fluctuations than during 2017-2019- since the market is behaving similarly, same types of transactions (close/long) are being performed during common time-intervals. Looking at the lower Figure (\ref{fig:lstm_results_sectors_years_covid}) it can be seen that practically only the games' sector manage to make relevant, positive returns relative to its initial capital.\\
Although the results were not so low as with approaches above, trading in 2020 using LSTM did not manage to make any reliable profits. Major drops when the pandemic in Poland started were potentially highlighted by the long-term memory of running recurrent network resulting in a lot of losses in the last months of the year Since transitions between LSTM's $\beta$s are smoother than with dimensionality reduction methods (due to an increasing increasing number of inputs which are then slowly and partially forgotten rather than a strict moving window), such disruptions were not destructive to equity level.
\subsubsection{Exchange Traded Funds of market indices approach}
Last but not least, let us see whether actual indices can perform better in recession scenarios after being the worst ones in standard market conditions.\\
\textbf{Existing ETFs sub-approach}\\
\begin{table}[H]
\small
\centering
\begin{tabular}{lr}
\toprule
{} &  2020 \\
\midrule
ARCHITECTURE &  -inf \\
BANKS        & -2.56 \\
CHEMISTRY    &  -inf \\
CLOTHES      &  0.90 \\
ENERGY       &  1.33 \\
FOOD         &  0.98 \\
FUELS        &  0.98 \\
GAMES        &  1.09 \\
INFORMATICS  &  1.52 \\
MEDIA        &  0.98 \\
MINING       &  0.82 \\
MOTO         &  -inf \\
PHARMA       &  1.18 \\
REAL EST     & -1.87 \\
\textit{other}        & -0.30 \\
\textbf{PORTFOLIO}    &  0.56 \\
\bottomrule
\end{tabular}
\caption{Sharpe ratios for existing ETFs' approach based on 2020 recession trading interval}
\label{tab:10}
\end{table}
\noindent Trading with existing funds during 2020 resulted in a positive Sharpe ratio of the overall portfolio (Table \ref{tab:10}). Two out of $15$ sectors (including \textit{other}) did not make any transactions resulting in no $\mathcal{S}$. Other groups were mostly outperforming the risk free-rate $r_f$ with an average ratio of $0.45$.
\begin{figure}[H]
\centering
\includegraphics[scale=0.3]{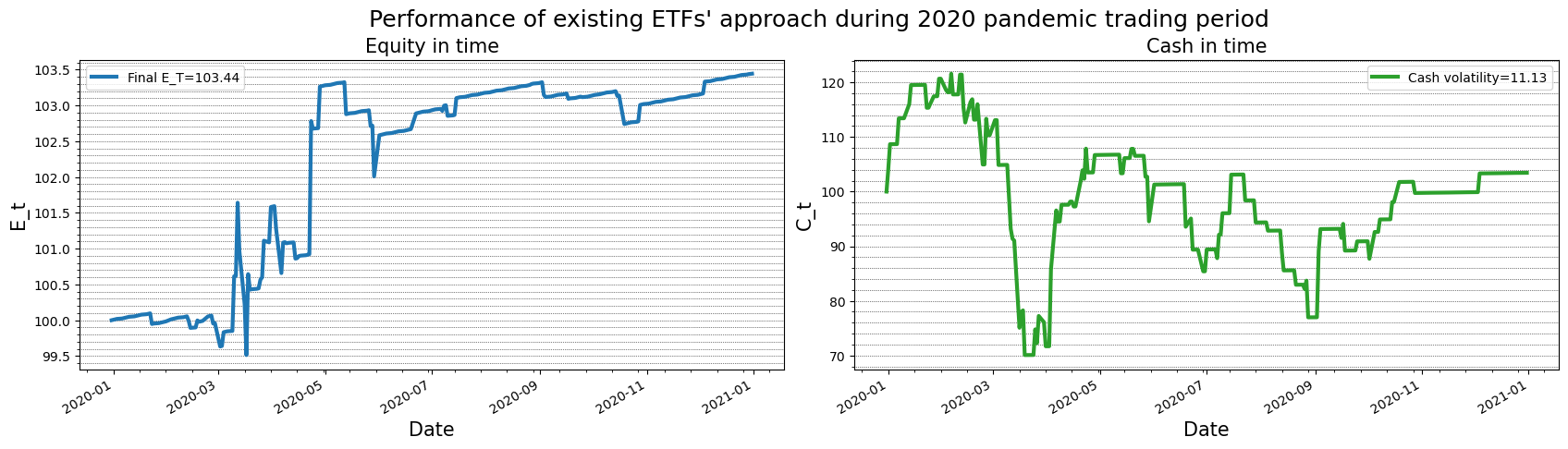}
\caption{Equity and cash levels throughout the 2020 recession trading interval for existing ETFs' approach}
\label{fig:etf1_results_years_covid}
\end{figure}
\begin{figure}[H]
\centering
\includegraphics[scale=0.3]{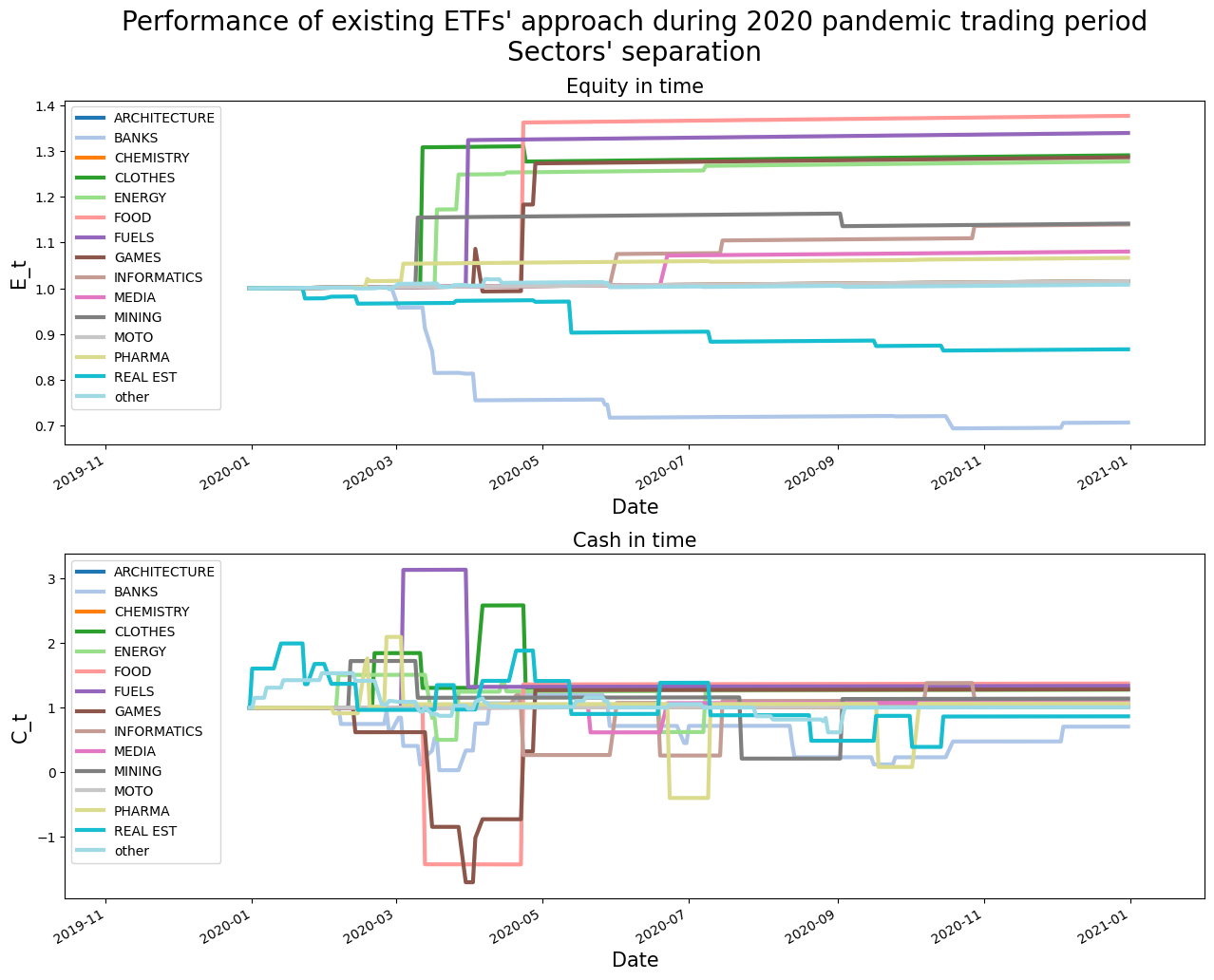}
\caption{Equity and cash relative levels throughout the 2020 recession trading interval for existing ETFs' approach- sectors' separation}
\label{fig:etf1_results_sectors_years_covid}
\end{figure}
\noindent As both Figures (\ref{fig:etf1_results_years_covid} and \ref{fig:etf1_results_sectors_years_covid}) show, technique of using just three explanatory variables to replicate individual stocks is the first profitable approach in 2020's conditions. Besides a fluctuation at the start, $E_t$ is increasing fairly stable resulting in $E_T=103.44$ (and thus a return of $3\%$). Also the cash level $C_t$ can be seen as not volatile. Within sectors, profit were mainly made in the first half of 2020 advantaging from major recession movements.\\
\textbf{Artificial ETFs sub-approach}\\
\begin{table}[H]
\small
\centering
\begin{tabular}{lr}
\toprule
{} &  2020 \\
\midrule
ARCHITECTURE &  0.98 \\
BANKS        &  0.73 \\
CHEMISTRY    & -0.98 \\
CLOTHES      &  0.76 \\
ENERGY       &  1.08 \\
FOOD         &  -inf \\
FUELS        &  1.64 \\
GAMES        &  0.11 \\
INFORMATICS  &  1.00 \\
MEDIA        &  0.81 \\
MINING       &  0.98 \\
MOTO         &  0.67 \\
PHARMA       & -1.51 \\
REAL EST     & -0.85 \\
\textit{other}        & -0.44 \\
\textbf{PORTFOLIO}    &  0.68 \\
\bottomrule
\end{tabular}
\caption{Sharpe ratios for artificial ETFs' approach based on 2020 recession trading interval}
\label{tab:11}
\end{table}
\noindent All ratios of the artificial ETFs made of sector indices are fairly similar to ones achieved by just $3$ explanatory variables (Table \ref{tab:11}). Again, daily returns of the overall portfolio manage to outrun government bonds' yield of $0.5\%$.
\begin{figure}[H]
\centering
\includegraphics[scale=0.3]{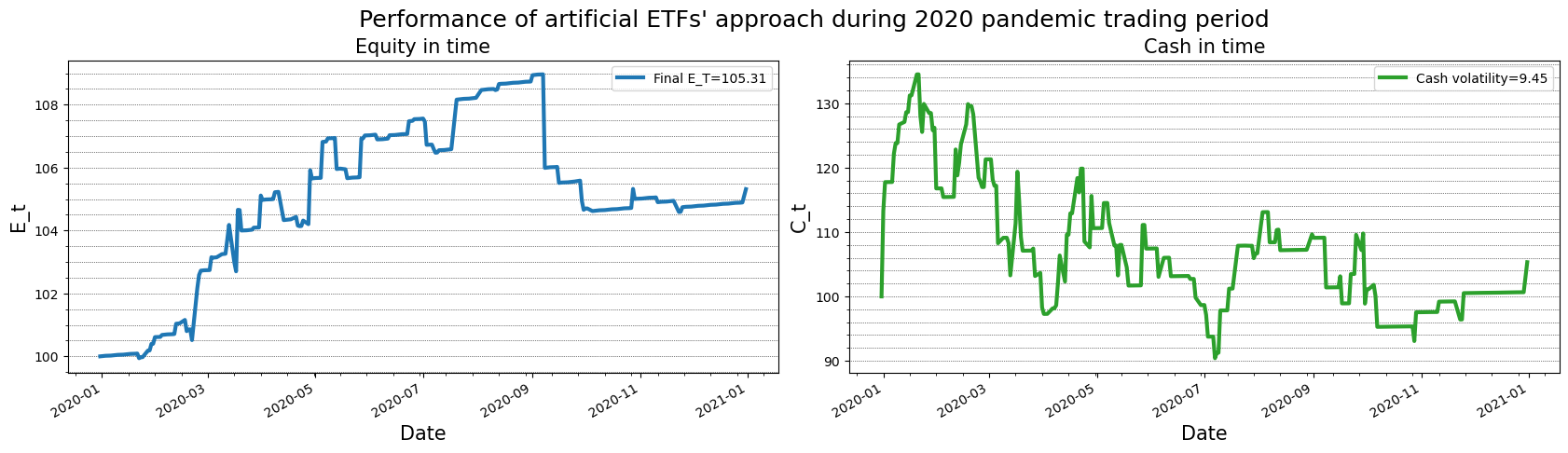}
\caption{Equity and cash levels throughout the 2020 recession trading interval for artificial ETFs' approach}
\label{fig:etf2_results_years_covid}
\end{figure}
\begin{figure}[H]
\centering
\includegraphics[scale=0.3]{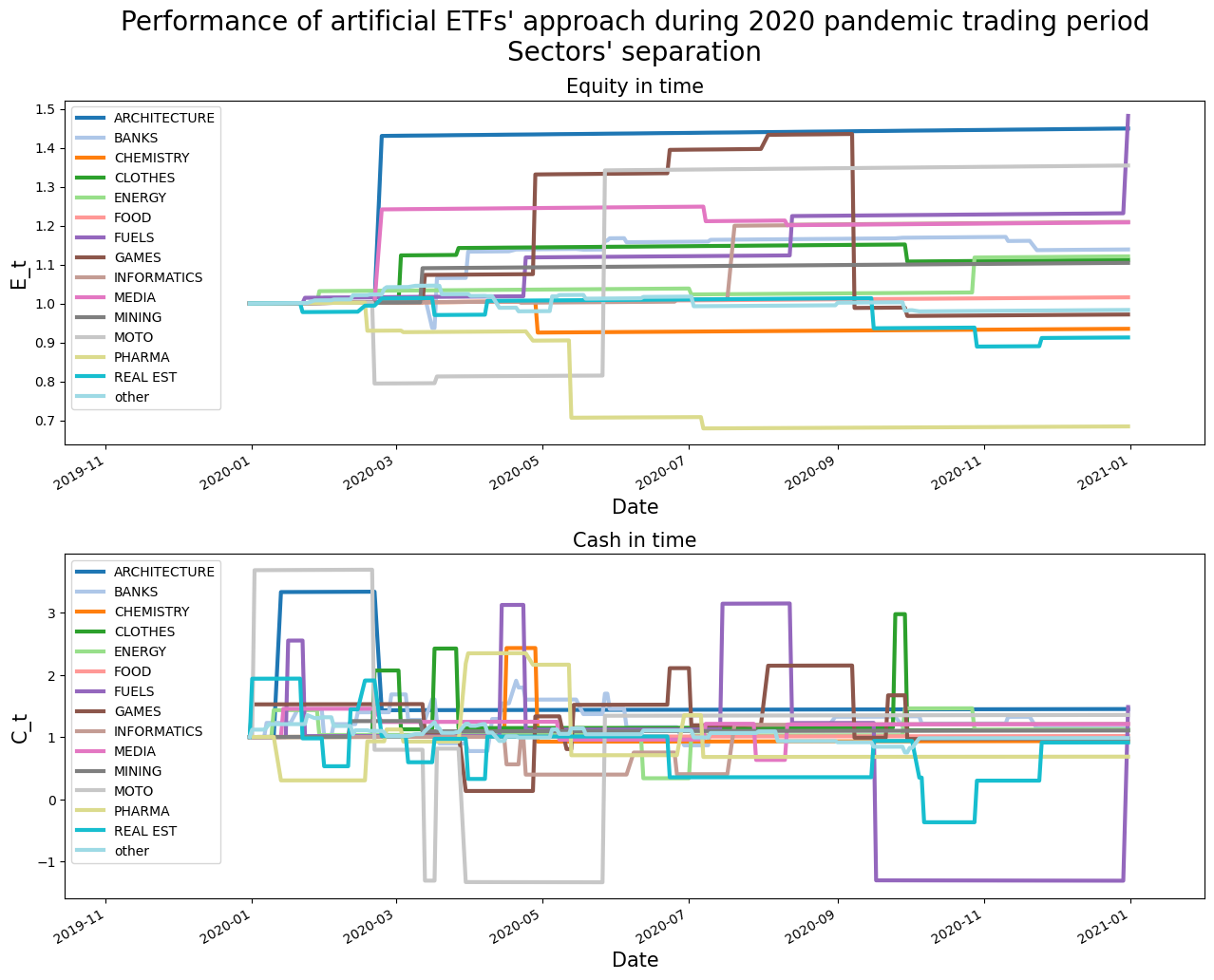}
\caption{Equity and cash relative levels throughout the 2020 recession trading interval for artificial ETFs' approach- sectors' separation}
\label{fig:etf2_results_sectors_years_covid}
\end{figure}
\noindent The use of more explanatory variables increased $E_t$ of ETFs-related method- final profit of around $5$ PLN was made within a year giving a $5\%$ return (Figure \ref{fig:etf2_results_years_covid}). In opposition to existing ETFs' sub-technique, this time bigger trades are made not only in the first half of the year but throughout entire 2020. As can be seen only the pharmaceutical stocks' ended up with negative annual return.\\
Since considered ETFs related approaches use market indices instead of individual stocks they are less sensitive to historical fluctuations. Their results are practically as good as in standard economic conditions which cannot be said for other techniques. Distinguishing between sub-techniques, the latter one with more factors seem to give slightly better.
\chapter{Conclusions}
We presented a systematic approach to Pairs Trading --- a popular Statistical Arbitrage technique. In the usual approach pair of similarly behaving assets is selected and trades are performed on an offset portfolio of the two- we decided to substitute the second equity with a linear combination of risk (systematic) actors replicating the first (main) one. The aim then was to analyse residuals of main asset and its corresponding systematic components' approximation. Based on their potential mean-reversion properties, trading signals were determined trying to profit from potential technical mispricings. The selection of risk factors was done using three techniques: creating them with Principal Components Analysis (PCA) from eigenvectors of returns' empirical correlation matrix, using real market indices, and as a new method that was not covered in the literature of the subject yet: considering Long short-term memory networks (LSTMs).\\
There are two main contributions made within the paper: re-defining and testing already discussed techniques of PCA and market indices~\cite{main_paper} on a far less developed equities market of Poland; and, as already mentioned, the introduction of new deep learning based approach of deriving the risk factors. Applying techniques of Statistical Arbitrage presented by Avellaneda and Lee~\cite{main_paper} to polish stock exchange required switching from around $500$ stocks' portfolios to just $60$ most influential companies gathered by two main indices: \textit{WIG20} and \textit{mWIG40}. Since the spectrum of exchange traded funds of market indices is also way narrower than in the US, some simplifications had to be made for the ETFs approach. Two sub-techniques were discussed: one using only existing funds to replicate and other relying on artificial ones created as direct copies of sector indices. Additional market factors such as transaction costs or risk free rate were also adjusted from 2008 paper\cite{main_paper} for better reality matching. Second contribution i.e. LSTM approach of replicating stocks' returns was constructed to check whether more flexibility can be gained by not limiting oneself to common market factors and instead creating unique portfolios on a basis of singular stocks. Network was trained to provide the most appropriate weightings of all companies' shares to approximate returns of the main one. Since it is possibly the first documented use of LSTMs in this, exact context- some steps such as the search for the most optimal hyperparameters were skipped to focus on initial potential of the approach. Any positive results coming from using the technique should be seen as a encouragement for the start of a new branch and further development of the strategy.\\
On the performance side, two trading intervals were considered for the backtesting of all $3$ approaches: 2017-2019 representing standard market conditions and highly recessive 2020. During the first, $3$ years period all three techniques manage to profit with PCA and LSTM approaches achieving up to $2.63$ and $2.09$ annualized Sharpe ratios during the first two years consecutively. Both ETF sub-methods performed worse ending up with around $5\%$ of a $3$-years return (comparing it to $>10\%$ achieved by previously mentioned approaches). Understandably, the use of just three real exchange traded funds provided the most stable results throughout the main trading period. Although a lot of assumptions were changed compared to ones made by authors we refer to\cite{main_paper}, both PCA and ETF approaches' Sharpe ratios can be seen as sufficiently comparable- worse performance of the latter method was also proven in their case. In the additional, 2020 trading PCA and LSTM strategies failed to make any profits with the first method ending up with significant losses. On the other hand, funds' sub-methods managed to made comparable returns (of around $5\%$) to ones from 2017-2019 thus outperforming other techniques. One of the reason for such switch may be that both PCA and LSTM techniques rely on all stocks directly and thus are far more sensitive to prices' rapid, recessive movements.\\
Focusing on the performance of new LSTM approach we can say that it did not manage to outperform PCA technique during 2017-2019 period but provided positive and comparable results. Both methods failed during 2020 recession period but recurrent neural network remained more stable. It is important to recall that the approach was purely constructed by the author of this paper and thus not optimized to its full potential. Note that in opposition to PCA, LSTM takes time into account in promising a new set of replication weightings and does not rely on returns' normality when determining potential dependencies. It is thus more flexible and can also be optimized on a basis of different, more sophisticated loss functions (we used regularized MSE). For all of the reasons above, we access LSTM approach as a promising alternative for the future of Statistical Arbitrage and highlight the need for a more detailed analysis of method's capabilities especially since concept of treating more recent historical data as more important but at the same time gathering some crucial informations from earlier periods is very intuitive for all trading activities.\\
There is also much more to the theory of systematic Pairs Trading approach: our predecessors successfully added volume as an additional factor making trading requirements more realistic; different processes such as mean-reverting jump-diffusion model can also substitute used Ornstein-Uhlenbeck one. But most importantly, already at this level of expertise we can see that the transition between theory and practical results is working admirable in various market conditions.


\newpage
\begin{thebibliography}{9}
\bibitem{pt_history} E. Gatev, \textit{Pairs Trading: Performance of a Relative Value Arbitrage Rule} (2006)
\bibitem{emh} E. F. Fama, \textit{Efficient Capital Markets: A Review of Theory and Empirical Work} (1970)
\bibitem{pedersen} L. H. Pedersen, \textit{Efficiently inefficient: how smart money invests and market prices are determined} (2015)
\bibitem{main_paper} M. Avellaneda and J-H. Lee- \textit{Statistical Arbitrage in the U.S. Equities Market} (2008)
\bibitem{fama} E. F. Fama and J. D. Macbeth- \textit{Risk, Return and Equilibrium: Empirical Tests} (1973)
\bibitem{capm} W. F. Sharpe- \textit{Capital Asset Pricing: a theory of market equilibrium under conditions of risk} (1964)
\bibitem{apt} S. A. Ross- \textit{The arbitrage theory of capital asset pricing} (1976)
\bibitem{bs} F. Black and M. Scholes- \textit{The Pricing of Options and Corporate Liabilities} (1973)
\bibitem{pca1} L. Laloux, P. Cizeau and M. Potters- \textit{Random matrix theory and financial correlations} (2000)
\bibitem{pca2} V. Plerou- \textit{Random matrix approach to cross correlations in financial data} (2002)
\bibitem{gd} S. Boyd and L. Vandenberghe — \textit{Convex Optimization} (2004)

\bibitem{nn} C. M. Bishop — \textit{Pattern Recognition and Machine Learning} (2006)

\bibitem{nn2} K. Hornik — \textit{Approximation Capabilities of Multilayer Feedforward Networks} (1991)


\end{thebibliography}
\end{document}